\documentclass[11pt,twoside]{report}
\usepackage{multicol,tabularx,array}

\pdfoutput=1
\usepackage[fit]{truncate}
\usepackage[T1]{fontenc}

\newcommand{\BLKP}{
  \ifthenelse{\isodd{\value{page}}}{\relax}{\mbox{}\thispagestyle{empty}\newpage}}
\usepackage{geometry}
\usepackage{fancyhdr}
\usepackage{refcount}
\setrefcountdefault{0}
\usepackage{pdfpages}
\usepackage[bookmarks, colorlinks=true, linktoc=page, pdftex, linkcolor=red, citecolor=red, urlcolor=red]{hyperref} 
\usepackage{hyperref}  
\fancypagestyle{ARTTITLE}{%
\fancyhf{} 
\lhead{\footnotesize{Study on the career trajectories of people with a working experience at CERN, C. Bianchin et al.\\ CERN Yellow Reports: Monographs, CERN-2019-004 (CERN, Geneva, 2019)}} 
\lfoot{\footnotesize{2519-8041-- \copyright\space CERN, 2019. Published by CERN under the Creative Common Attribution CC BY 4.0 Licence.}}
\rfoot{\thepage\hspace*{3mm}}
 
} 
\fancypagestyle{ARTBODY}{%
\fancyhf{} 
\fancyfoot[C]{\thepage\hspace*{3mm}}
\fancyhead[RE]{\textsc{\ARTauthor}\hspace*{3mm}}
\fancyhead[LO]{\hspace*{3mm}\truncate{.95\headwidth}{\textsc{\ARTtitle}}}
 
} 

\begin{document}
\setlength{\fboxsep}{0pt}
\setlength{\fboxrule}{0.0pt}
\include{title}
\pagestyle{plain}
\pagenumbering{roman}
\thispagestyle{empty}
\setlength{\unitlength}{1mm}
\begin{picture}(0.001,0.001)
\put(-8,8){\large CERN Yellow Reports: Monographs}
\put(120,8){\large CERN-2019-004}

\put(-5,-60){\LARGE\bfseries
                                              Study on the career trajectories}
\put(-5,-70){\LARGE\bfseries  of people with a working experience at CERN}

\put(-5,-90){\Large C. Bianchin}
\put(0,-94){\small Volkswagen Financial Services AG, Gifhorner Str. 57, Braunschweig, Germany,}
\put(0,-98){\small previously at Wayne State University, Detroit, USA}

\put(-5,-105){\Large P. Giacomelli}
\put(0,-109){\small INFN, Sezione di Bologna, viale B. Pichat 6/2, 40127 Bologna, Italy}

\put(-5,-120){\Large L. Iconomidou-Fayard}
\put(0,-124){\small LAL, CNRS/IN2P3, Universit\'{e} Paris-Sud, Paris-Saclay, 91440 Orsay, France}

\put(-5,-135){\Large J. Niedziela}
\put(0,-139){\small CERN, CH-1211 Geneva 23, Switzerland}

\put(-5,-150){\Large B. Sciascia}
\put(0,-154){\small INFN, Laboratori Nazionali di Frascati, via E. Fermi 33, 00044 Frascati, Italy}

\put(65,-250){\includegraphics{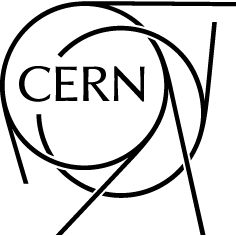}}

\end{picture}
\newpage

\thispagestyle{empty}
\mbox{}
\vfill

\begin{flushleft}
CERN Yellow Reports: Monographs\\
Published by CERN, CH-1211 Geneva 23, Switzerland\\[3mm]

\begin{tabular}{@{}l@{~}l}
  ISBN & 978-92-9083-543-1 (paperback) \\
  ISBN & 978-92-9083-544-8 (PDF) \\
  ISSN & 2519-8068 (Print)\\ 
  ISSN & 2519-8076 (Online)\\ 
  DOI & \url{http://dx.doi.org/10.23731/CYRM-2019-004}\\
\end{tabular}\\[3mm]
Accepted for publication by the CERN Report Editorial Board (CREB) on 8 September 2019\\[1mm]
Available online at \url{http://publishing.cern.ch/} and \url{http://cds.cern.ch/}\\[3mm]

Copyright \copyright{} CERN, 2019\\[1mm]
\raisebox{-1mm}{\includegraphics[height=12pt]{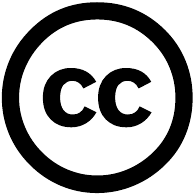}}
Creative Commons Attribution 4.0\\[1mm]
Knowledge transfer is an integral part of CERN's mission.\\[1mm]
CERN publishes this volume Open Access under the Creative Commons Attribution 4.0 license\\
(\url{http://creativecommons.org/licenses/by/4.0/}) in order to permit its wide dissemination and use.\\
The submission of a contribution to a CERN Yellow Report series shall be deemed to constitute the contributor's agreement to this copyright and license statement. Contributors are requested to obtain any clearances that may be necessary for this purpose.\\[5mm]

This volume is indexed in: CERN Document Server (CDS)\\[5mm]

This volume should be cited as:\\[1mm]

Study on the career trajectories of people with a working experience at CERN, C. Bianchin et al.\\  CERN Yellow Reports: Monographs, CERN-2019-004 (CERN, Geneva, 2019), \url{http://dx.doi.org/10.23731/CYRM-2019-004}\\[3mm]

\end{flushleft}

\pagenumbering{roman}

\pagestyle{plain}
\setcounter{page}{3}

\begin{center}
\mbox{}\\[3cm]
\bfseries\LARGE Abstract\\[1cm]
\end{center}

\noindent
This document describes the results of a study, aiming to measure the impact of CERN and of its environment on the career of people who worked at the laboratory. The data was collected using two on-line questionnaires, launched in 2016 and 2017, targeting experimentalists and theorists, respectively. The mandate, the methodology followed, the questionnaires and the analysis of the data collected are presented.
\vspace*{2.0cm}

\providecommand{\Titi}{Table of contents}
\providecommand{\Auti}{}
\providecommand{\Refi}{pdffinal/1_TOC}

\providecommand{\Titii}{Study on the career trajectories of people with a working experience at CERN}
\providecommand{\Autii}{}
\providecommand{\Refii}{pdffinal/2_Text}

\cleardoublepage

\newpage\BLKP

\tableofcontents

\chapter{Introduction}
\label{sec:introduction}
\pagenumbering{arabic}
\setcounter{page}{1}
\section{Overview}
In the last few decades, several thousands of people have worked at CERN, the European Organization for Nuclear Research, or in collaboration with CERN departments. Especially in the past 15 years, the advent of the Large Hadron Collider (LHC) boosted this population, with the big LHC experiments (ALICE~\cite{alice-1,alice-2}, ATLAS~\cite{atlas-1,atlas-2}, CMS~\cite{cms-1,cms-2}, and LHCb~\cite{lhcb-1,lhcb-2}) representing the largest fraction of CERN Users~\footnote{In the following we will use \textit{CERN users} to refer generically  to both CERN staff members and CERN Users.}.


The purpose of the study described in this document is to investigate the effect of CERN on the careers of the people who are or have been in close contact with it. Many of these persons are relatively young in terms of both age and working experience and it is therefore very interesting to monitor the evolution of their career trajectories. The specificity of their education and training on cutting-edge fundamental research and technologies gives this community an original feature, as far as their integration to the broad job market is concerned~\cite{careers-1,careers-2}. 
This study is part of a broader action that CERN management started in 2016 to establish and maintain links~\footnote{The \textit{CERN Alumni} programme, \url{https://alumni.cern}.} between people who have worked at CERN and left to pursue a career in a different domain and those who still work in connection with CERN. Of the former, a large fraction has gone to work in many different fields of activity outside of research in particle physics. 

In the following, details of the mandate of a study group to collect this information will be given, together with the tools developed to harvest the data.

\section{Mandate}

In 2016 the Director-General of CERN, Fabiola Gianotti, appointed a {\it{Students Career Study Group}} (referred to as the Study Group in the following) composed of a representative of each of the four main LHC experiments, appointed by the spokespersons of the ALICE, ATLAS, CMS and LHCb collaborations. The goal was to collect and provide information about the careers of students who worked on their theses in the LHC experiments. Similar studies had been made in the past in some of the LEP and LHC experiments; these were mainly based on questionnaires sent to the team leaders of the different institutes of the collaborations. The reason for redoing the work was that most of the previous surveys were quite old or based on small statistical samples, or both. The new study was aimed at collecting a large sample of up-to-date information from all the experiments. The mandate suggested that the Study Group define a list of questions to distribute through an anonymous questionnaire to the members of all LHC collaborations, analyse and interpret the collected data, present the final results to the CERN Council and document them in a written report. 

\section{Preparation of the surveys}\label{sec:survey}

Initially, the study aimed to address only young physicists, PhD students, who had left the high-energy physics (HEP) community, to give a quantitative measurement of  the value of the  education and skills acquired at CERN in finding jobs in other domains. This information is of prime importance in order to evaluate the impact and role of the CERN culture on the wider jobs market.
A first version of the online questionnaire was prepared, sent and answered by 282 people (out of about 470 contacts). The results were presented at the CERN Council meeting in December 2016.

The experience of the first questionnaire indicated the potential interest of collecting information from a wider population and, also, deepen and customize the proposed options in some questions.
Consequently, in agreement with CERN's Director-General, it was decided to enlarge the study to all persons who had been or were still involved with CERN, without any particular restrictions. Moreover, in addition to questions asked to all respondents, sets of questions depending on whether the person who answered was still working at CERN or not were proposed when filling the questionnaire.

The questionnaire addressed various professional and sociological aspects: age, nationality, education, domicile and workplace, time spent at CERN, current position, and satisfaction with the CERN environment. Additional points were specific to those who were no longer CERN users, in relation to their current situation and types of activity.  The second version of the questionnaire was opened for about 4 months and obtained 2692 answers. 

During the phase in which the second survey was opened, in agreement with the CERN Theory Department, it was decided to prepare a dedicated questionnaire specifically designed to match the typical career paths of theoretical physicists. In addition to the main stream of questions discussed, this questionnaire addressed to theorists included topics that helped to investigate the role that CERN plays for this community. 

Overall, the delicate point of this poll is the inherent risk of bias. These may originate from the formulation of the questions, or might also depend on the number of respondents to the questionnaire, if not all the concerned people respond. In practice, only 20-30\% of the targeted populations answered, depending on the community addressed (experiment or theory; current or past users). Therefore, this risk may be substantial. As no corrections for possible biases have been applied to the data,  the results of the poll cannot be considered representative of the whole CERN population but only a snapshot of opinions of the relatively large sample of people who responded to the questionnaire. 
In the following chapters, the paths followed in order to contact the largest possible number of people for each community will be detailed.

Chapter 2 describes the structure of the questionnaire sent to the experimentalists and the analysis of the results. Chapter 3 presents the Theory questionnaire and its results. Chapter 4 draws some conclusions from both studies and presents some ideas on how to improve similar studies in the future. More information from the surveys is reported in the appendices. 

\chapter{Analysis of the questionnaire for experimental physicists}
\label{ch:experimental}

\section{Collecting the contact information}
The experimental community at CERN is today mainly concentrated around the LHC collaborations. With the support of the spokespersons of the ALICE, ATLAS, CMS,  LHCb and NA62~\cite{na62-1,na62-2} collaborations, access to the email lists of all members of these experiments (including physicists, engineers, technicians and administrative personnel) was made possible and these were used to distribute the questionnaire.
It was a non-trivial task to collect the largest possible set of email addresses from former CERN users, remaining in compliance with privacy regulations. 
This was achieved thanks to the co-operation of the institute leaders of the experiments, who usually keep contact information of their former personnel. When the institutes were not allowed to provide  the contacts directly, for privacy reasons, they consented to send the questionnaire internally on CERN's behalf. 
The questionnaire was also posted on social networks, such as LinkedIn~\cite{linkedin}, and the preliminary list of CERN Alumni.
Reminders were sent at regular intervals to encourage and motivate the largest participation possible and the questionnaire was finally closed in October 2017. The largest number of replies came from advertising the questionnaire through the mailings lists of the LHC and NA62 collaborations.

\section{Content of the questionnaire}
Answers to the questionnaire were anonymous. The first part of the questionnaire was intended to collect biographical information such as age, sex, nationality and workplace. Following this, there were other questions concerning the level of education  (Master's, PhD, postdoc, etc.) and  background  (physics, engineering, etc.), giving the possibility to select subsamples of the full population with different experiences.

Next, a set of questions was aimed at gathering information on the career paths of the polled persons, such as permanent or temporary position, duration of affiliation with CERN and degree of satisfaction with their experience at CERN. Next, questions on the current working situation were proposed. One particular question distinguished between people still in HEP or outside of it. For the two groups, there were different follow-up questions.

Persons in HEP were asked how they saw their future in HEP, whether they want to stay in the field and whether their CERN experience was considered useful for eventually finding a job in a different area.
Persons who had left HEP were asked both about their career at CERN and their current situation. These questions addressed the primary goal of the study, namely to learn about the career of those who built a fraction of their working experience at CERN. The current field of employment, type of position and their working location have been investigated. They were asked their reasons for leaving CERN, the position they last occupied as a CERN user, and when and after how many contracts they had left. The impact the CERN experience had on their following careers in terms of developed skills, practical help for finding a new position and the relevance of CERN's reputation on their CVs was then investigated. 
The analysis of the collected answers is presented in the following sections.

\section{Demographic profile of the sample}

The total number of people who replied to the second questionnaire was 2692. Each question of the questionnaire could have a different number of answers since most of them were not mandatory or were not addressed to all cases. Moreover, some questions allowed more than one answer. 

\begin{figure}
\centering\includegraphics[page=54,width=.9\linewidth]{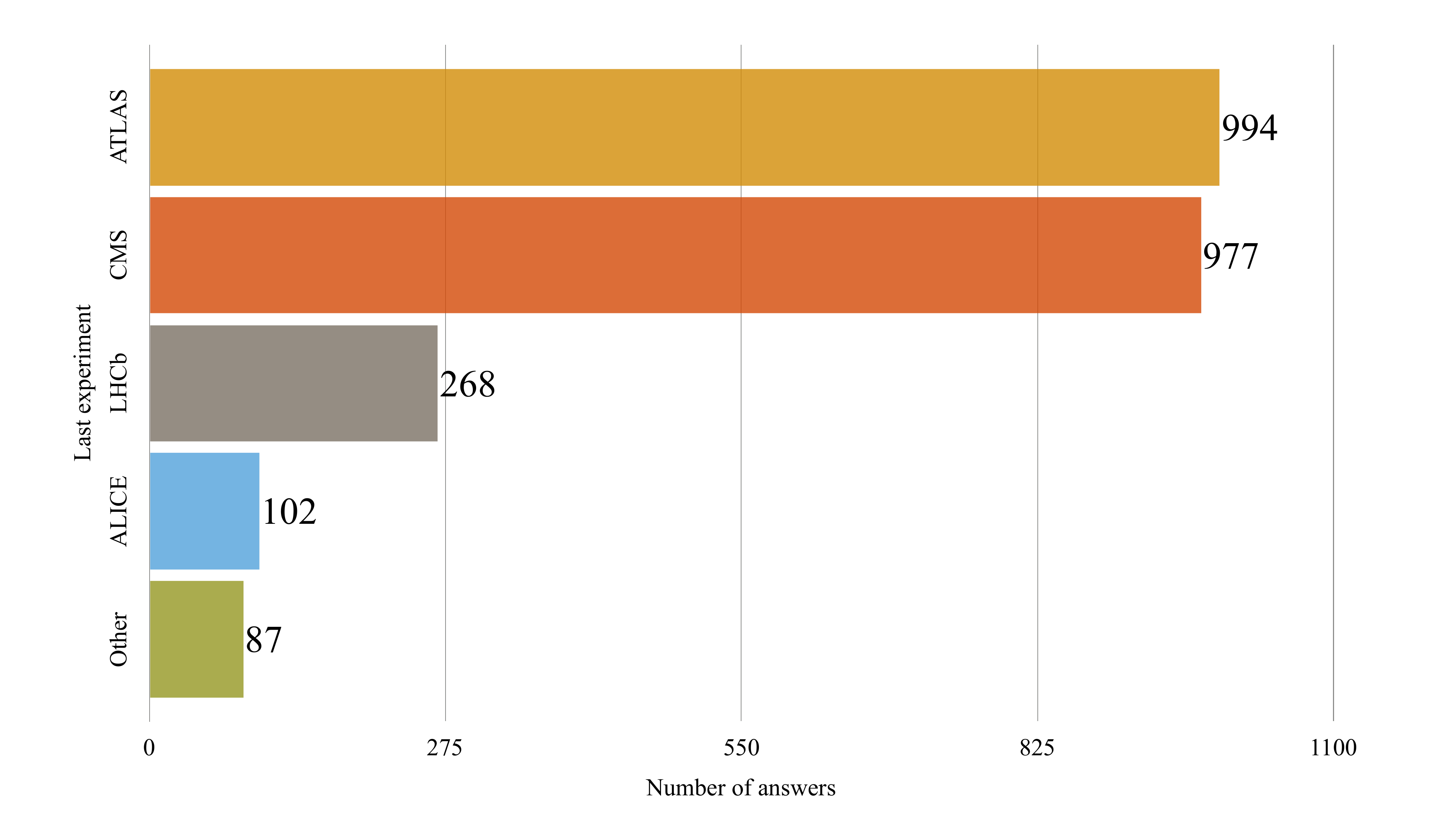}
\caption{Distribution of respondents among the various CERN experiments. The category 'Other' contains mainly members of the NA62 collaboration.}
\label{fig:experiments}
\end{figure}

\begin{figure}
\centering\includegraphics[trim={20cm 0 20cm 0},clip,page=1,width=.45\linewidth]{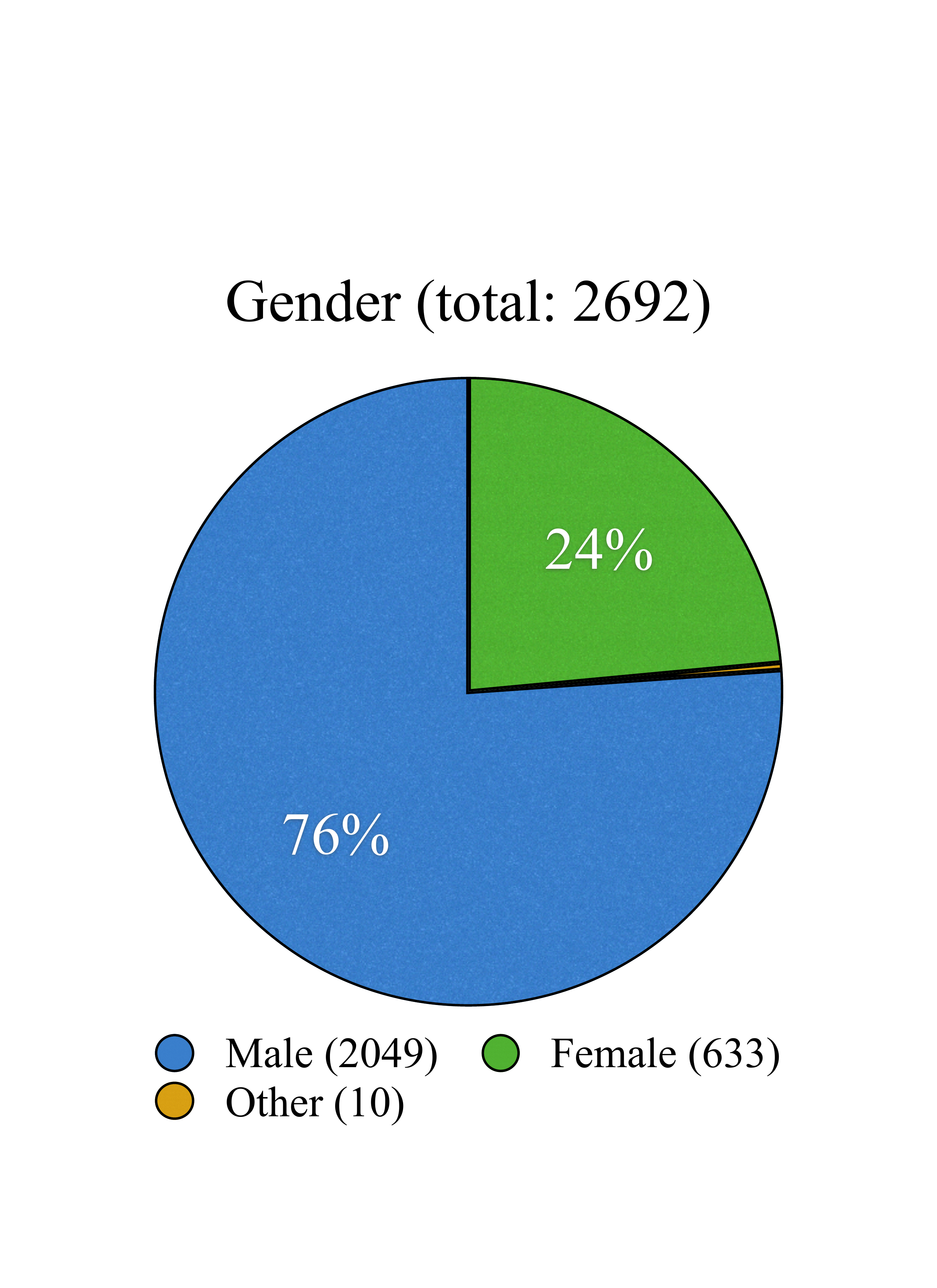}
\centering\includegraphics[trim={20cm 0 20cm 0},clip,page=58,width=.45\linewidth]{careers_plots.pdf}
\caption{Left: gender distribution of the people who completed the questionnaire. Right: fraction of people still working in the high-energy physics domain and people who left.}
\label{fig:gender_hep}
\end{figure}

Figure~\ref{fig:experiments} shows the number of respondents to the questionnaire from each experiment. As mentioned in Chapter \hyperref[sec:introduction]{1}, the answers received from the different experiments are not proportional to the size of the collaborations, which are approximately ordered as ATLAS $\approx$ CMS > ALICE > LHCb.  Almost 3/4 of the answers were from men and the remaining 1/4 from women (left panel of Figure~\ref{fig:gender_hep}). This subdivision reflects closely the gender distribution at CERN and in the LHC experiments. Because of the relatively easier connection with current CERN users than former users, the two groups represent 72\% and 28\% of the answers, respectively. In particular we have 757 answers from the latter group (right panel of Figure~\ref{fig:gender_hep}). As expected, most of the people who replied (85\%) already hold a PhD, or are in the process of obtaining one.

\begin{figure}
\centering\includegraphics[page=6,width=.9\linewidth]{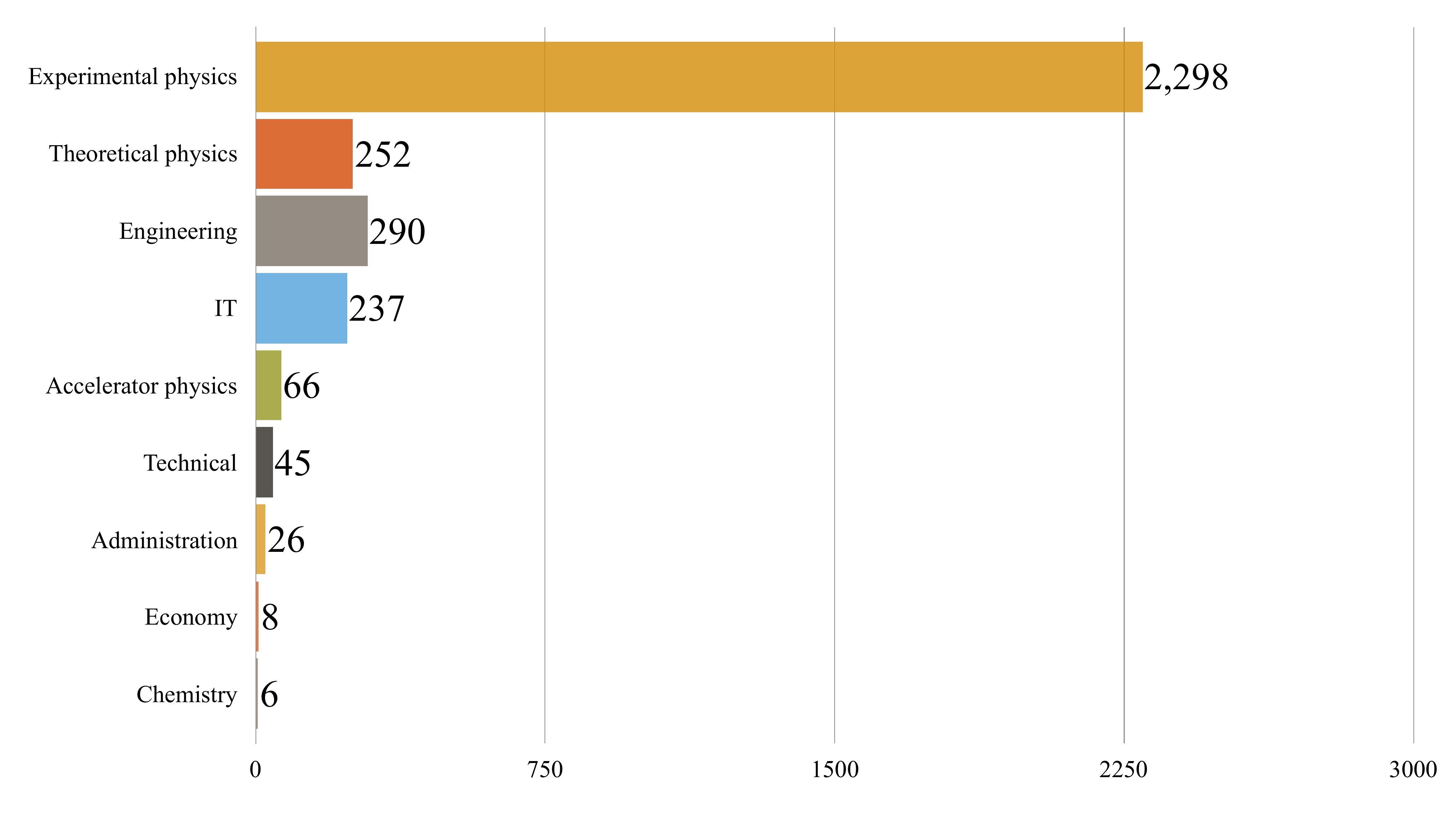}
\caption{Educational background of the people in the sample: several answers were allowed.}
\label{fig:background}
\end{figure}

For the question concerning educational background (Figure~\ref{fig:background}), several answers could be selected. The greatest proportion of respondents had a background in experimental physics. This reflects the fact that the questionnaire was distributed among contacts coming from the experiments. Of 252 people who answered that they also had a theory background, 217 were attached to one collaboration and 40\% had left HEP, compared with the 28\% found for the full sample.

\begin{figure}
\centering\includegraphics[page=3,width=.9\linewidth]{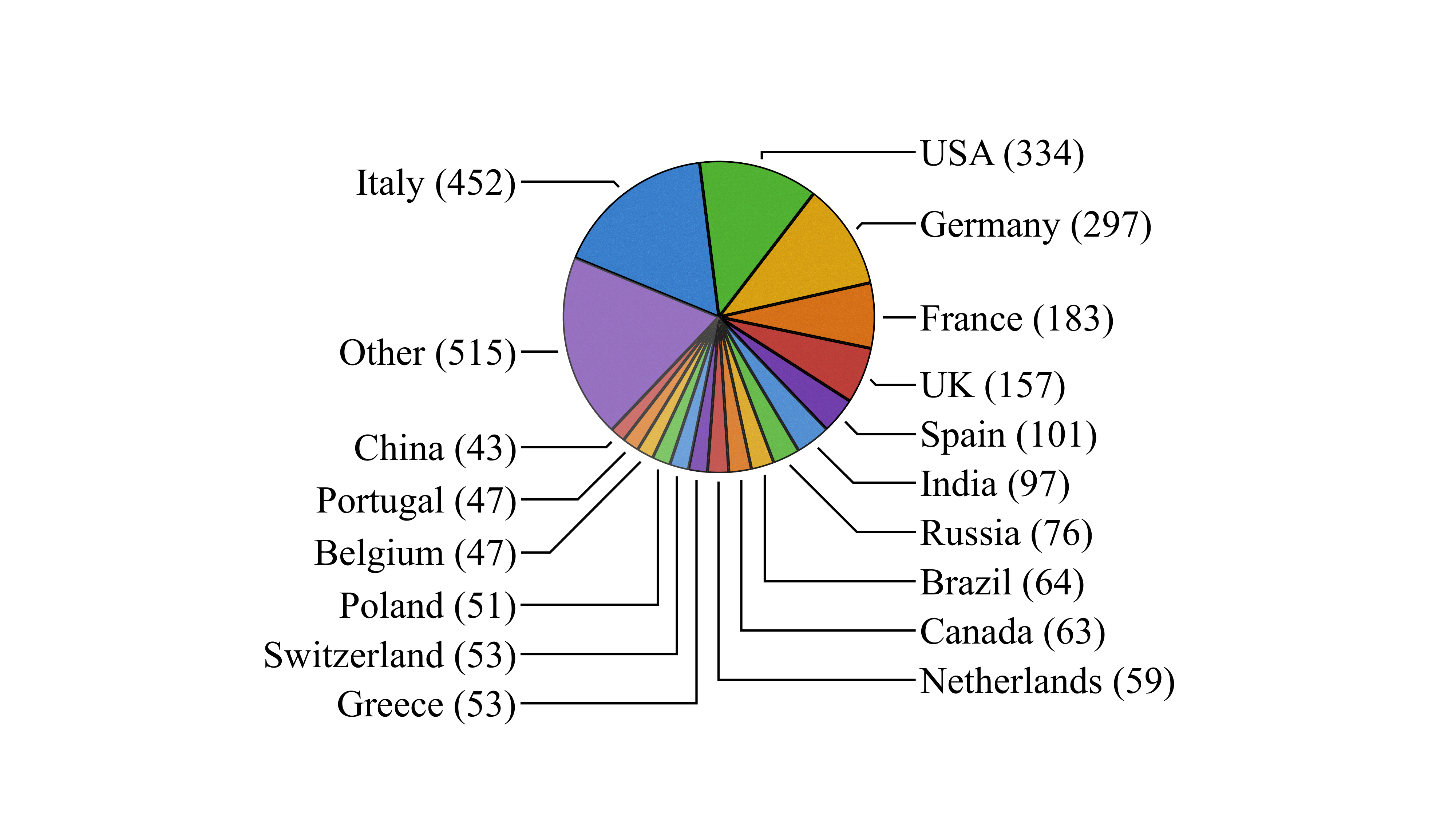}
\caption{Nationalities of the respondents to the questionnaire. Only countries with at least 40 respondents  are shown. 'Other' includes all countries with fewer respondents. (A full list of countries is given in Appendix A.)}
\label{fig:nationality}
\end{figure}

\begin{figure}
\centering\includegraphics[page=4,width=.9\linewidth]{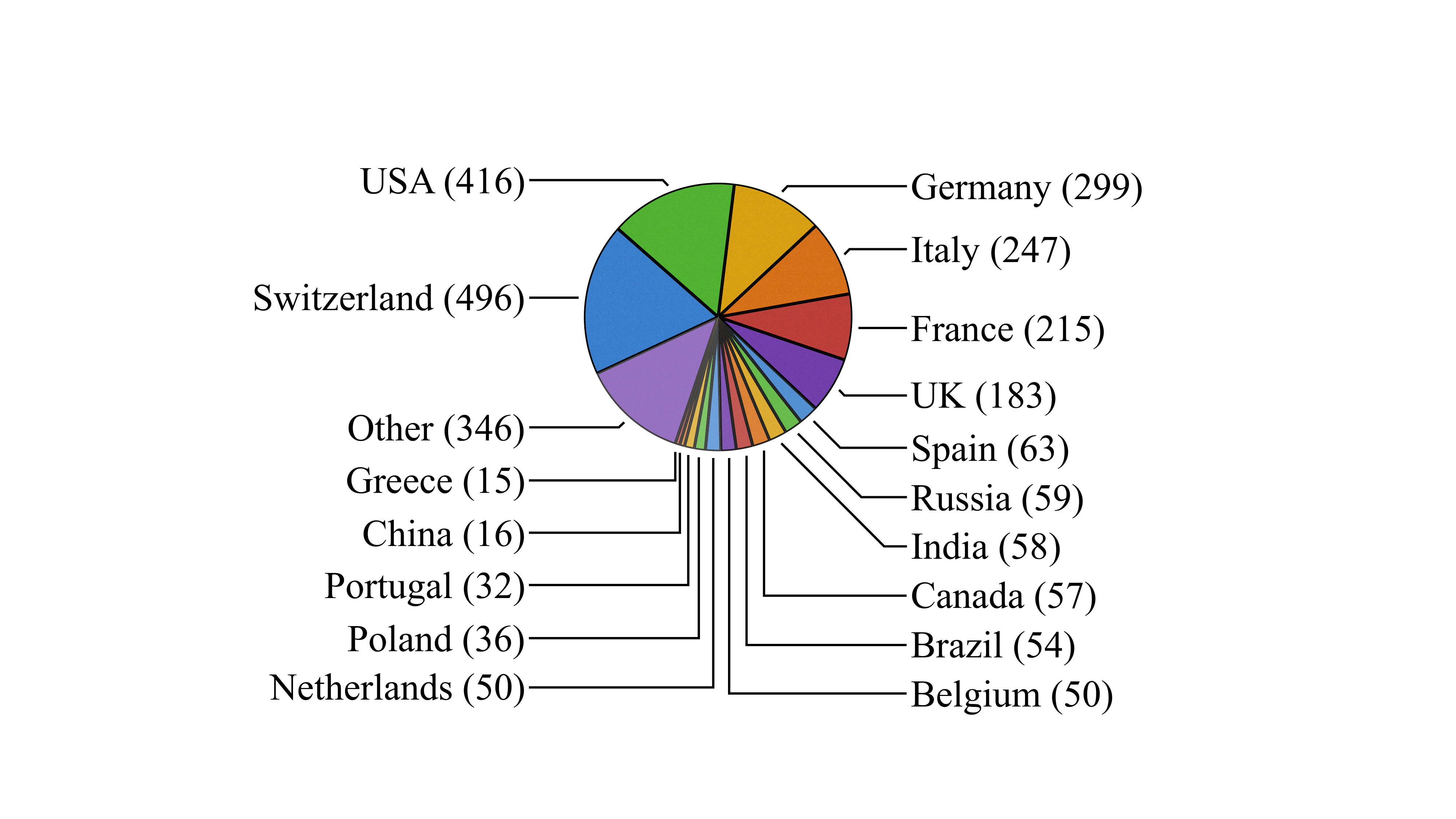}
\caption{Current countries of residence of all the respondents, at the time of responding. Only countries of residence for nationalities listed in Fig. \ref{fig:nationality} are shown. 'Other' includes all other countries. (A full list of countries is given in Appendix A.)}
\label{fig:residence}
\end{figure}

Eighty-four nationalities were represented in our sample; those with more than 40 answers are shown in Figure~\ref{fig:nationality}. About half of the sample was made up of Italians, Americans, and Germans. Figure~\ref{fig:residence} shows the country of current residence and work. 

\section{Age of the respondents}

\begin{figure}
\centering
\includegraphics[width=.9\linewidth]{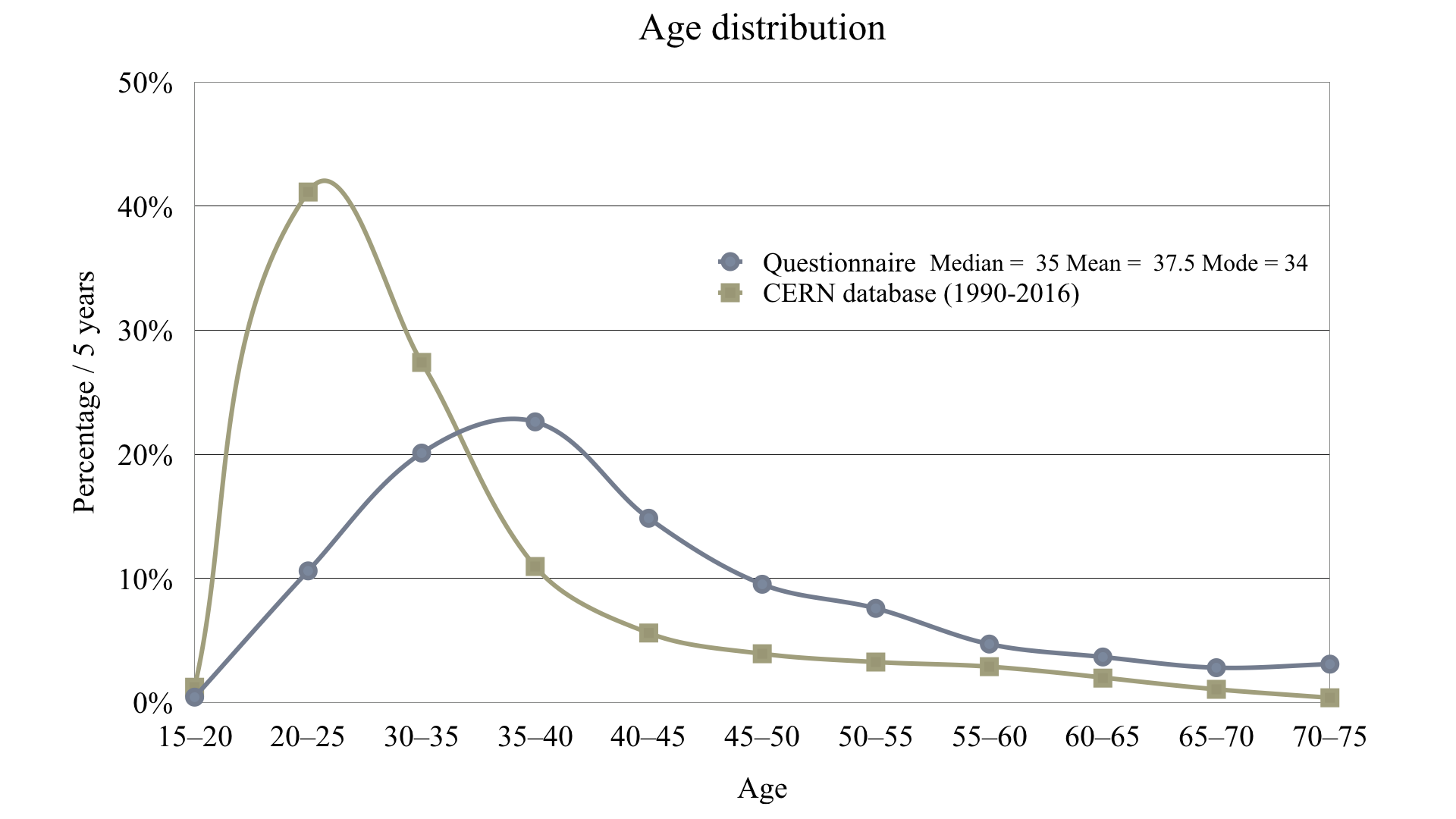}
\caption{Age distribution of the respondents to the questionnaire compared with the age distribution of CERN users and former users as taken from the CERN database for 1990-2016.}
\label{fig:age_study_database}
\end{figure}

\begin{figure}
\centering
\includegraphics[width=.9\linewidth]{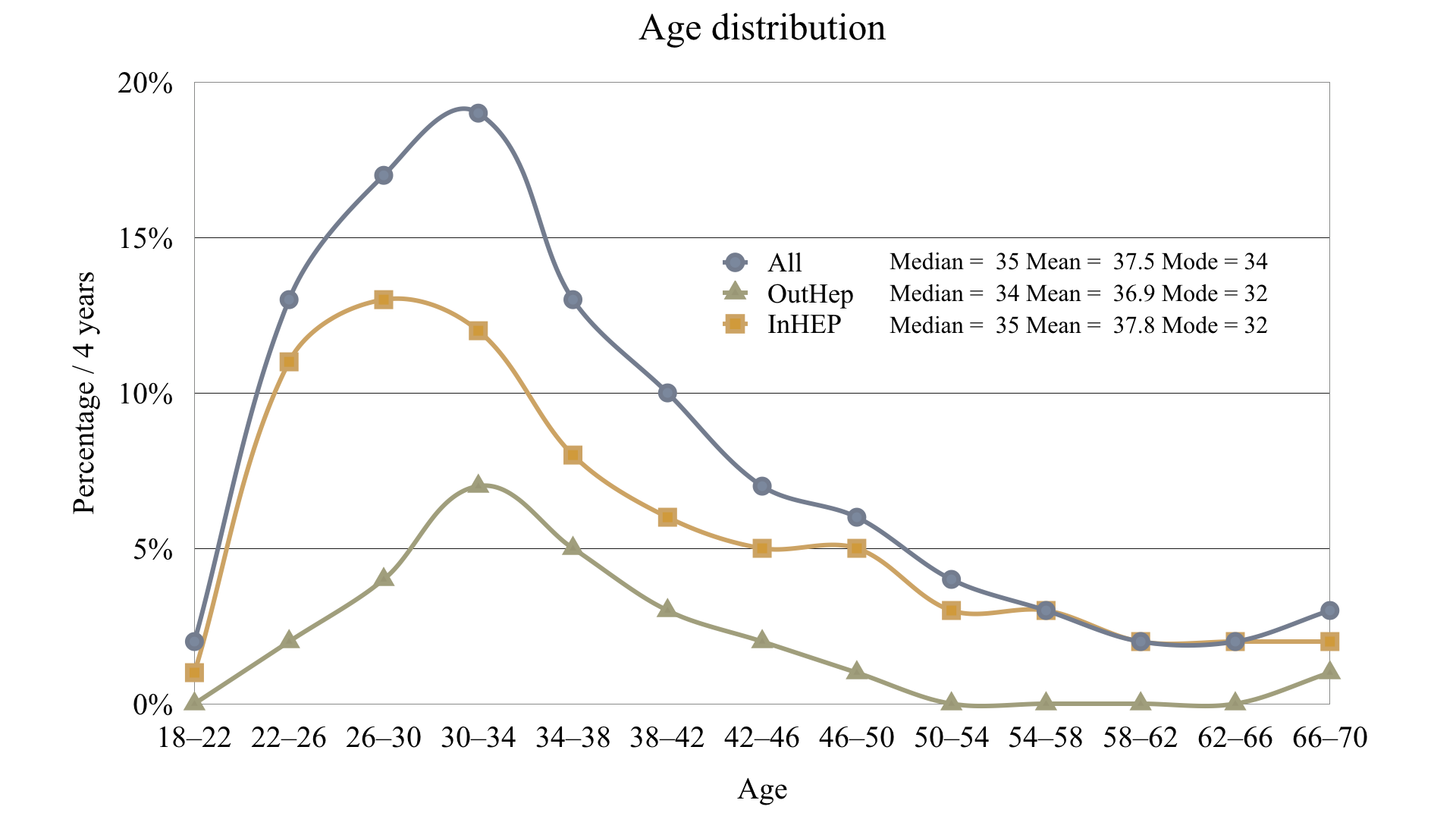}
\caption{Age distribution of the respondents to the questionnaire. Circle markers represent all respondents, triangles represent the people who left HEP, and square markers represent those who are still in HEP.}
\label{fig:age_in_out}
\end{figure}

Figure~\ref{fig:age_study_database} shows the age distribution of the respondents in comparison with the age distribution from the database of CERN users. This figure indicates that the age distribution of the respondents does not follow that of the CERN database. In particular, the questionnaire was preferentially completed  by people in their 30s or 40s. 
Figure~\ref{fig:age_in_out} shows a comparison of the age distribution of the people in HEP and outside HEP. The distribution for people outside HEP peaks at slightly older ages. Both figures also contain retired people, populating the region of ages beyond 65.

\begin{figure}
\centering
\includegraphics[page=66, width=.9\linewidth]{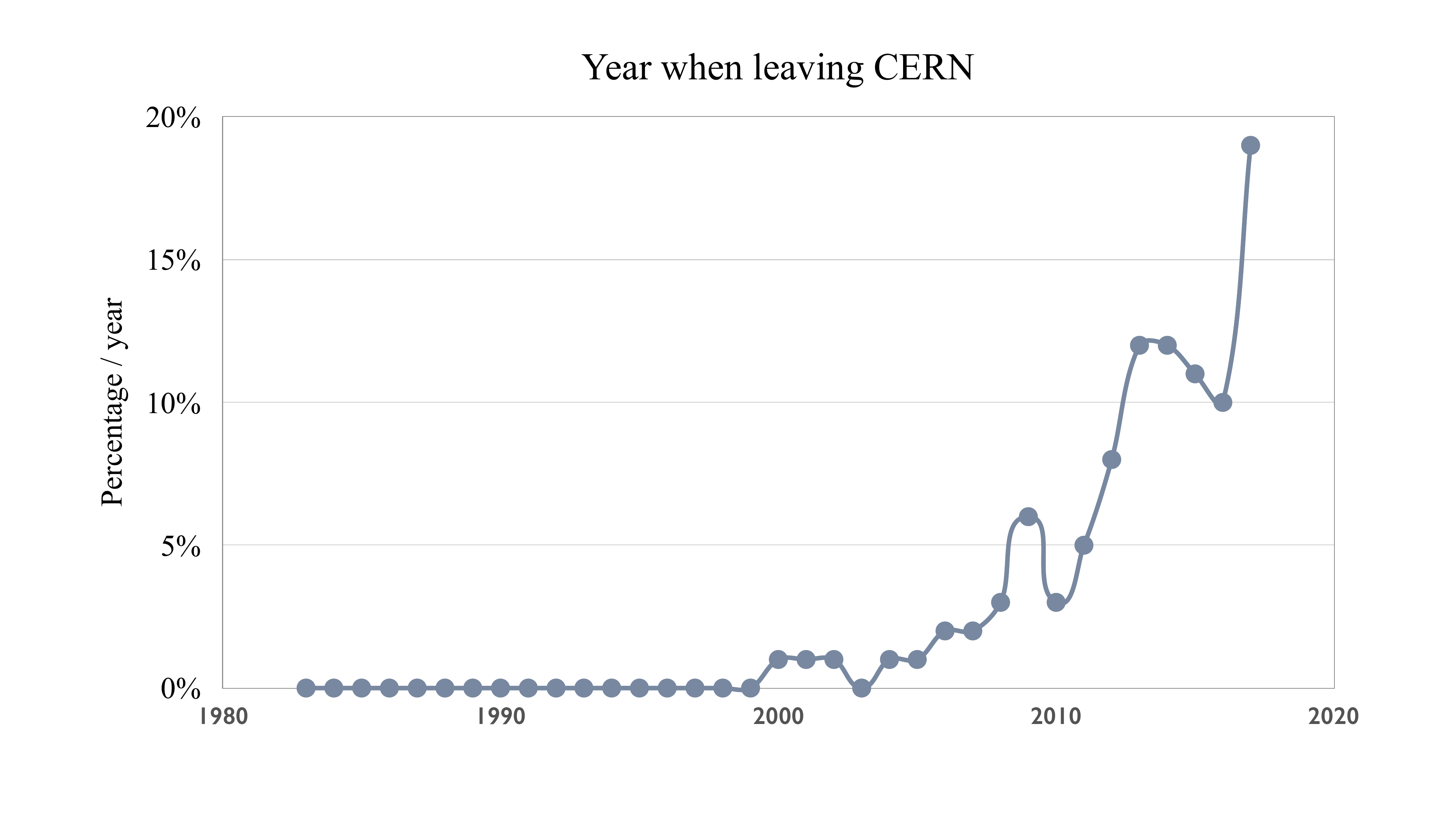}
\caption{Years in which respondents left CERN.}
\label{fig:year_left}
\end{figure}

For people who left HEP, Figure~\ref{fig:year_left} shows the year in which they left. Obviously, most respondents left HEP in the past 6-7 years; this is an expected bias, owing to the enhanced difficulty of reaching people who left CERN many years ago.

\section{Geographical and professional mobility}\label{sec:mobility}

\begin{figure}
\centering
\includegraphics[page=52,width=.9\linewidth]{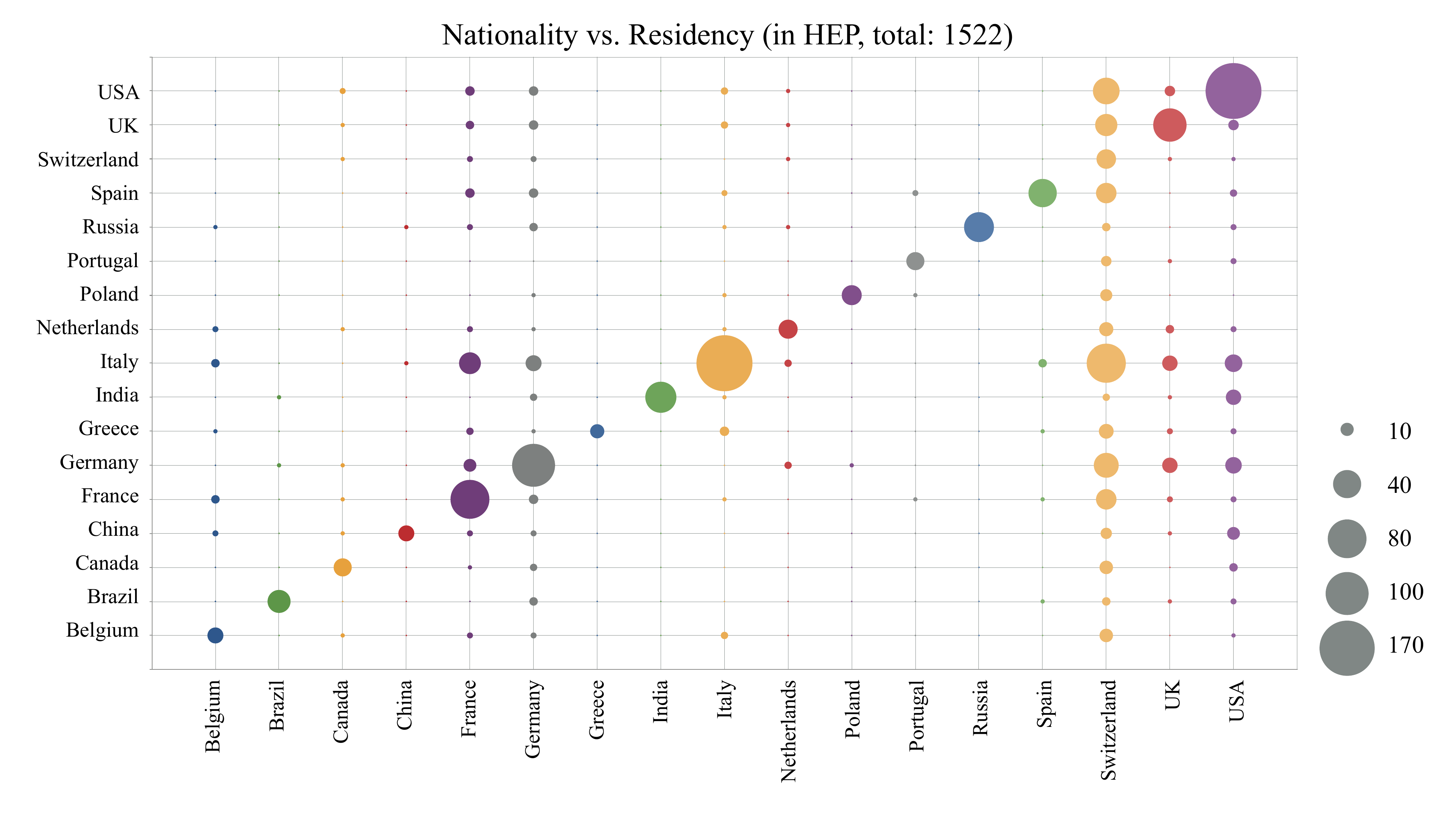}
\caption{Country of nationality ($y$-axis)  as a function of residence ($x$-axis). The bubble size is proportional to the number of people in HEP in a given country. Only nationalities with at least 40 entries are shown. (A full list of countries is given in Appendix A.)}
\label{fig:hep_nationality_residence}
\end{figure}

\begin{figure}
\centering
\includegraphics[page=53,width=.9\linewidth]{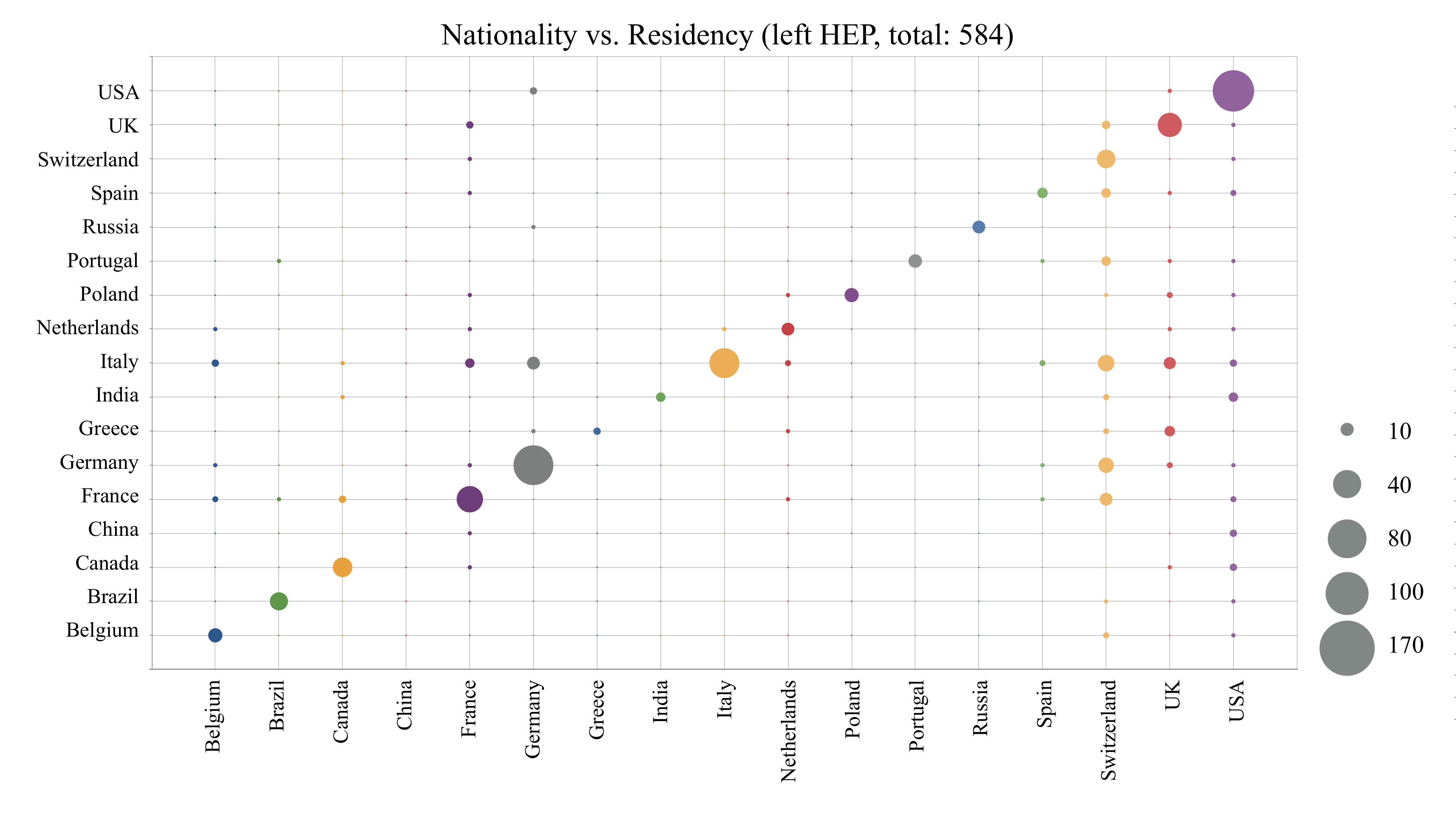}
\caption{Country of nationality ($y$-axis) as a function of residence ($x$-axis). The bubble size is proportional to the number of people that left HEP. Only nationalities with at least 40 entries are shown. (A full list of countries is given in Appendix A.)}
\label{fig:non_hep_nationality_residence}
\end{figure}

\begin{figure}
\includegraphics[page=46, width=.9\linewidth]{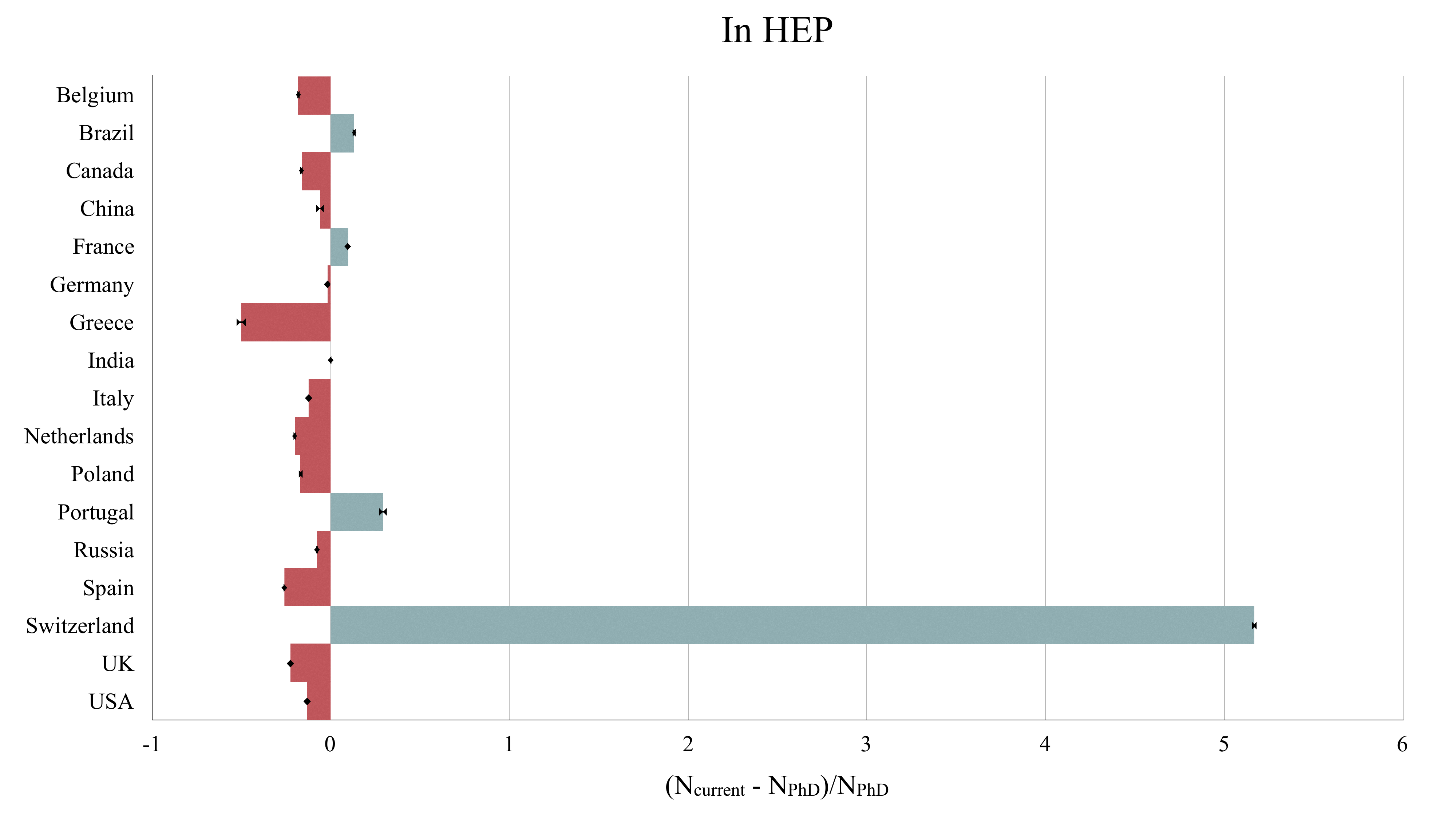}
\caption{Relative difference between the number of people in HEP who currently work in a country with respect to those who obtained a PhD in that country. Only responses for nationalities with at least 40 entries are shown. (A full list of countries is given in Appendix A.)}
\label{fig:reldiffcountry_hep}
\end{figure}

\begin{figure}
\centering\includegraphics[page=45, width=.9\linewidth]{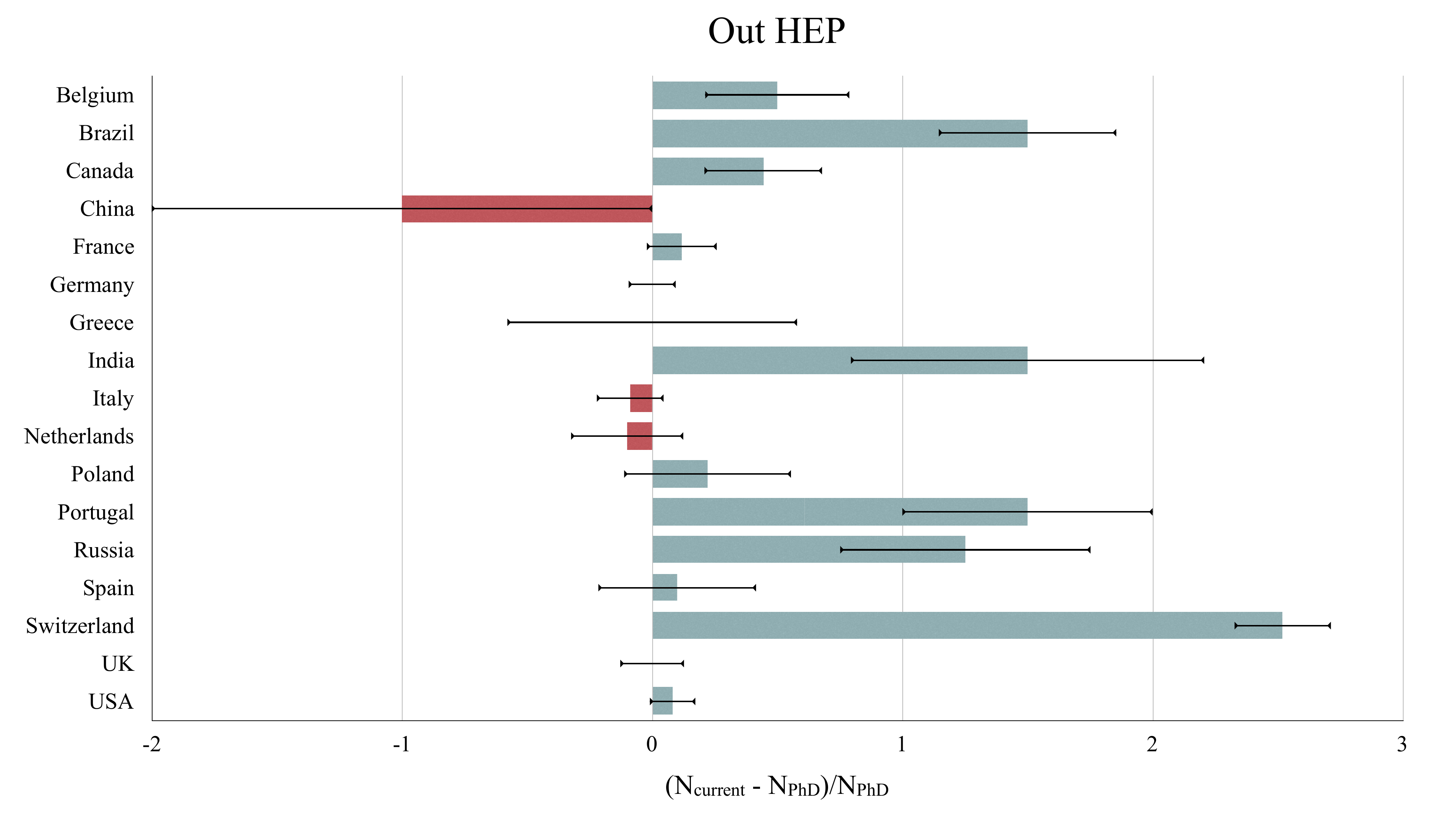}
\caption{Relative difference between the number of people who left HEP with respect to those who obtained a PhD in that country. Only responses for nationalities with at least 40 entries are shown. (A full list of countries is given in  Appendix A.)}
\label{fig:reldiffcountry_outhep}
\end{figure}

The geographical mobility of the respondents was studied as a function of their nationalities, the countries where they obtained their PhDs and their current places of work. Figures~\ref{fig:hep_nationality_residence} and \ref{fig:non_hep_nationality_residence} show the number of persons of a given nationality ($y$-axis) plotted by their country of residence ($x$-axis), for those in HEP and out of HEP, respectively.
Figure~\ref{fig:hep_nationality_residence} confirms that the largest fraction of people continue to work in their country of origin. It can be noticed that Switzerland has an extremely diverse set of nationalities, owing to the frequent temporary or permanent relocation of HEP researchers to CERN. A non-negligible relocation of persons with respect to their nationality can be observed for many other countries, such as Italy, Germany, France, UK, USA and the Netherlands.  
Figure~\ref{fig:non_hep_nationality_residence}, showing the relocation for people outside HEP, indicates that Switzerland is a very 
attractive country. Germany, France, the UK, and the USA behave in a similar way. In general, many persons go back to their home countries when seeking a job outside research.
The incoming-outgoing fractions in HEP and out of HEP are given in Figures~\ref{fig:reldiffcountry_hep} and ~\ref {fig:reldiffcountry_outhep}, respectively. Both graphs show the difference between the number of people who currently work in a country with respect to those who obtained a PhD in that country, normalized to the number of PhDs obtained in that country, $(N_{current} - N_{PhD}) / N_{PhD}$ (relative difference).  Switzerland hosts five times as many HEP members as the PhDs it produces and 2.5 times as many former-HEP members with respect to the PhD produced.


\begin{figure}
\centering\includegraphics[page=16,width=.9\linewidth]{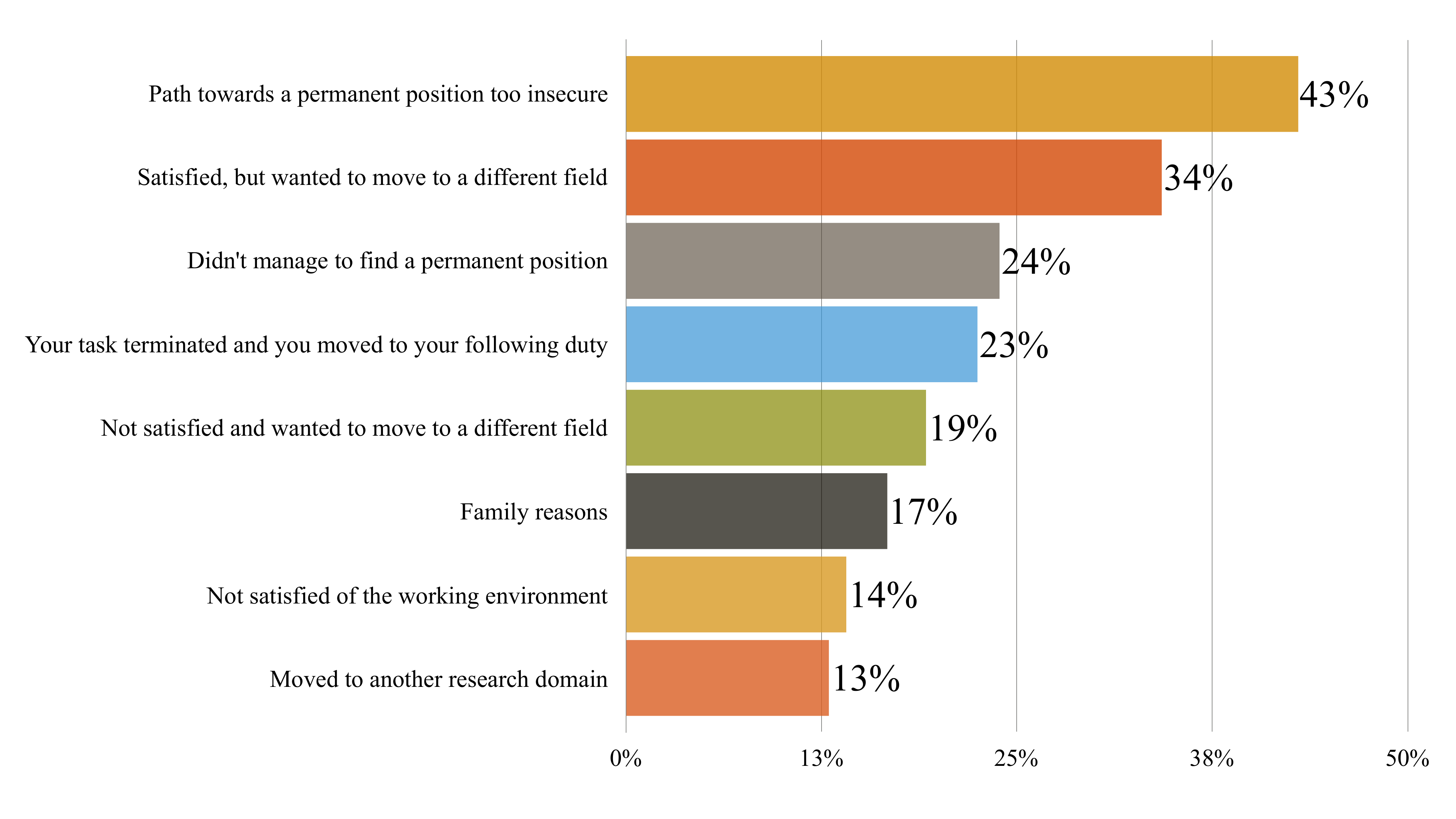}
\caption{Reasons for leaving HEP, with the percentage of respondents choosing each: several  answers were allowed.}
\label{fig:reason_to_leave}
\end{figure}

Professional mobility was investigated through questions put to people who had stopped working in HEP. A number of common causes for looking for another profession were proposed, as answers, of which more than one could be selected. Figure~\ref{fig:reason_to_leave} shows the percentage of respondents who indicated a specific option as the reason for leaving. The most common (43\% of the respondents) is that the path towards a permanent position in HEP was too insecure. Failure to find a permanent position was the reason for leaving for 24\% of the respondents. For 34\% the job in HEP was satisfactory, but they wished to move to another field. About 19\% wanted to change because they were not satisfied with their occupation and 14\% were not satisfied with the work environment. Family reasons were cited by 17\% of the respondents. In 23\% of cases, the respondents declared that HEP was a temporary step and that they moved to a different domain after completing their temporary assignment.

Overall, for almost 70\% of the answers, career insecurity or failure to find a permanent position in research were the main reasons for leaving HEP. Another 34\% of answers indicated that people left HEP, although they were satisfied, because they wanted to move to another field. Finally, 33\% left because they were not satisfied and wanted to move to a different field or were not satisfied by the work environment.

\section{Career evolution}

\begin{figure}
\centering\includegraphics[page=50,width=.9\linewidth]{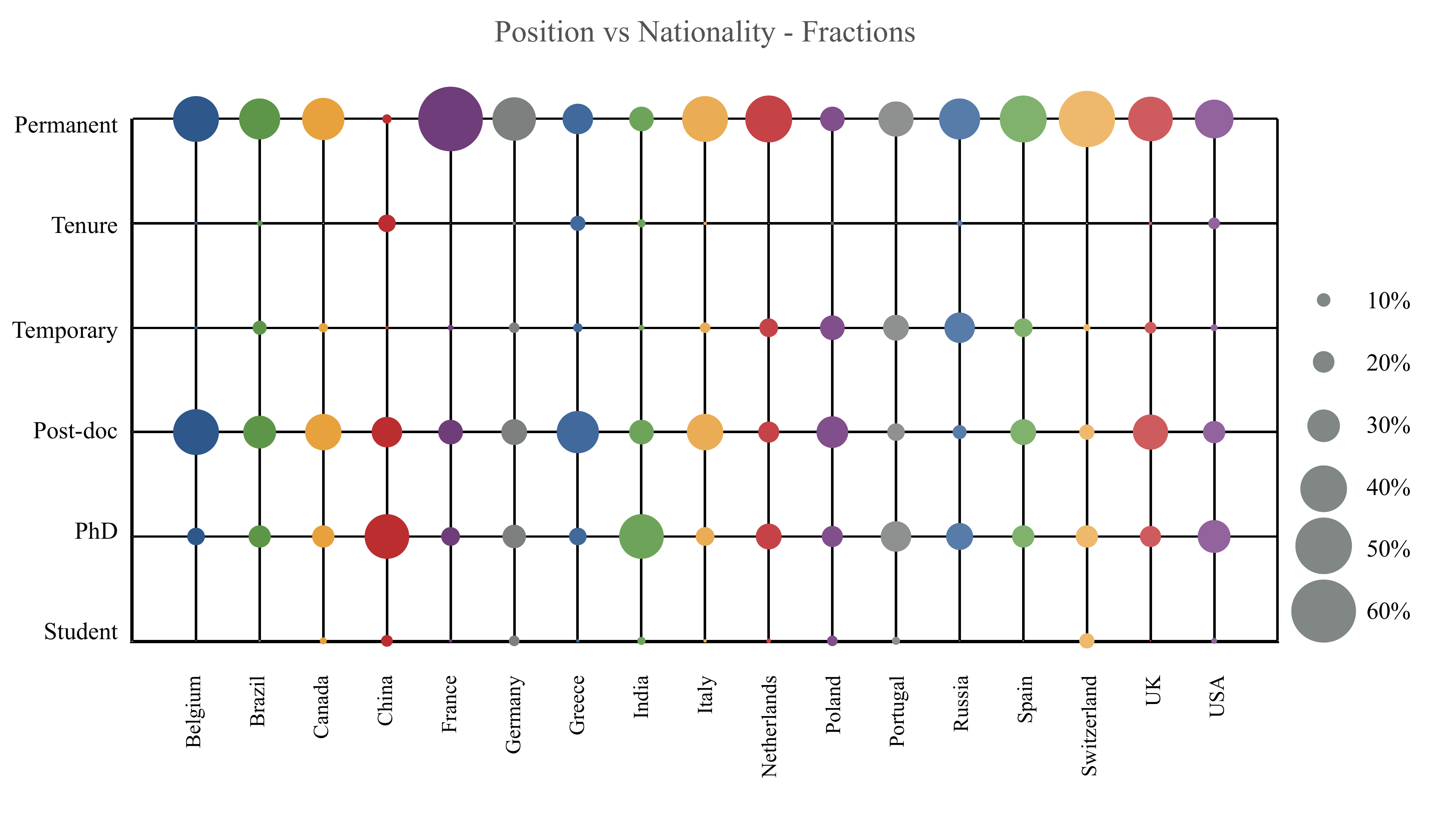}
\caption{Fraction of respondents still working in HEP who are: student, PhD, post-doc, tenure track or permanent contract by country of nationality. A larger marker size indicates a larger percentage. Only responses for nationalities with at least 40 entries are shown. (A full list of countries is given in Appendix A.)}
\label{fig:hep_position_nationality}
\end{figure}

\begin{figure}
\centering\includegraphics[trim={20cm 0 20cm 0},clip,page=20,width=.45\linewidth]{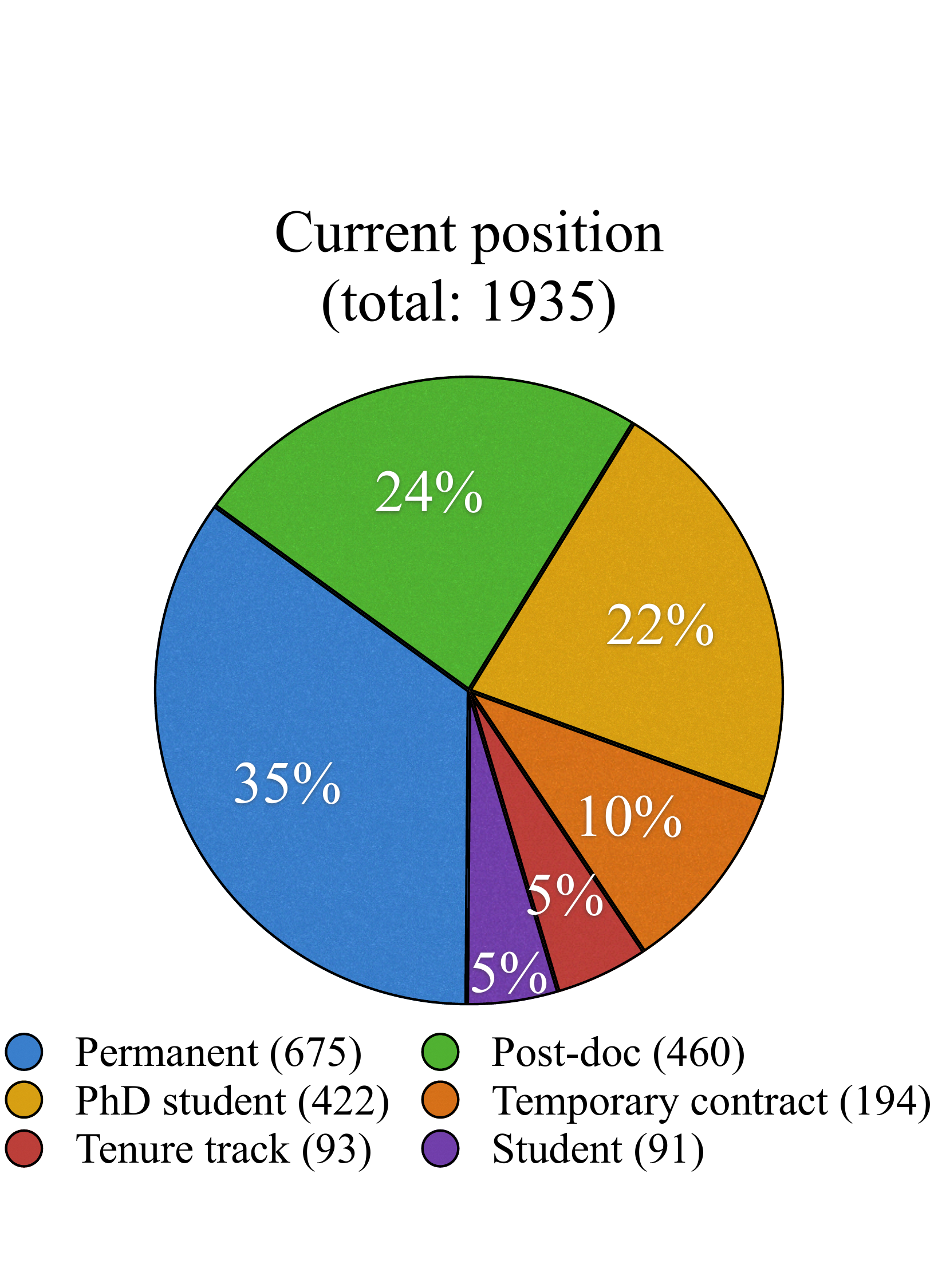}
\caption{Current positions of the respondents who were still in HEP}
\label{fig:hep_position}
\end{figure}

\begin{figure}
\centering\includegraphics[page=73,width=.9\linewidth]{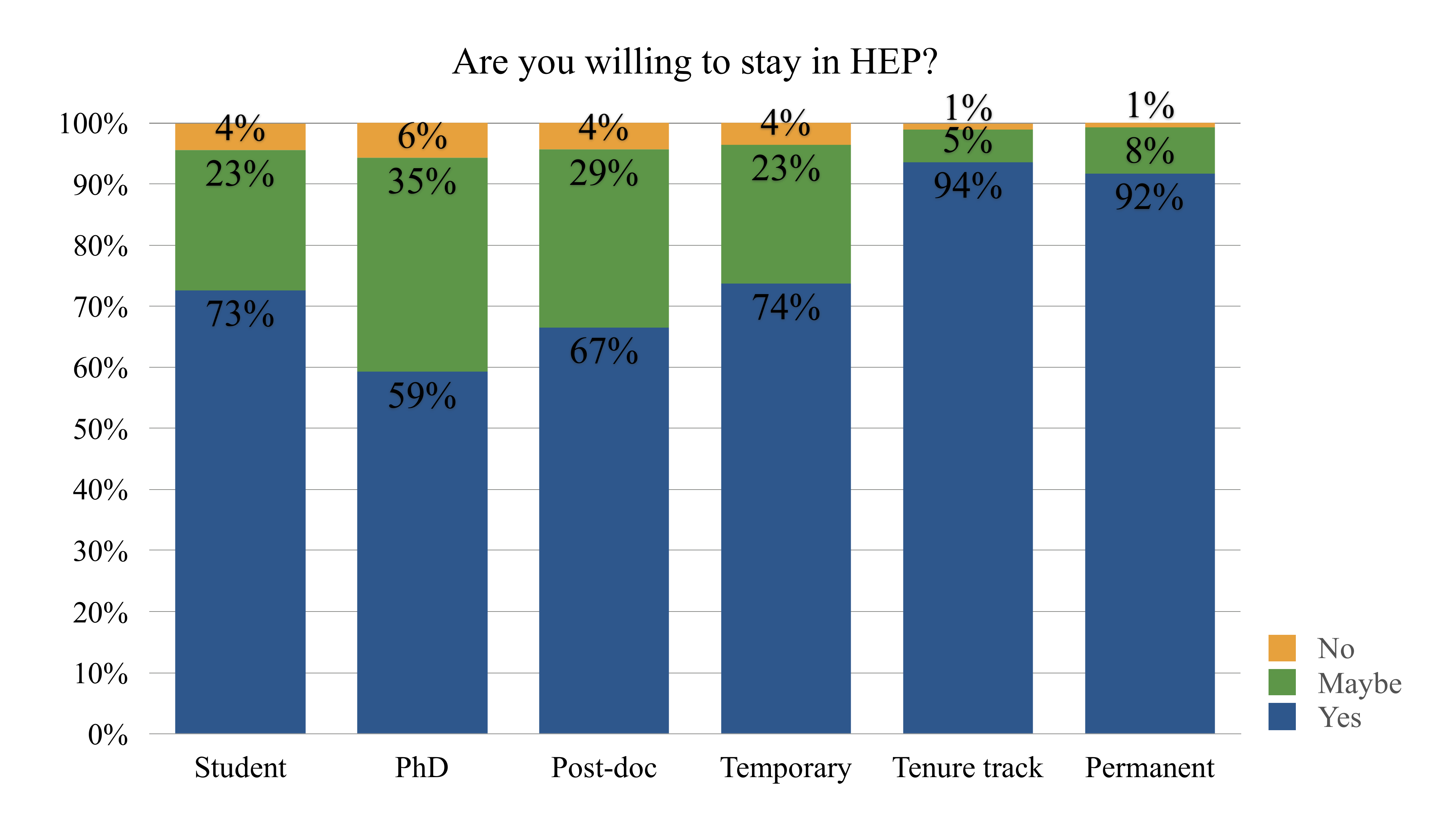}
\caption{Willingness of respondents to continue a career in HEP, as a function of their current position.}
\label{fig:stay_in_hep}
\end{figure}

The sample of respondents still working in HEP is made up of 1937 people with permanent, tenure track, post-doc or temporary contracts.
Figure~\ref{fig:hep_position_nationality} shows the percentages of people in the various positions, computed for each nationality. In most countries, people with permanent contracts form the majority, but some countries (e.g., China, Greece and India) show a larger fraction of people with non-permanent contracts.  The contract types held by the respondents were the following:  permanent position (35\%), post-doc (24\%), PhD student (22\%), temporary contract (10\%), tenure track (5\%), undergraduate student (5\%), as shown in the left panel of Figure~\ref{fig:hep_position}. The majority of the respondents (76\%) expressed a desire to continue to work in HEP  (Fig.~\ref{fig:stay_in_hep}).

\begin{figure}
\centering\includegraphics[trim={20cm 0 20cm 0},clip,page=10,width=.45\linewidth]{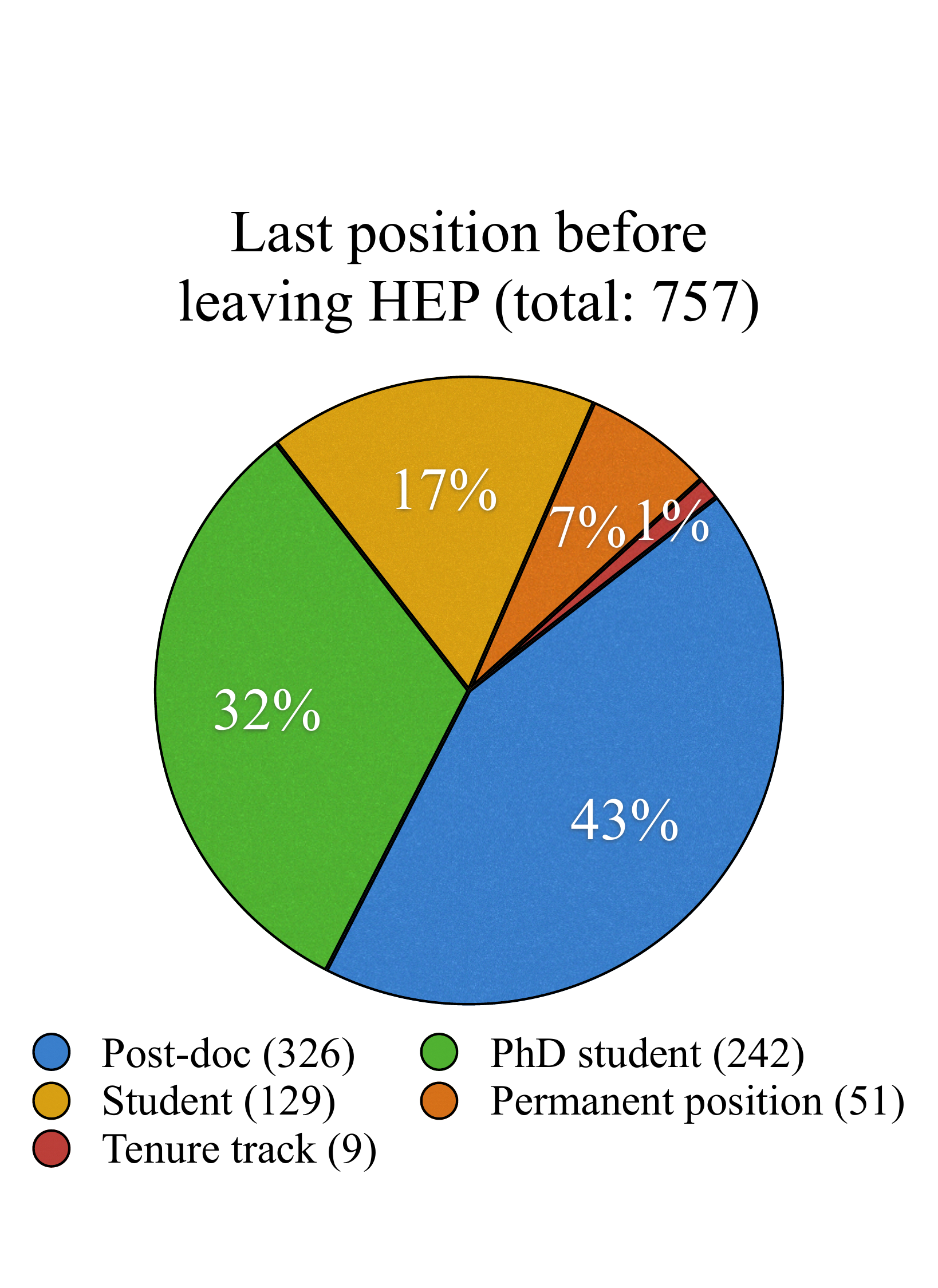}
\centering\includegraphics[trim={20cm 0 20cm 0},clip,page=60,width=.45\linewidth]{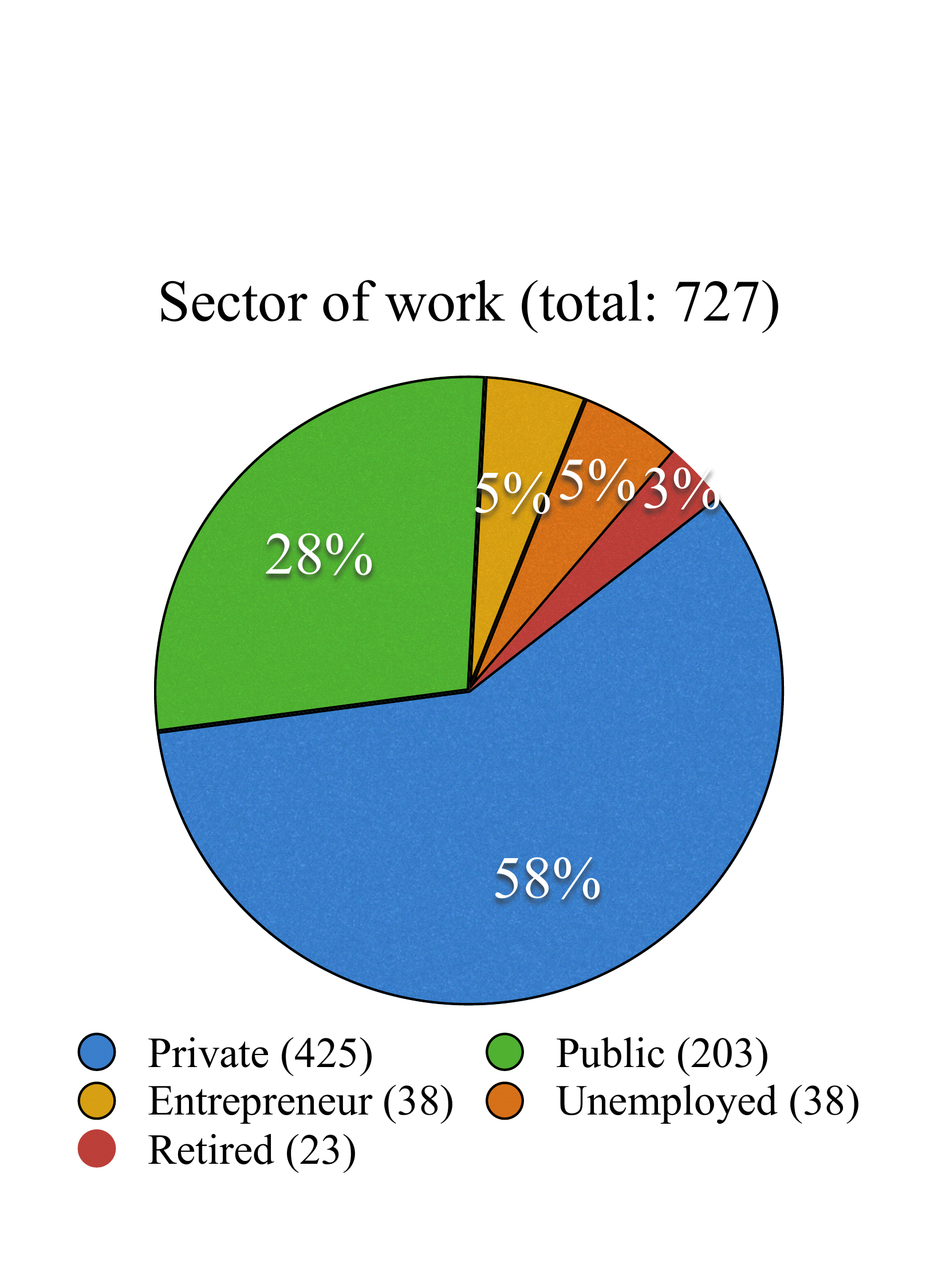}
\caption{Left: last position held in HEP. Right: current employment sector. The number of entries for each category is shown in parenthesis.} 
\label{fig:last_position}
\end{figure}

The last type of contract held while working in HEP by respondents who subsequently left HEP was mostly post-doc and PhD student, as shown in the left panel of Figure~\ref{fig:last_position}. The majority were currently employed in the private sector (58\%), while 28\% were employed in the public sector (Figure~\ref{fig:last_position}, right panel). A further 5\% of the sample declared themselves to be  unemployed, with an average unemployment period of a few months.

\begin{figure}
\centering\includegraphics[page=72,width=.9\linewidth]{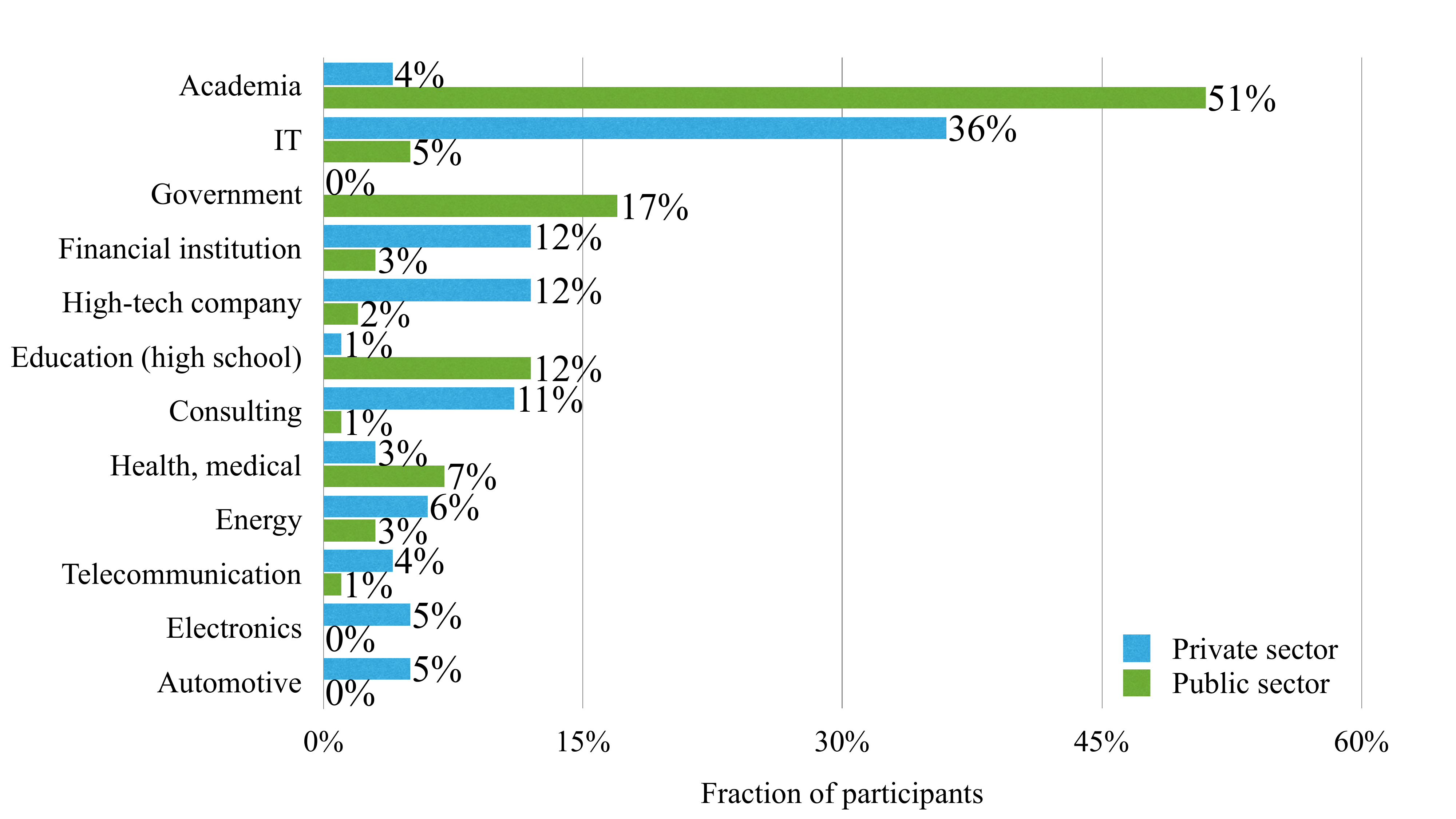}
\caption{Sectors in which former CERN users continue their careers. The percentages correspond to the fractions of answers. Blue, results for the private sector; green, results for the public sector.}
\label{fig:sector}
\end{figure}

The variety of sectors in which the persons were involved and the job type covered by former CERN users was quite wide. The majority of people continued their careers in the information technology (IT) sector and in academia (Fig~\ref{fig:sector}). Considering only employees in the private sector and entrepreneurs, the leading sectors were IT, finance, high-tech companies, and consulting. Those working in the public sector were mostly in academia, government, and education.

\begin{figure}
\centering\includegraphics[page=12,width=.9\linewidth]{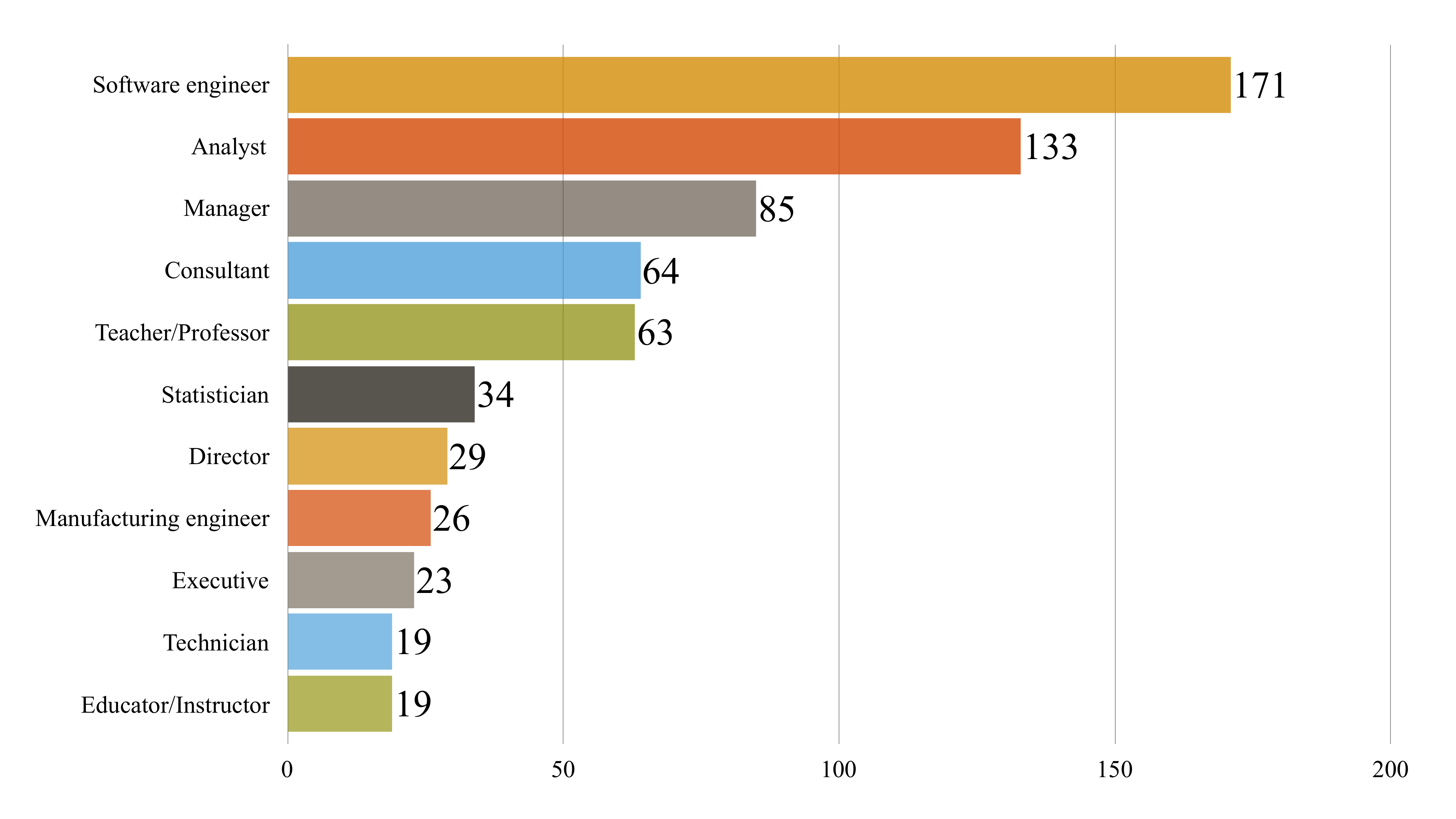}
\caption{Current job titles of people who left HEP}
\label{fig:job_title}
\end{figure}

\begin{figure}
\centering\includegraphics[page=43,width=.9\linewidth]{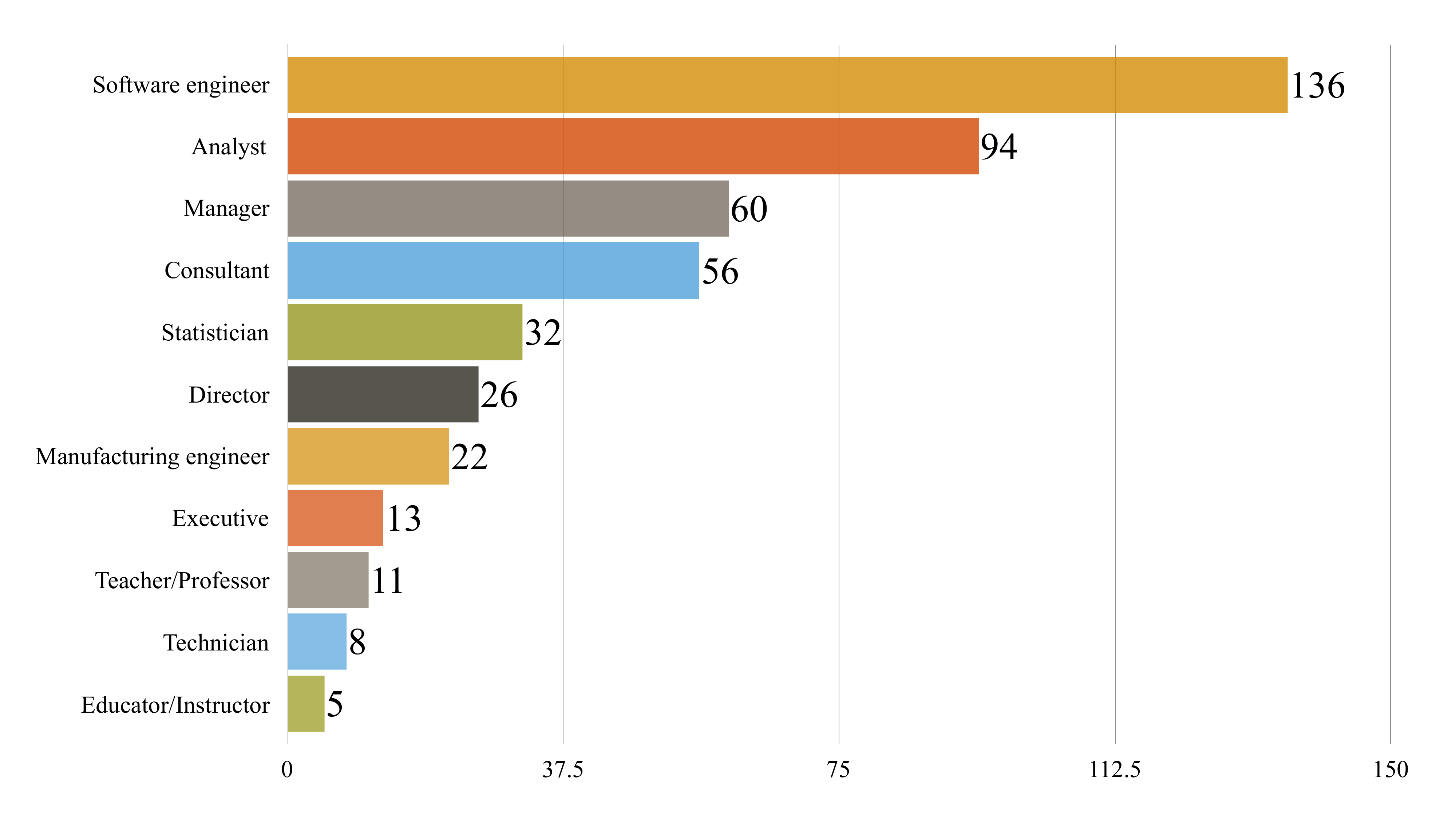}
\caption{Current job titles of people working in a private company or who had become entrepreneurs}
\label{fig:job_title_private}
\end{figure}

\begin{figure}
\centering\includegraphics[page=44,width=.9\linewidth]{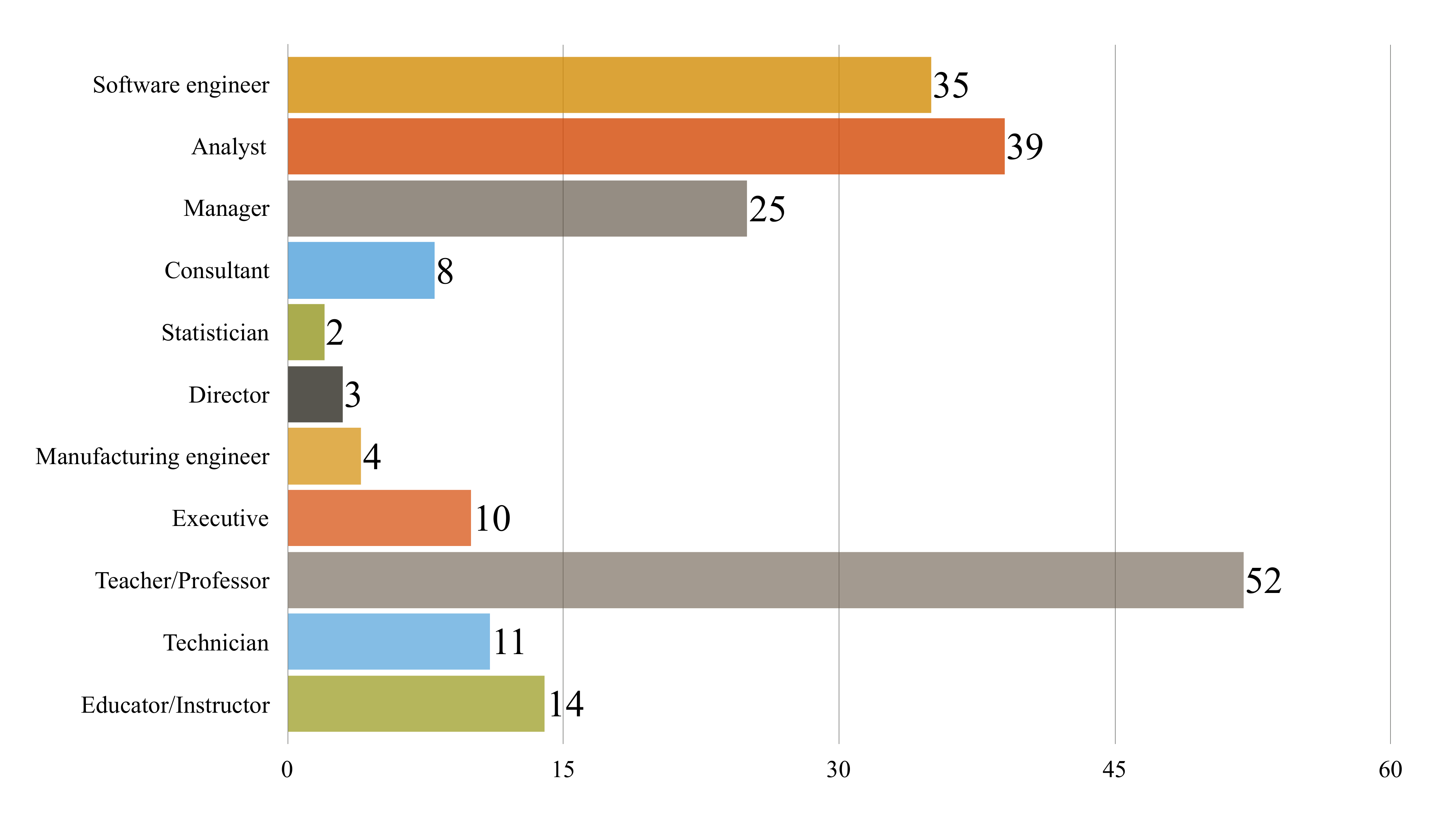}
\caption{Current job titles of people who now work in the public sector}
\label{fig:job_title_public}
\end{figure}

The job types most commonly held by former CERN users were: software engineer, analyst, manager and consultant (Figure~\ref{fig:job_title}). The skills needed in the actual work of a researcher, namely analytical thinking, programming, and problem solving, just to mention a few, are the basis for being competitive in such professions. Figures~\ref{fig:job_title_private} and \ref{fig:job_title_public} show the professions covered in the private and public sectors, respectively.

\begin{figure}
\centering\includegraphics[page=13,width=.9\linewidth]{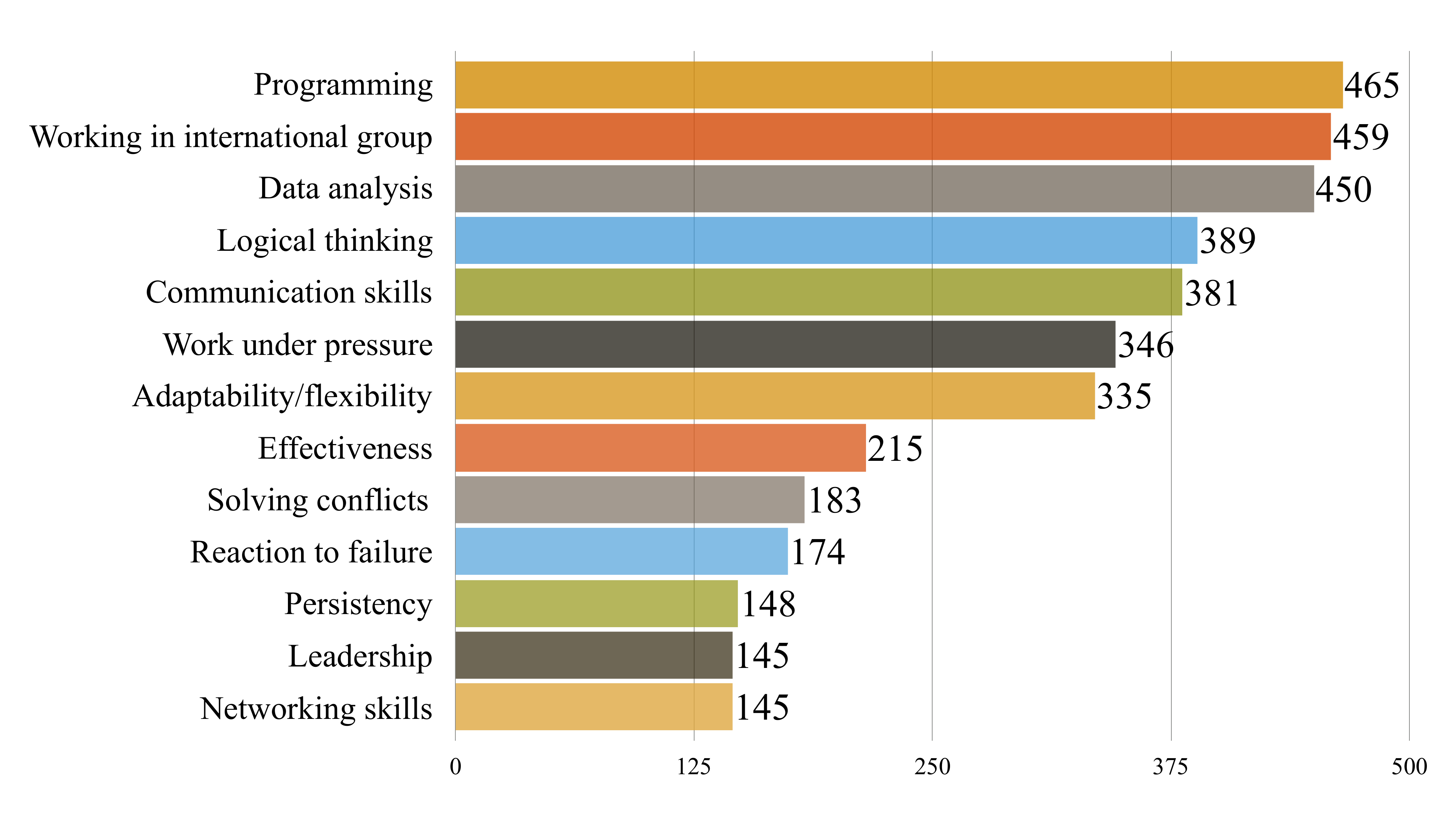}
\caption{Skills acquired at CERN which turned out to be important for working outside HEP. Multiple answers were allowed.}
\label{fig:skills}
\end{figure}

Finally, Figure~\ref{fig:skills} shows a list of the skills that people acquired at CERN and that they consider important for their current jobs. Several answers were proposed and allowed. The most popular skills (programming, data analysis, logical thinking) were correlated to some of the professions shown in Figure~\ref{fig:job_title}. The skills acquired, such as 'working in international groups', 'communication skills', 'work under pressure', and 'adaptability or flexibility', are useful for many highly qualified jobs and allow a wide variety of professions to be approached, with excellent prospects for career development.

\section{Satisfaction with CERN experience}

\begin{figure}
\centering\includegraphics[page=14,width=.9\linewidth]{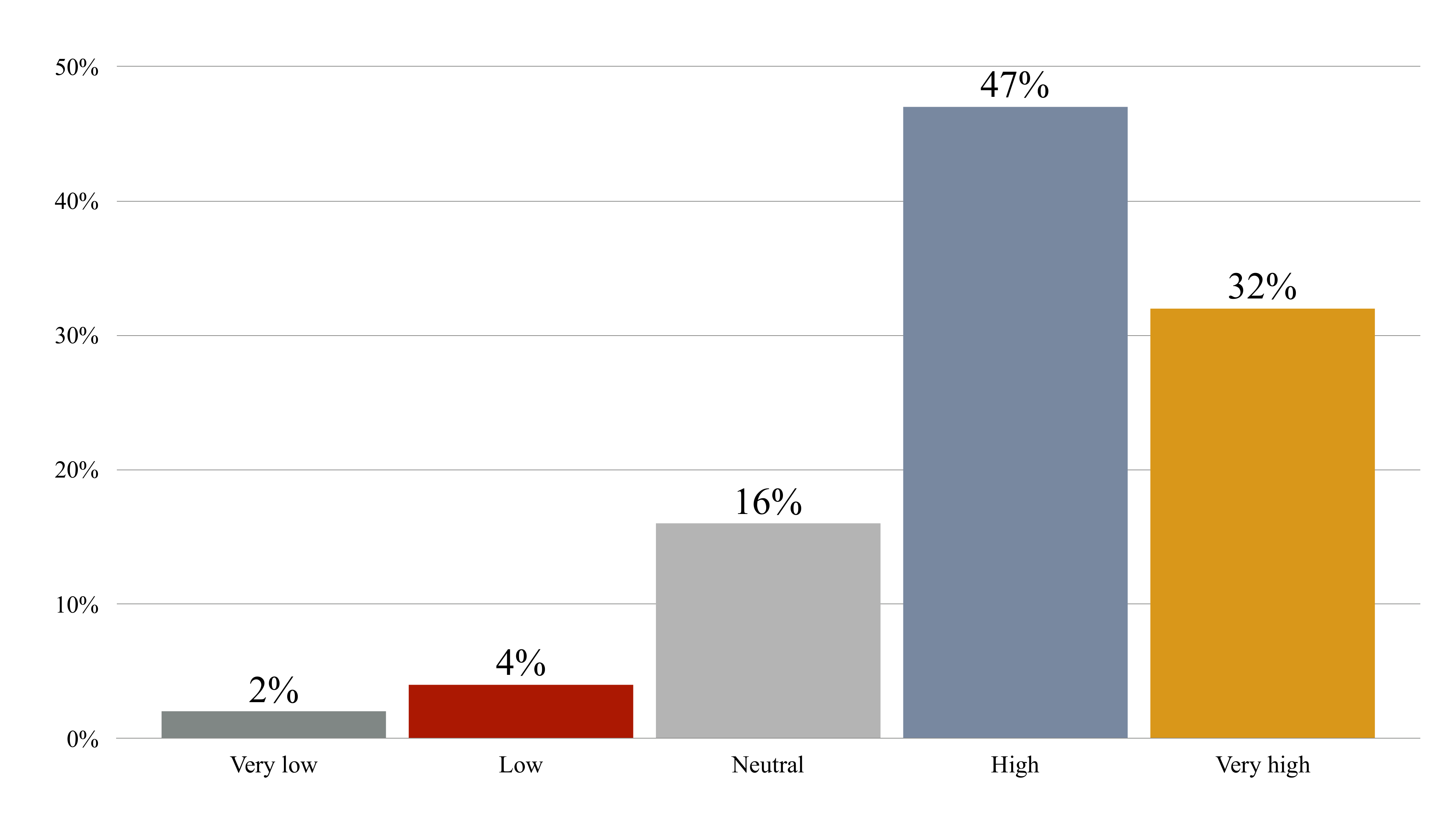}

\caption{Satisfaction with current job, for people who left HEP}
\label{fig:current_satisfaction}
\end{figure}

\begin{figure}
\centering
\includegraphics[page=65,width=.9\linewidth]{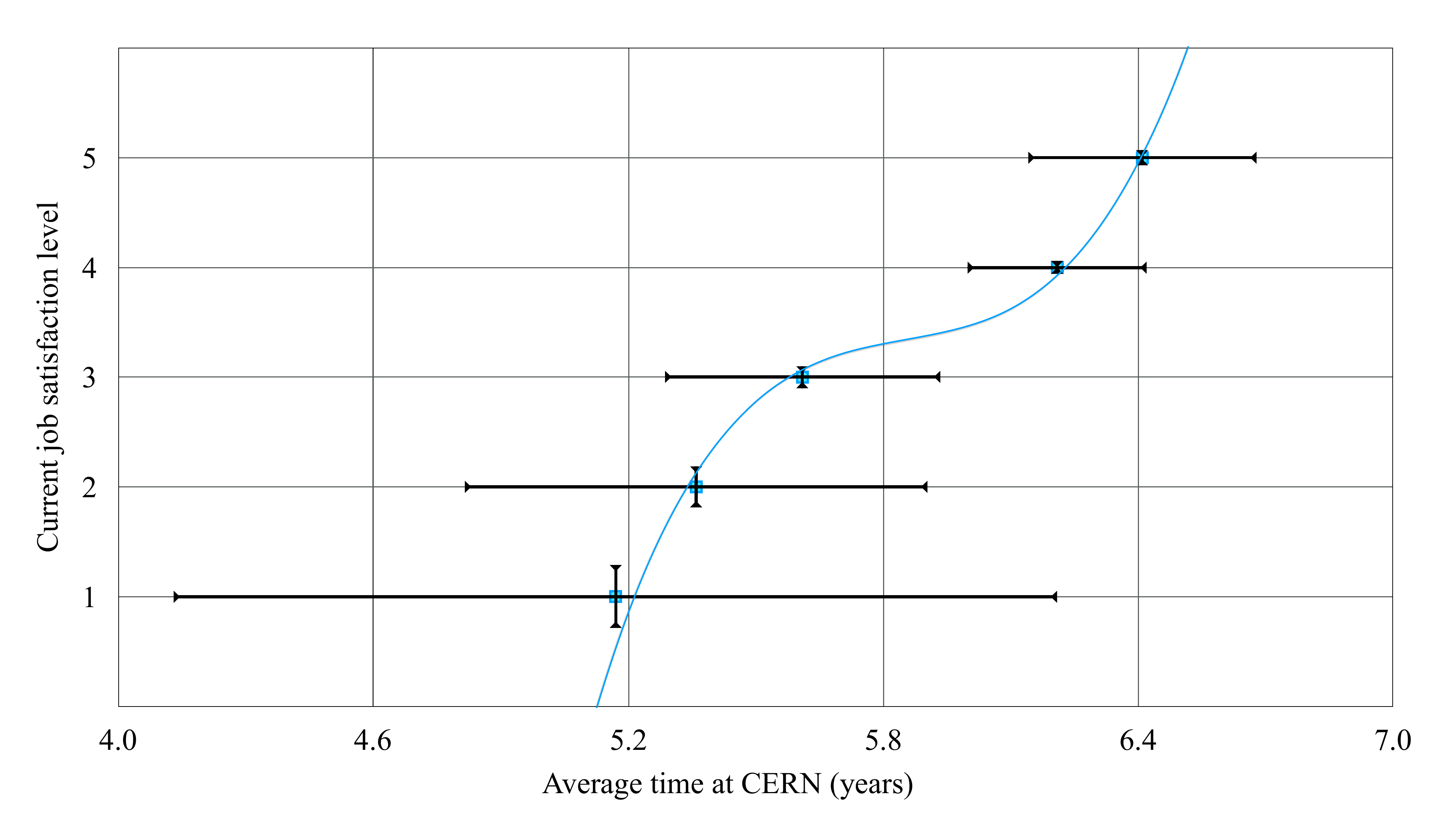}
\caption{Average satisfaction level (1 is low, 5 is very high) with current job, for persons who left HEP, as a function of time spent at CERN. Error bars represent standard deviations of the distributions.}
\label{fig:current_satisfaction_vs_years_cern_average}
\end{figure}

Figure~\ref{fig:current_satisfaction}
shows satisfaction with current employment of the respondents who left HEP. In this case, the satisfaction was high for a large majority of the respondents. When looking at the average number of years spent at CERN for the different levels of satisfaction (1 is low, 5 is very high), a higher satisfaction was obtained from those who had worked at CERN for longer periods (see Figure~\ref{fig:current_satisfaction_vs_years_cern_average}).

\begin{figure}
\centering\includegraphics[page=19,width=.9\linewidth]{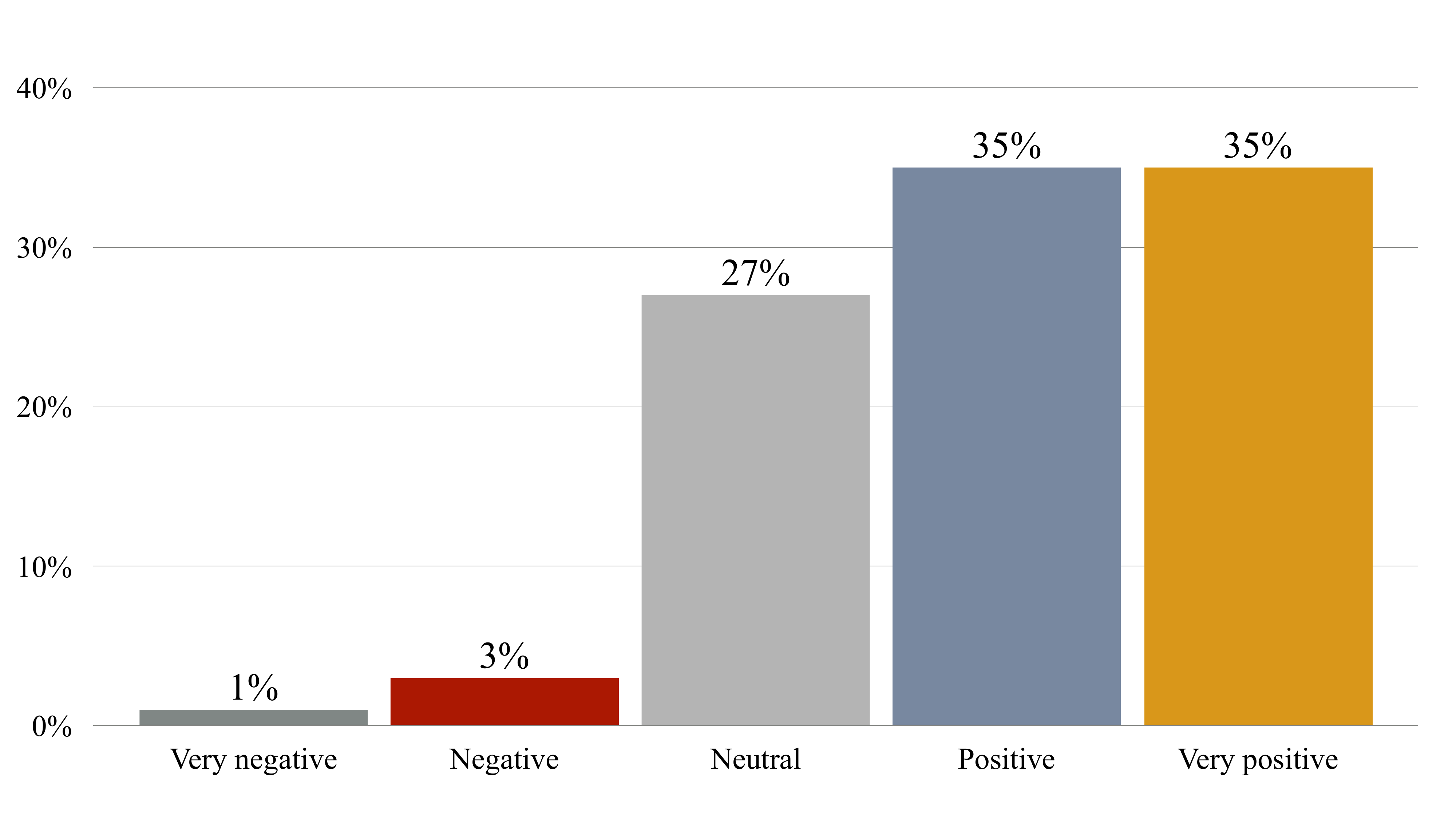}
\caption{Impact of experience at CERN in obtaining the current job for respondents who left HEP}
\label{fig:cern_impact}
\end{figure}

\begin{figure}
\centering\includegraphics[page=18,width=.9\linewidth]{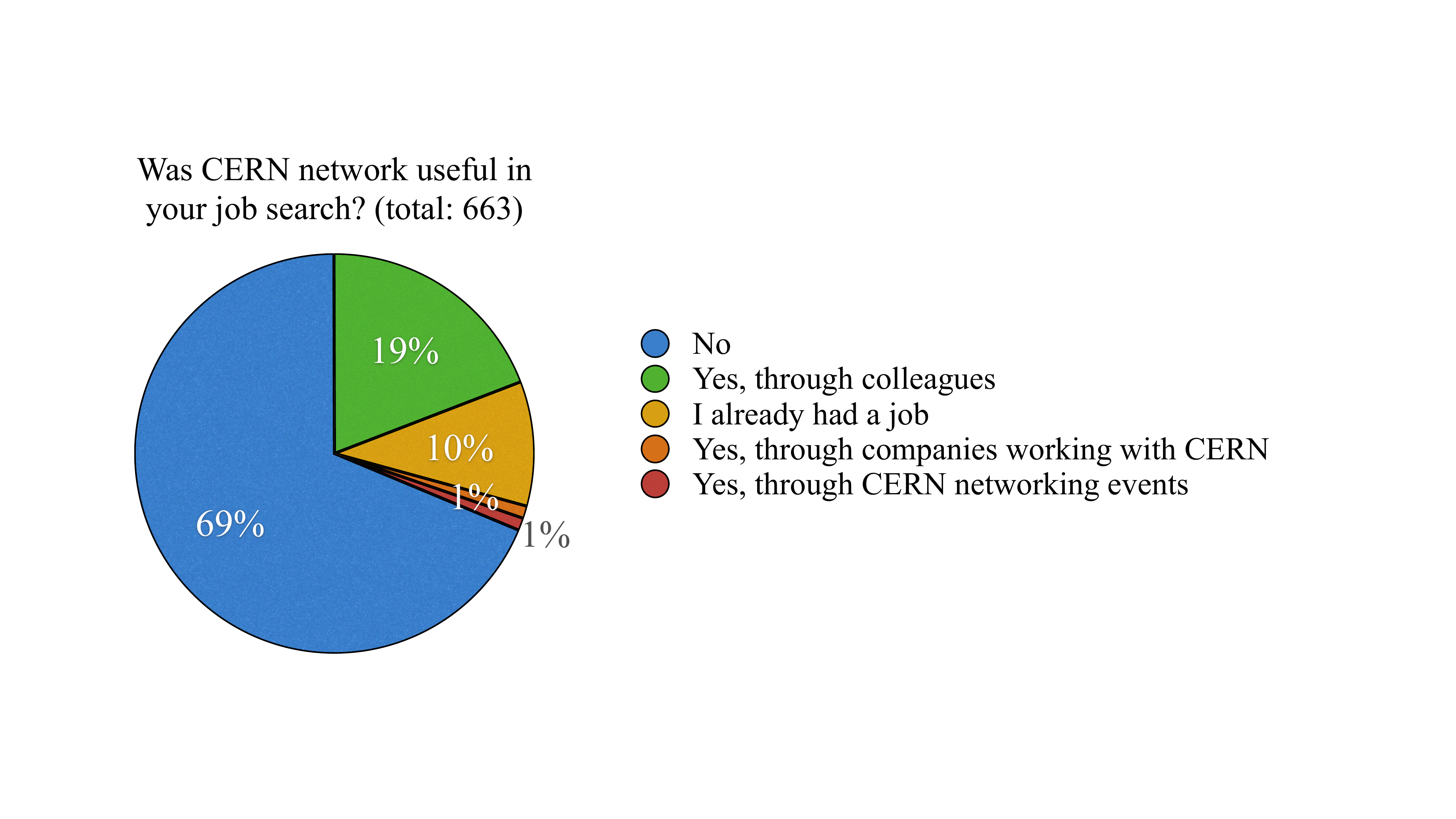}
\caption{Usefulness of the CERN network in finding a job outside HEP}
\label{fig:cern_network}
\end{figure}

The respondents' feeling on the impact of their working experience at CERN in finding a new job is shown in Figure~\ref{fig:cern_impact}. For 27\% of the respondents the impact was neutral, while for almost all the others the effect of their CERN experience was considered either positive or very positive.
While the overall experience at CERN is positive, the practical help provided by CERN and its services in finding a job outside HEP was considered insufficient by 69\% of the respondents, as shown in  Figure~\ref{fig:cern_network}. A fraction of 19\% of former CERN users declared that they managed to find a job through colleagues.

\begin{figure}
\centering\includegraphics[page=7,width=.9\linewidth]{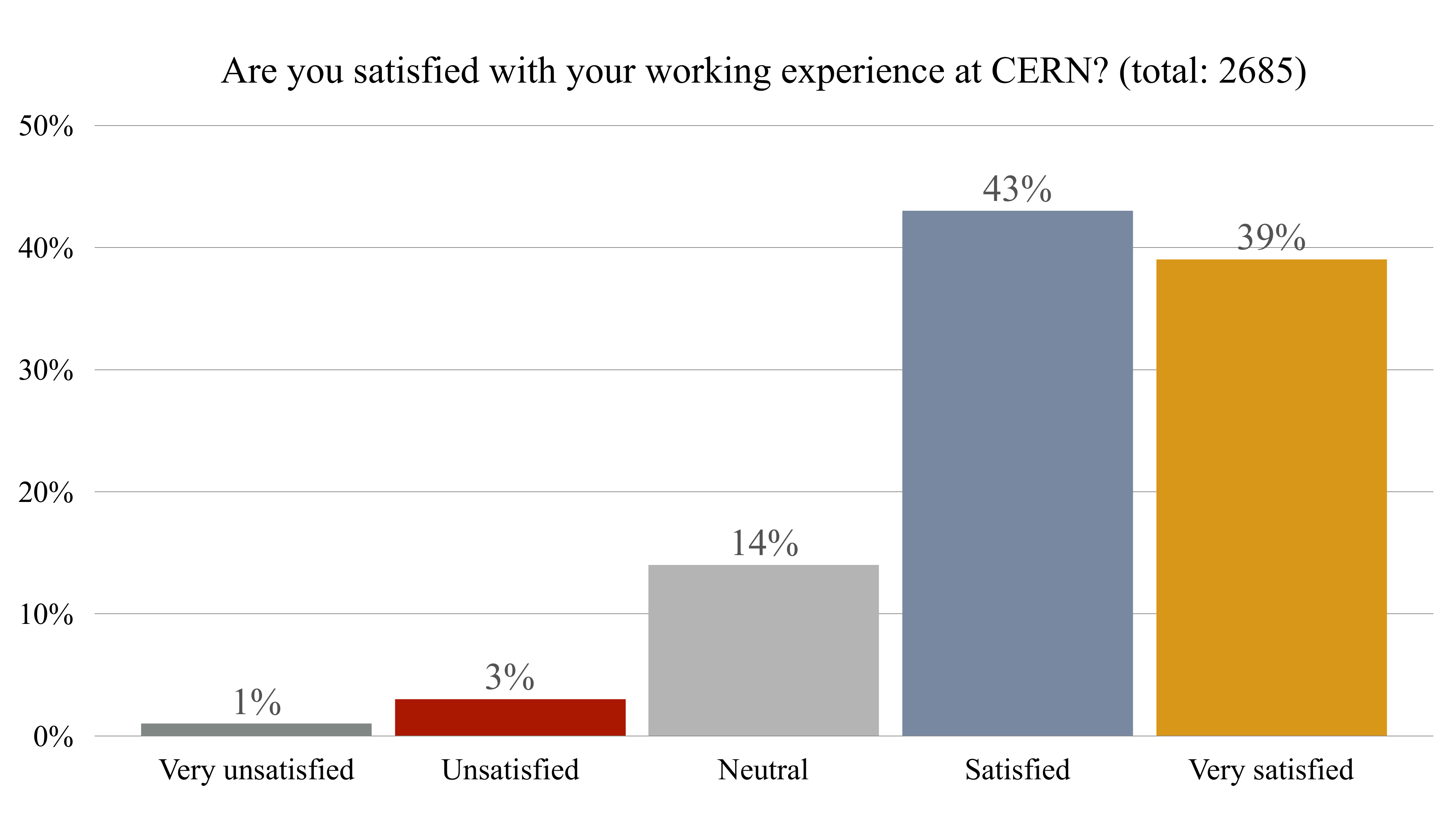}
\caption{Degree of satisfaction of the working experience at CERN for the full sample}
\label{fig:satisfaction_cern}
\end{figure}

\begin{figure}
\centering\includegraphics[page=8,width=.9\linewidth]{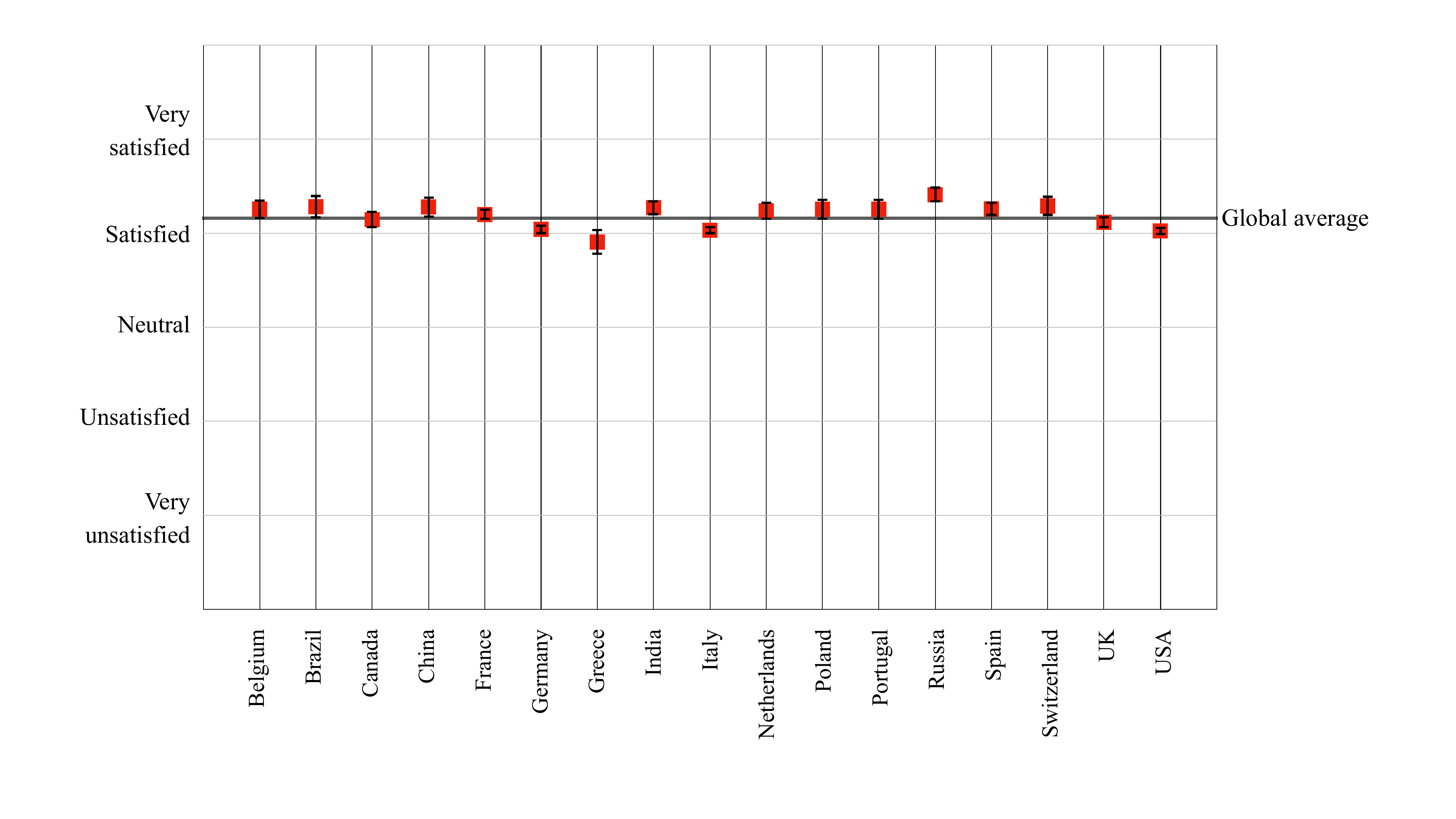}
\caption{Average satisfaction subdivided by nationality. Only countries with at least 40 entries are shown. (A full list of countries is given in Appendix A.)}
\label{fig:satisfaction_cern_country}
\end{figure}

Figure~\ref{fig:satisfaction_cern} shows the degree of satisfaction for the whole sample of respondents with respect to their working experience at CERN. More than 
80\% of them were either satisfied or very satisfied. The average satisfaction subdivided per country of nationality was uniform at a level of 4 in a scale between 1 and 
5 (Figure~\ref{fig:satisfaction_cern_country}).


\section{Additional comments}

The online questionnaire also contained an entry for free comments by the respondents. In total, 331 persons used this opportunity to provide additional information. The comments received can be classified in four general categories:
\begin{enumerate}
\item{Clarifications on some parts of the questionnaire or information relating to some of the questions.}
\item{Suggestions on possible additional topics to be added to the questionnaire. The items in question related mostly to the specific circumstances of the respondents.} 
\item{Personal impressions on the CERN environment, work and life. In this category, the positive comments underlined the impact of the CERN experience on the person's career, the experience obtained within CERN and the feeling of contributing to important scientific results. The negative comments showed issues with the frequent affiliation changes, the limited help available at CERN in finding jobs outside HEP, the compatibility between work and family life, the working conditions and the organization, the administration, the opportunities offered, the CERN recruitment, and the salary inequalities. In some cases, these remarks are mitigated by the CERN satisfaction ('CERN is great but..'). }
\item{Impressions related to the internal life and dynamics of the experimental collaborations. A few very positive comments refer to the enriching experience obtained within the collaborations. The majority of the negative remarks underline the pressure felt in some groups, the inadequate team management performed by some conveners, the hard internal competition, and the gaps in the organization of the work. }
\end{enumerate}
 
The full list of comments will be provided to CERN and to the experimental collaborations managements.

\chapter{Analysis of the questionnaire for theoretical physicists}
\label{ch:theory}

\section{The Theory Department at CERN}
Since CERN's foundation, a Theory group has existed with the mandate to provide fundamental contributions on the various aspects of theoretical physics, to support the needs, and to guide and profit from the findings of the experiments on site, as well as to offer a high-quality training environment~\cite{history-of-cern}.
Today, the Theory Department comprises a small core of  
 staff and a much larger community of current or past fellows, associate physicists and visitors. Students registered in the Theory Department are members of external Institutes.

\section{The data sample}
A dedicated questionnaire was prepared, taking into account the special characteristics of the Theory Department, on top of the main topics discussed in the case of the experimental collaborations. The questionnaire was addressed to the mailing list of the Theory Department, which contained the details of about 1000 current and past users. The questionnaire remained open for 10 weeks in the autumn of 2017. A total of 169 answers were received and analysed. This sample is too small (about 17\% of the address list) to give a representative picture of the total sample of theorists; the usual warnings concerning possible biases, described in Chapter 1, must also be applied also for the conclusions of this study.

\section{Demographic profile of the sample}
\begin{figure}
\centering\includegraphics[page=25,width=.9\linewidth]{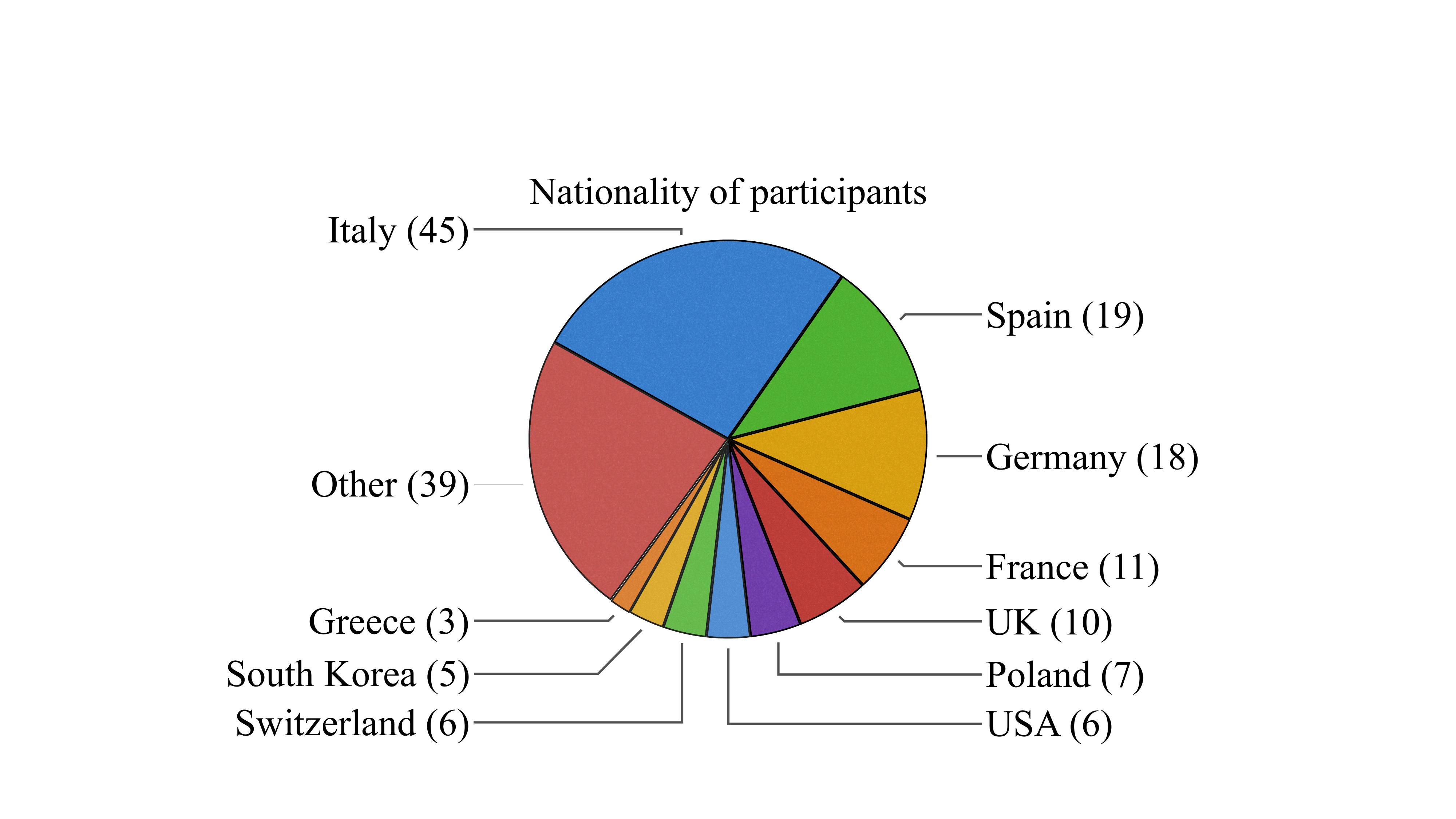}
\caption{Distribution of the main nationalities of users of the Theory Department who responded to the questionnaire.}
\label{fig:Theory_nationalities}
\end{figure}

\begin{figure}
\centering\includegraphics[width=.9\linewidth]{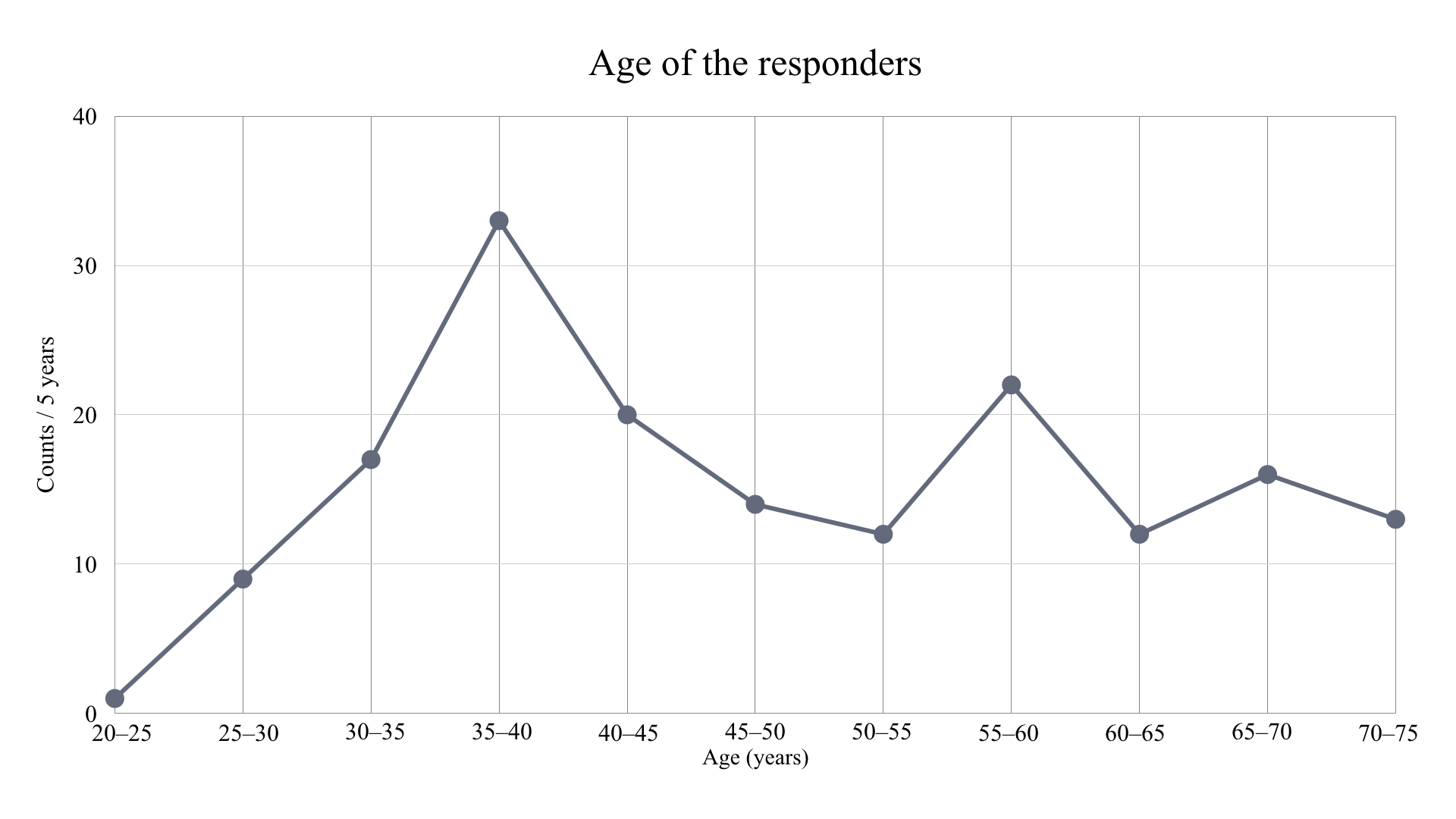}
\caption{Age distribution of the respondents}
\label{fig:Theory_age}
\end{figure}

\begin{figure}
\centering\includegraphics[trim={20cm 0 20cm 0},clip,page=26,width=.45\linewidth]{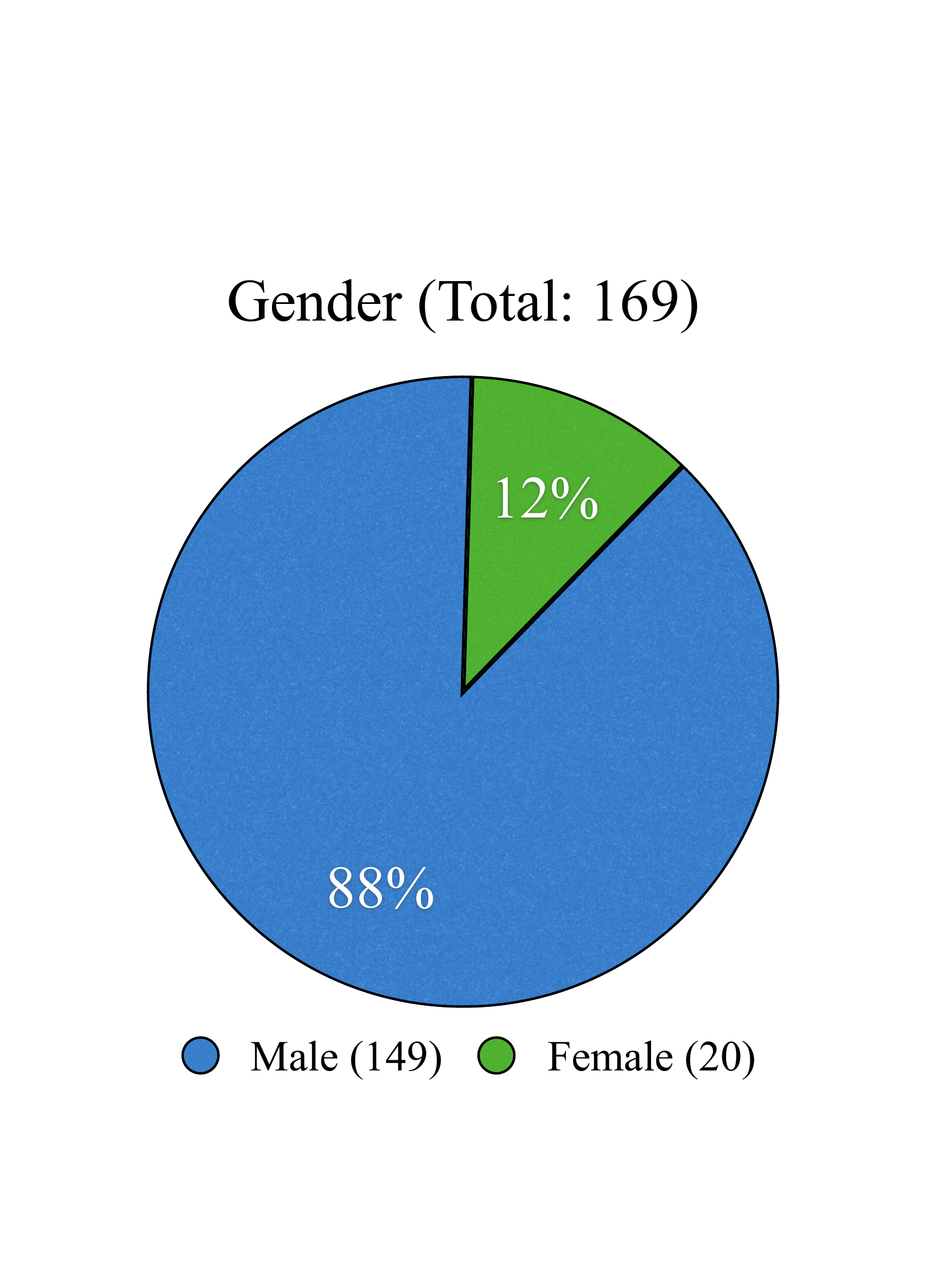}
\centering\includegraphics[trim={20cm 0 20cm 0},clip,page=62,width=.45\linewidth]{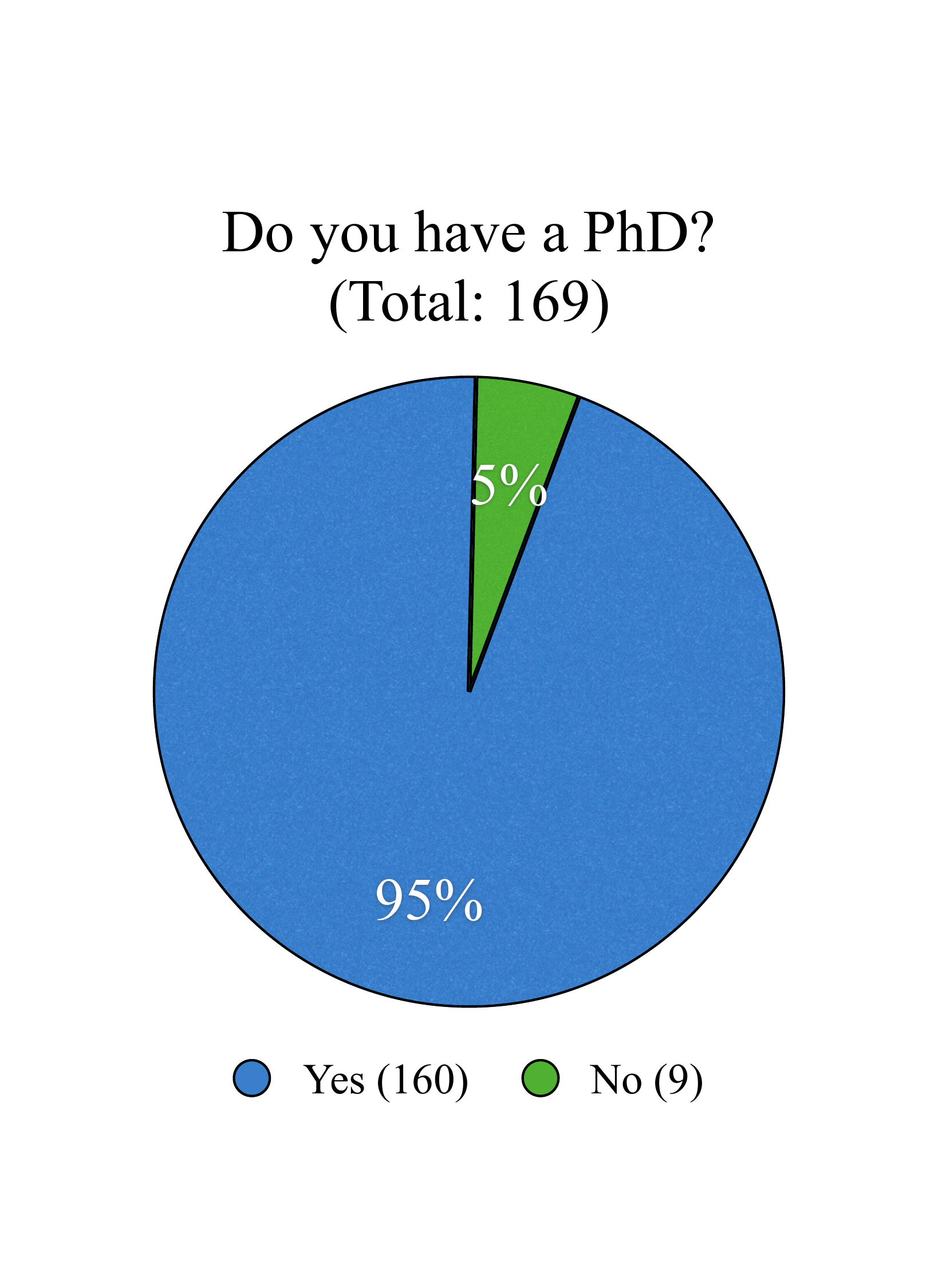}
\caption{Gender distribution (left) and fraction of respondents who obtained a PhD (right)}
\label{fig:Theory_gender&PhD}
\end{figure}

The left plot of Figure~\ref{fig:Theory_nationalities} illustrates the main nationalities of the analysed sample. Other nationalities with fewer than three entries do not appear in the pie chart: these include  Russia, India, Argentina, Canada, Portugal, Venezuela, Belgium,  Ukraine,  Denmark, Armenia, Brazil, Slovenia, Norway,  Sweden, Romania, Israel, Austria, Finland, Pakistan, Serbia, Georgia,  Luxembourg, Vietnam, South Africa and Slovakia. The age distribution of the respondents is shown in Figure~\ref{fig:Theory_age}.
The left panel of Figure~\ref{fig:Theory_gender&PhD} shows the gender distribution of the analysed sample. A fraction of 12\% of the respondents are women, in fair agreement with the average percentage of women in theoretical physics, which is lower than in experimental physics. 
A very large (95\%) fraction of the respondents have obtained their PhD (Figure~\ref{fig:Theory_gender&PhD} right).

\begin{figure}
\centering\includegraphics[trim={20cm 0 20cm 0},clip,page=29,width=.45\linewidth]{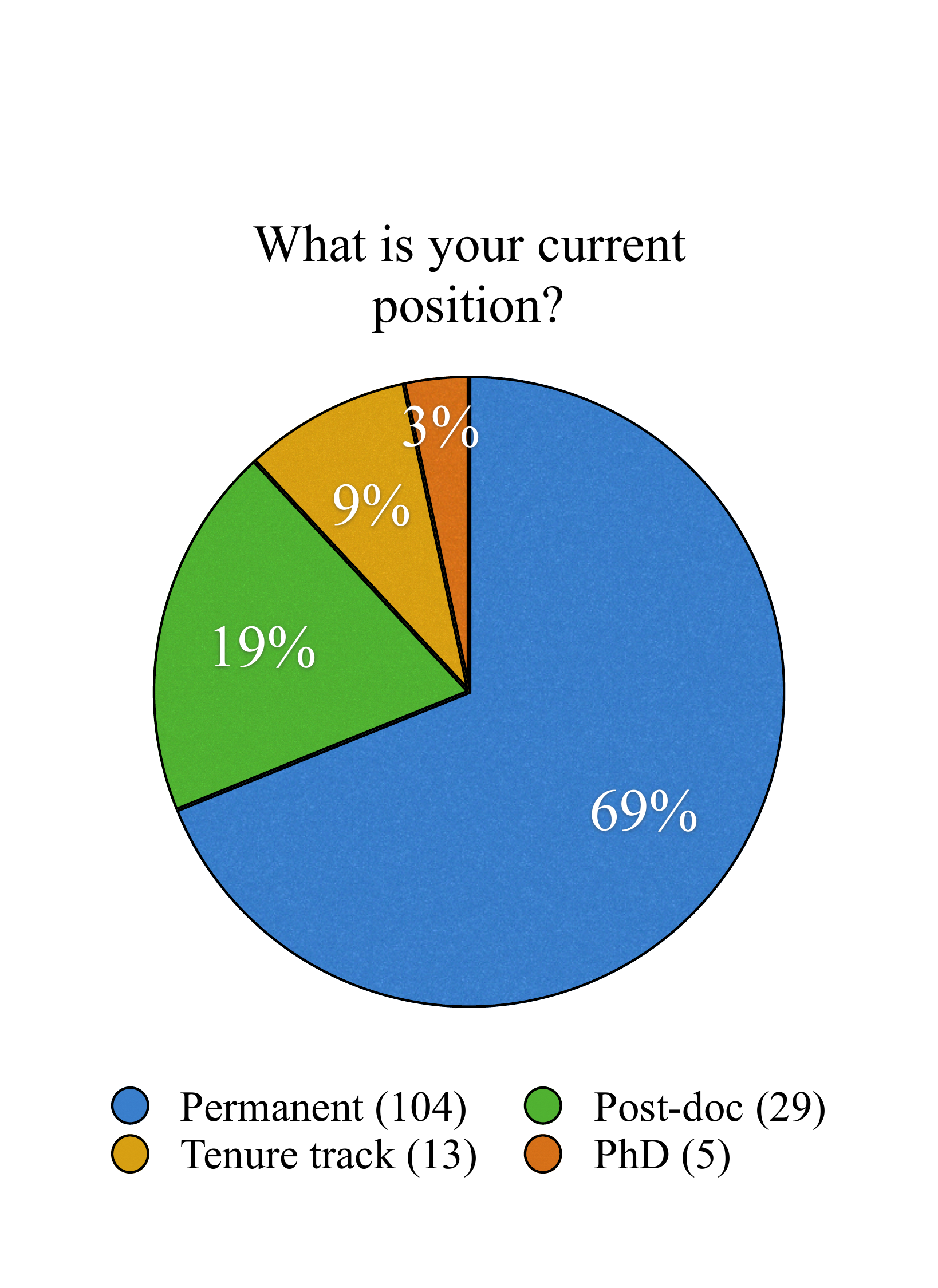}
\centering\includegraphics[trim={20cm 0 20cm 0},clip,page=64,width=.45\linewidth]{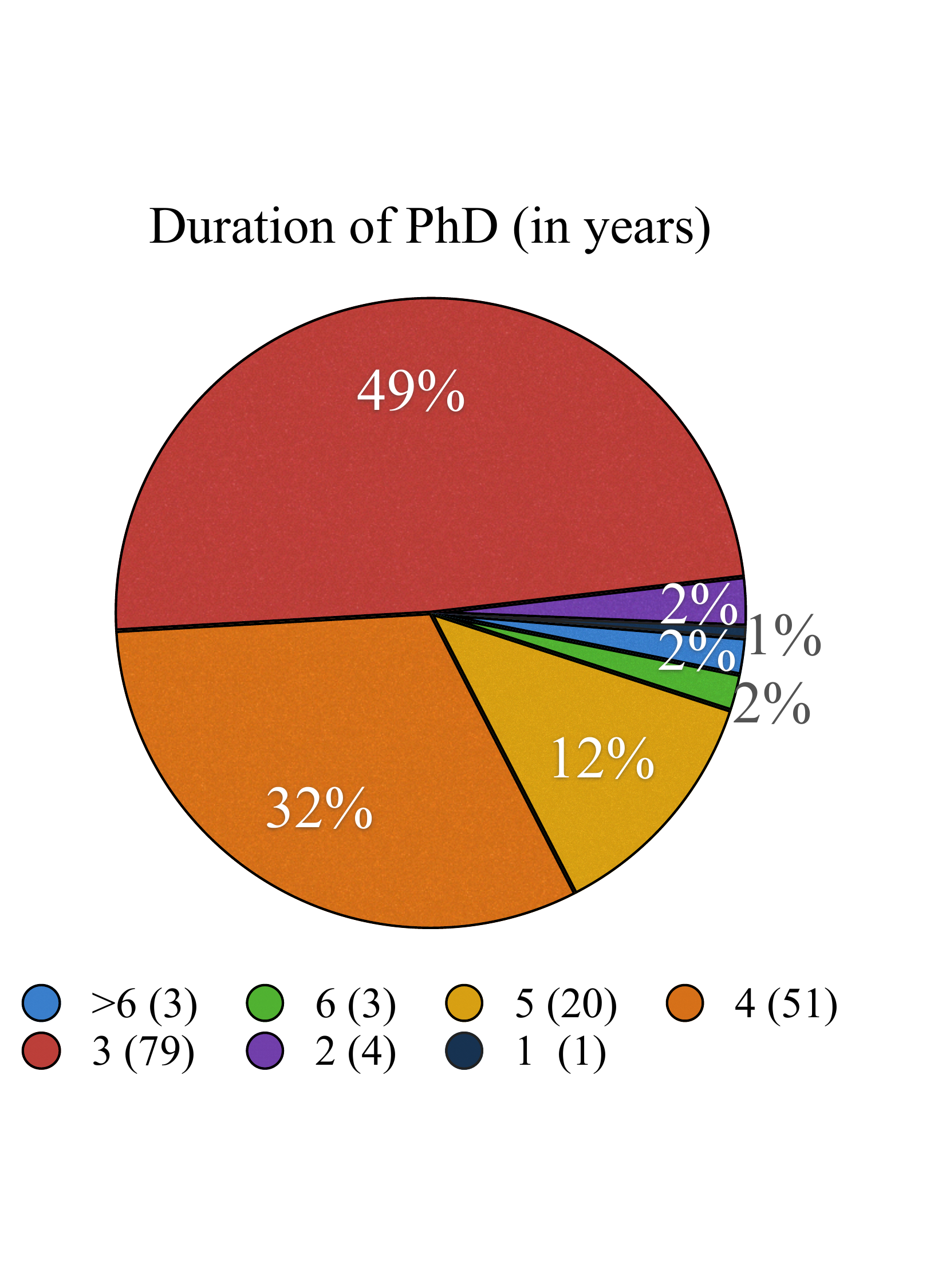}
\caption{Left: current position for the analysed sample. Right: duration (in years) of the PhD period.}
\label{fig:Theory_currentPosition_durationPhD}
\end{figure}

The time duration of the PhD is variable as for the experimentalists, from 3 to 6 years and more, as shown in the right plot of Figure~\ref{fig:Theory_currentPosition_durationPhD}. 

\begin{figure}
\centering\includegraphics[page=33,width=.9\linewidth]{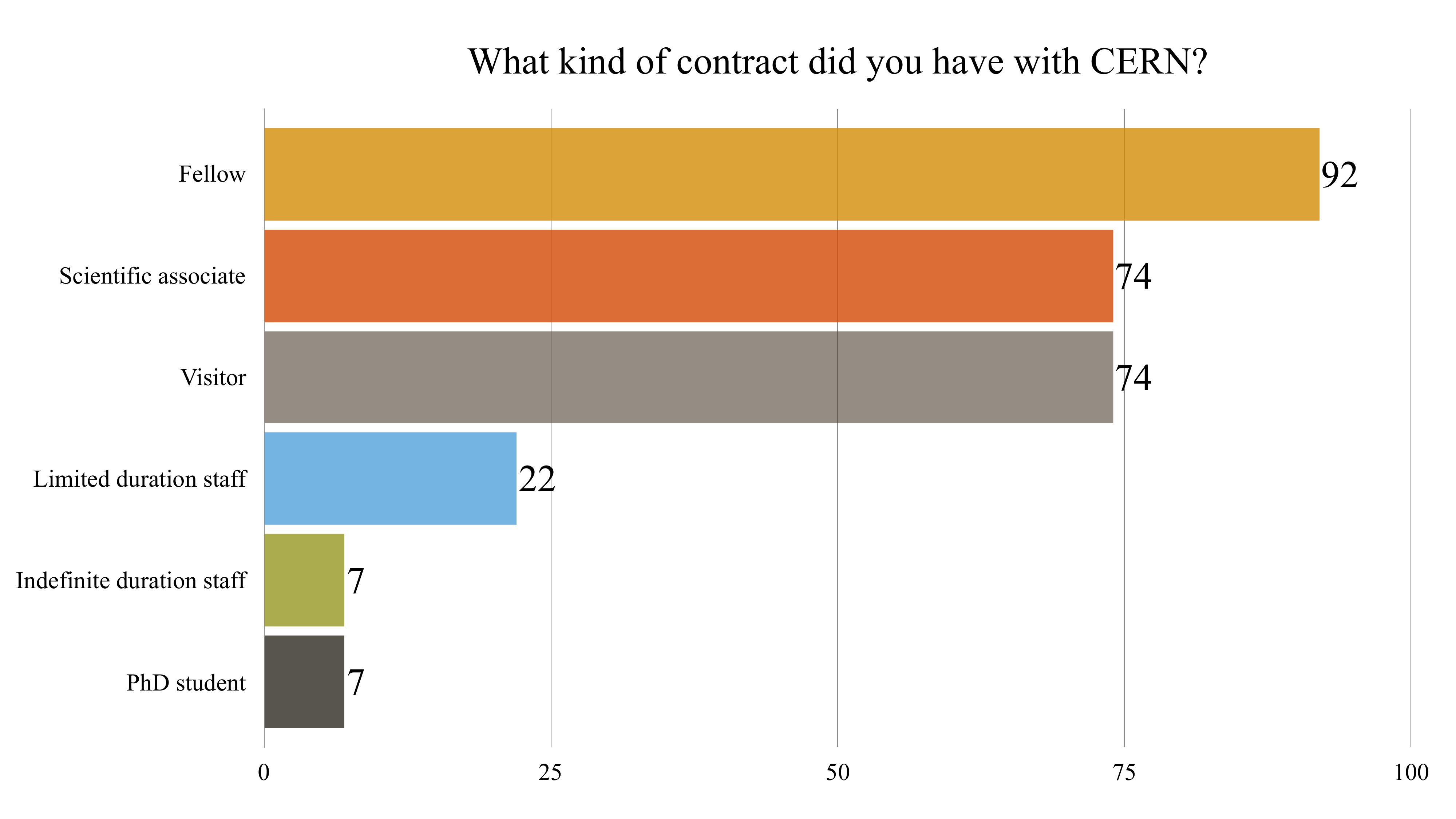}
\caption{Types of contracts that the respondents had with CERN: several answers were allowed}
\label{fig:Theory_contracts}
\end{figure}

The respondents had various types of contracts with CERN, as users in the Theory Department. Figure~\ref{fig:Theory_contracts} shows the distribution of contracts , where several answers were allowed. The majority of contracts were 2 year long CERN fellowships, which are usually awarded to young physicists who do not yet have a permanent position in outside institutions. Associate and visitor positions, foreseen mostly for more senior physicists, come next.     

\section{Career evolution}
\begin{figure}
\centering\includegraphics[page=35,width=.9\linewidth]{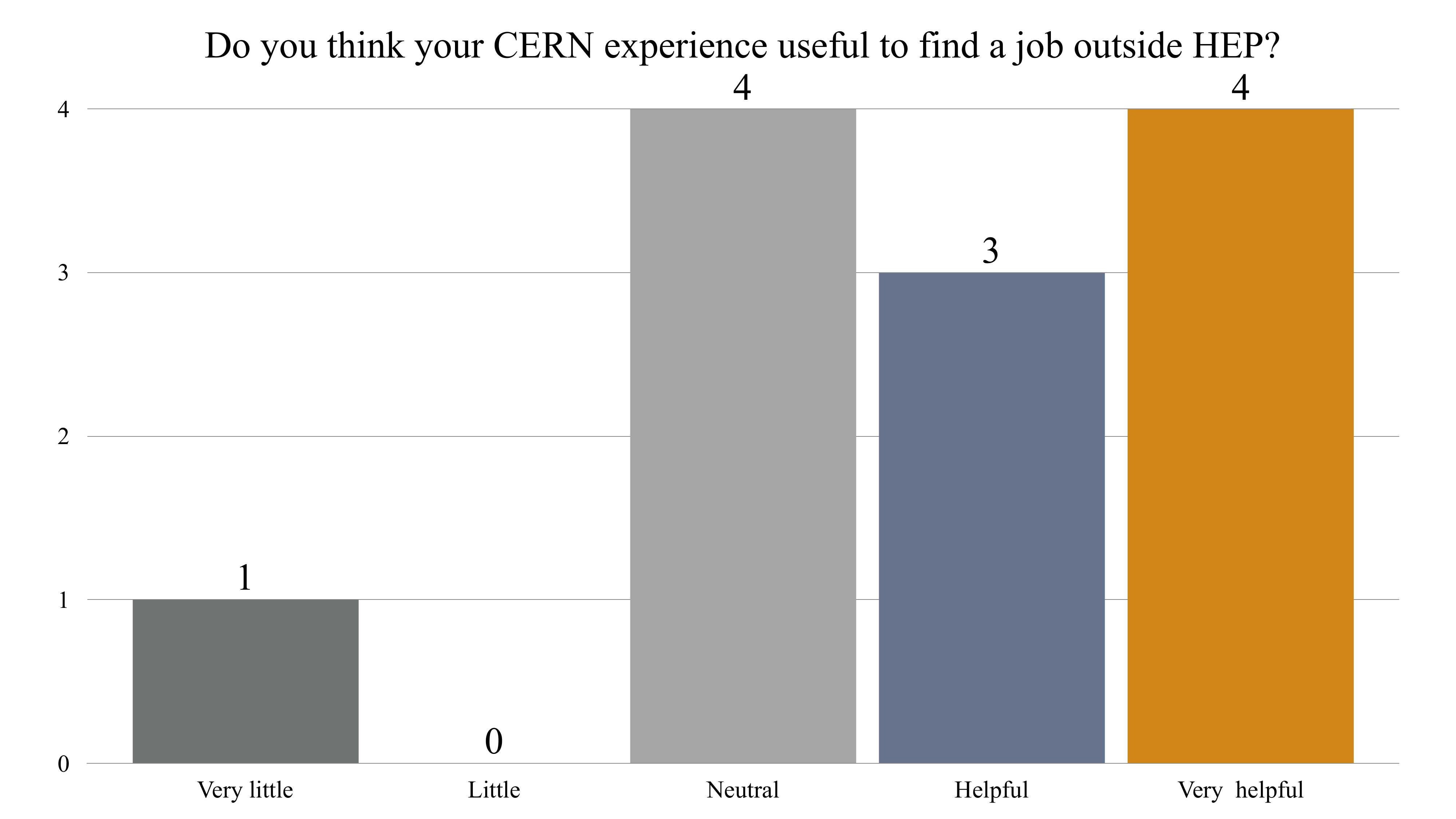}
\caption{Impact of CERN experience in finding a job outside HEP for former CERN theorists still working in HEP.}
\label{fig:Theory_Out_12}
\end{figure}

Over the total sample of 169 respondents, four persons left high-energy physics.  They had become teachers, professors and analysts. They left HEP after post-doc contracts (3) or a tenure track position (1).
Twelve other people still worked in HEP but were no longer CERN users. They spent from 1 to 9 years as users of the Theory Department. 
The majority currently had permanent or tenure track positions. Two had post-doc contracts, while two others were still students.  
 Among them, 11 wish to continue in HEP and 9 do not consider leaving the field at all. Eleven persons out of twelve believed that their CERN experience was decisive for their career, while ten considered it positive in general. The opinions on CERN's impact for finding a job outside HEP were diverse (Figure~\ref{fig:Theory_Out_12}).

\begin{figure}
\centering\includegraphics[trim={20cm 0 20cm 0},clip,page=28,width=.45\linewidth]{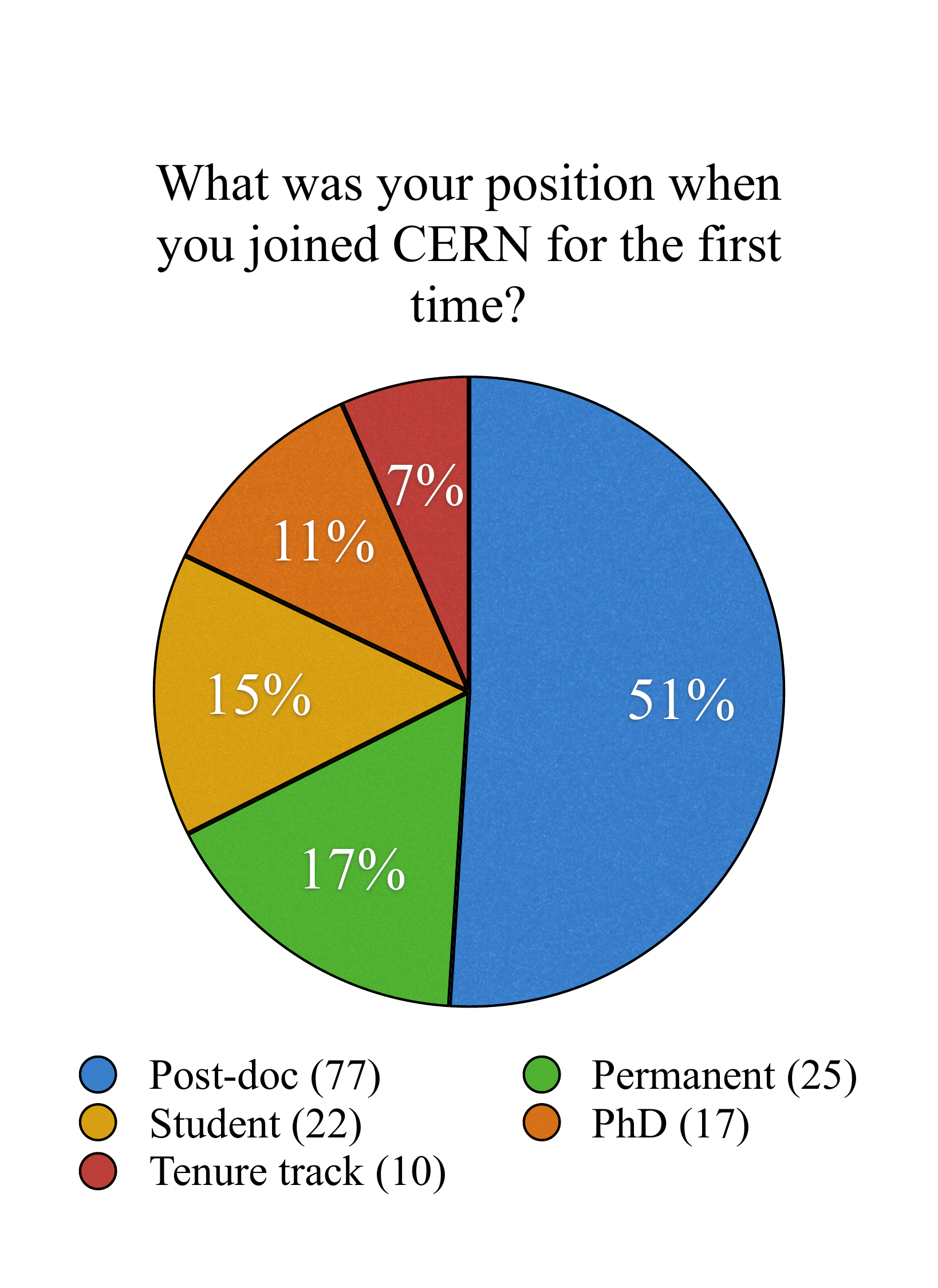}
\centering\includegraphics[trim={20cm 0 20cm 0},clip,page=63,width=.45\linewidth]{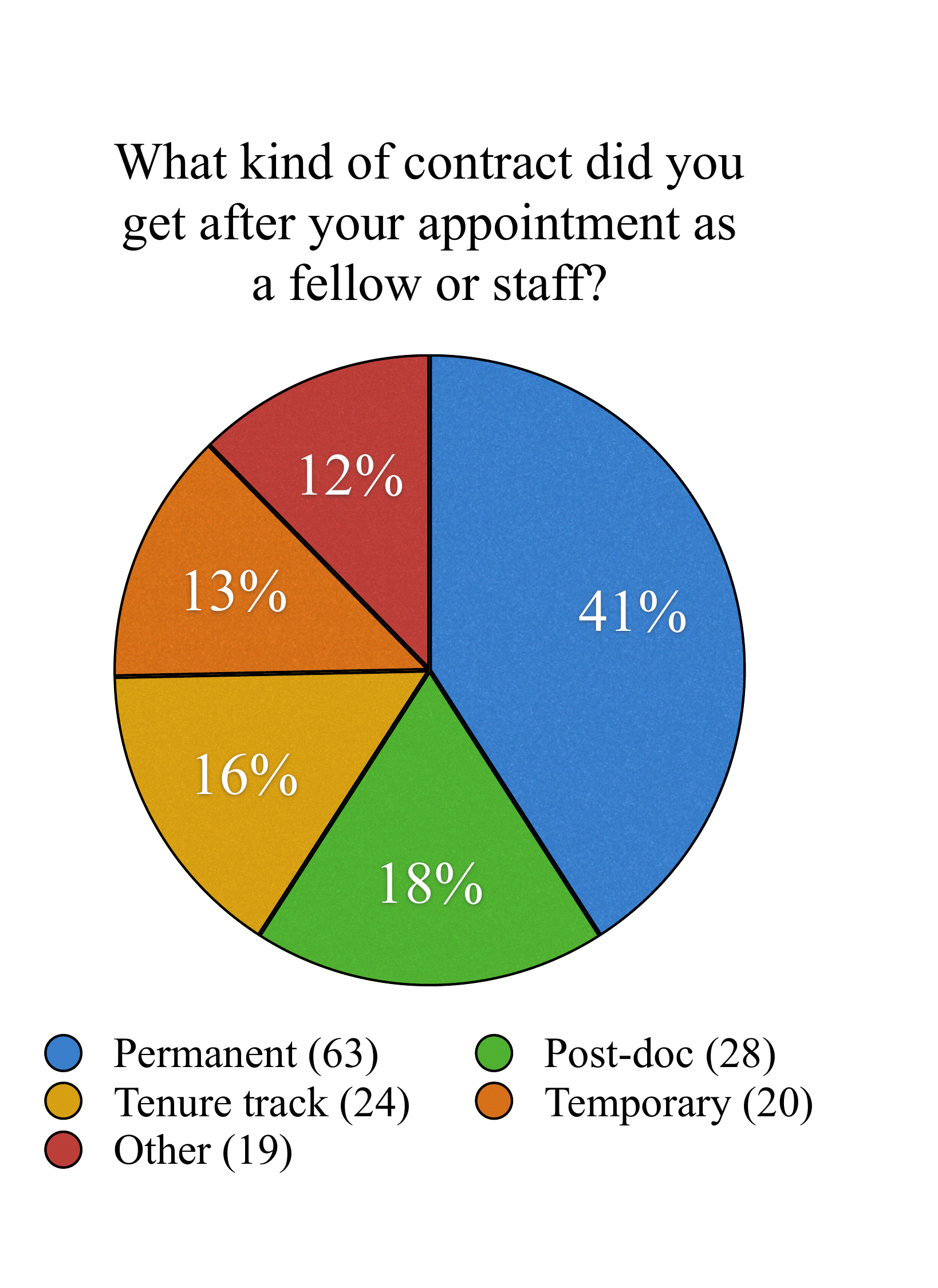}
\caption{Left: first contract type with CERN. Right: type of position obtained after the end of a CERN contract (fellowship or limited duration staff).}
\label{fig:Theory_FirstPosition_CurrentPosition}
\end{figure}

The vast majority of the respondents joined CERN at an early stage of their careers, as post-docs (51\%), PhDs (11\%) or undergraduate students (15\%). Only 24\% joined CERN while having a tenure track or permanent position and therefore as associate or visitor members of the Theory Department (Figure~\ref{fig:Theory_FirstPosition_CurrentPosition} left). A fraction of 41\% of people who left the Theory Department after a fellowship or a short-term staff contract were hired on a permanent position at the end of this contract, while an additional 16\% were hired with tenure track contracts.  (Figure~\ref{fig:Theory_FirstPosition_CurrentPosition} right). This demonstrates the positive impact of CERN short-term contracts on the future careers of young theorists. Notice that 78\% of the respondents currently occupied a permanent tenure track position, as shown in right-hand panel of Figure~\ref{fig:Theory_currentPosition_durationPhD}.

\section{Being a theorist at CERN}
\begin{figure}
\centering\includegraphics[page=30,width=.9\linewidth]{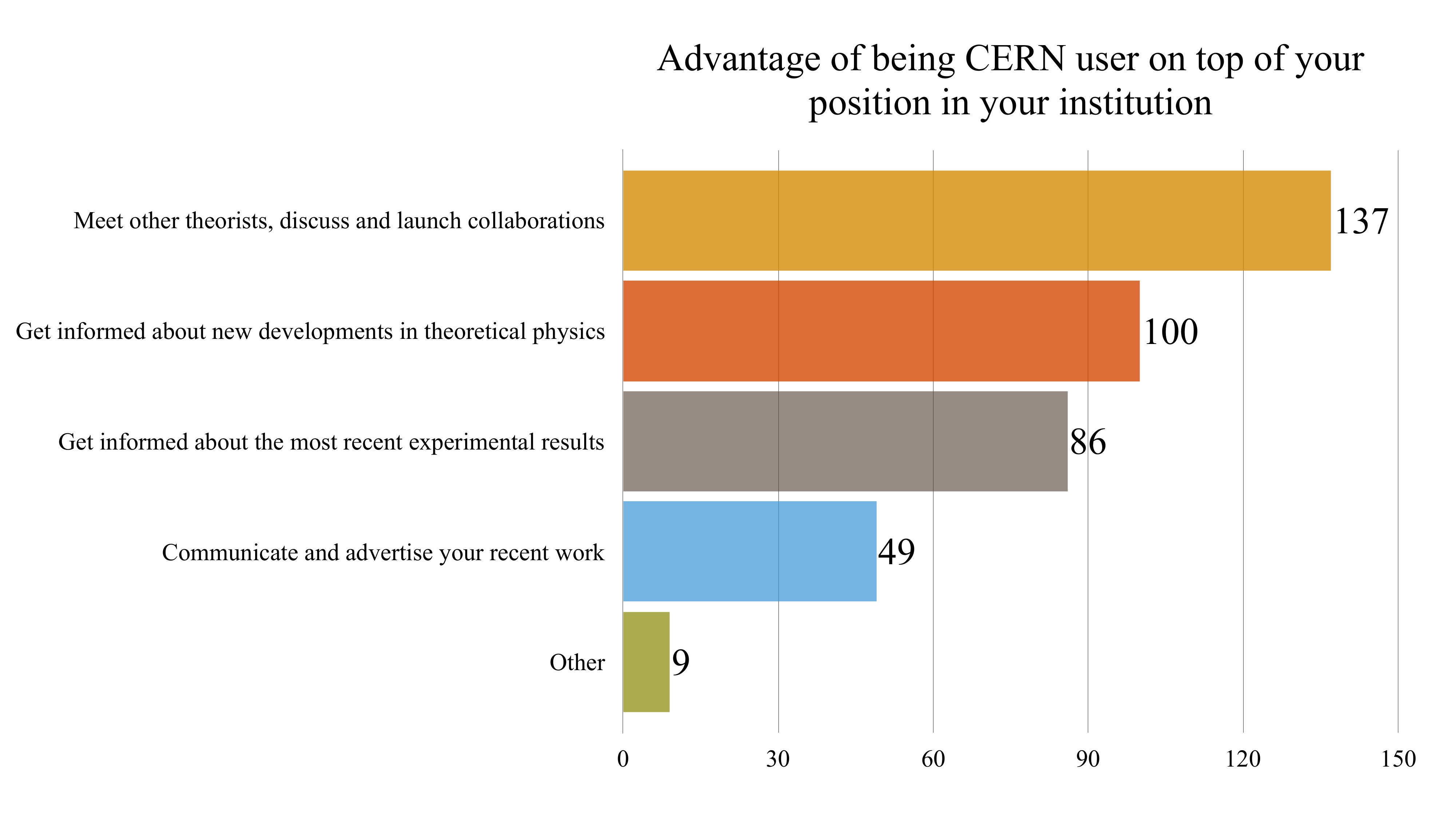}
\caption{Respondents' opinions on the advantages of being a CERN user in addition to their job in their institution: several answers were allowed.}
\label{fig:Theory_Advantage}
\end{figure}

A question was asked about the particular interest for a theorist of being user in the CERN Theory Department in addition to a position in a national institution. The aim was to investigate the possible difference compared with experimentalists, who are often members of a CERN-based experiment, and are therefore  required to have a frequent presence at CERN. Several reasons were proposed and several answers were allowed. The results are shown in Figure~\ref{fig:Theory_Advantage}. The main reasons expressed by the respondents were the possibility of meeting other theorists and starting collaborations and becoming informed about theoretical developments and the latest experimental results. Promotion of individual work was ranked afterwards. The answers noted  'Other' were detailed in the free comments area available in the questionnaire. They referred to the enhanced possibility of taking part in workshops, deriving benefit from the CERN environment, being involved in the combined theorist-experimentalist working groups and having access to the CERN Library.
More specific information regarding their interactions with the experimental community and their  scientific publications is given in Appendix B.

\section{Satisfaction with the CERN experience}
\begin{figure}
\centering\includegraphics[page=34,width=.9\linewidth]{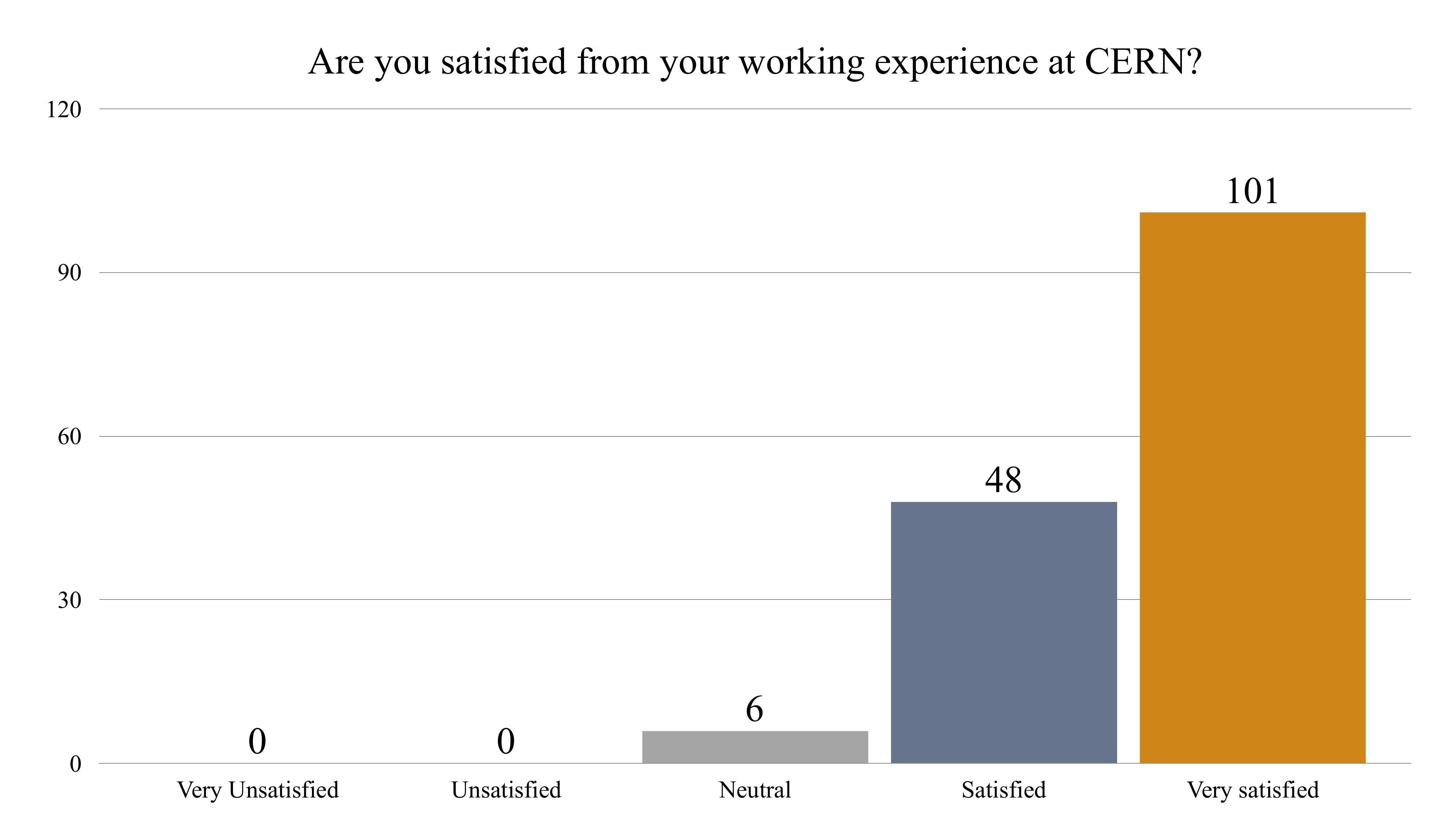}
\caption{Satisfaction of respondents with their working experience at CERN}
\label{fig:Theory_satisf}
\end{figure}

The final question concerned the overall satisfaction with the work experience at CERN. As was the case for the experimental community, satisfaction was graded from 1 (very unsatisfied) to 5 (very satisfied). The answers (Figure~\ref{fig:Theory_satisf}) show that 96\% of the respondents had a high level of satisfaction, distributed as 66\% at grade 5 and 31\% at grade 4; only 4\% gave a neutral answer (grade 3).

\chapter{Conclusions and appendices}

\section{Conclusions}

In this report, the results of a study on the training, career, age, mobility, expertise, and satisfaction of 
scientists who had a working experience at CERN are presented.
This study was based on an online questionnaire that was answered by 2692 people.
A separate questionnaire aimed at theoretical physicists was answered by 169 people.
The overall sample was composed of people who still worked in high-energy physics (1935) and people who left HEP (757), mostly in the previous 6-7 years, after their experience at CERN. 
It should be emphasized that, despite the large collected numbers of answers, the samples correspond to 
20-30\% of the addressed communities. 
This implies that some conclusions represent only a fraction of the CERN population and may not be totally unbiased.
From this sample of respondents, the following can be noted:
\begin{itemize}
\item{People from 84 (35) nationalities responded to the questionnaire for experimentalists (theorists), the majority of then coming from European countries.}
\item{Experimentalists currently working at CERN pursued their research in their home institutes in general, and only a small fraction were based at CERN for extended periods of time. The CERN environment and work experience were considered satisfactory or very satisfactory by 82\% of the respondents, evenly distributed across all the different nationalities.}
\item{ In 70\% of cases, people who left HEP mainly did so because of the often long and uncertain path towards obtaining a permanent job in the field. Other reasons, quoted by a smaller fraction of respondents, were interest in other fields, a lack of satisfaction at work and family reasons. The majority of the respondents (63\%) who left HEP were currently working in the private sector.
They occupied a wide range of positions and responsibilities, often close to IT, advanced technologies, and finance. Those in the public sector were mainly involved in academia or education.}
\item{For persons who left HEP, several skills developed during their experience at CERN were considered important in their current work. The overall satisfaction with the current position was high or very high for 78\% of the respondents, while CERN's impact for finding a job outside HEP was considered positive or very positive in 70\% of the answers. However, CERN's services and network were not found to be very effective in helping finding a new job outside HEP. This is an important point that certainly CERN will want to improve. A step in this direction has been made in creating the CERN Alumni programme; the human resources department can also play a crucial role in establishing and developing links between CERN users and  various private or public institutions and companies.}
\item{Theorists who responded to the questionnaire described in Chapter~\ref{ch:theory} mainly had permanent or tenure track positions. A large majority of them spent some time at CERN with a short- or medium-term contract in the Theory Department. This experience seemed to improve their career when leaving CERN for a national institution. Being CERN users, theorists could more easily start collaborations with other colleagues and get informed about the latest experimental results and the most recent theoretical developments. On average, about 35\% of the respondents scientific publications originate from collaborations started at CERN. 
A large fraction of respondents (96\%) were satisfied or highly satisfied with their experience at CERN.}
\end{itemize}

Despite the limited size of the sample analysed, this study allowed useful information to be extracted about people who had a working experience at CERN; a large majority of respondents gave positive feedback on the impact of their experience on their careers. It is also evident that there is room for improvement in some training areas, as well as in supporting the careers of those who choose to leave CERN and the high-energy physics field.
In the future, this study could be made more significant by collecting similar information from larger samples of people, especially former CERN users. The recently started CERN Alumni programme could help in building a continuously updated database of CERN current and former users and also provide more support for people who decide to leave HEP. 

\newpage
\appendix

\section{Appendix: Additional information about the experimental questionnaire results}

\begin{figure}[h]
\centering\includegraphics[trim={20cm 0 20cm 0},clip,page=2,width=.45\linewidth]{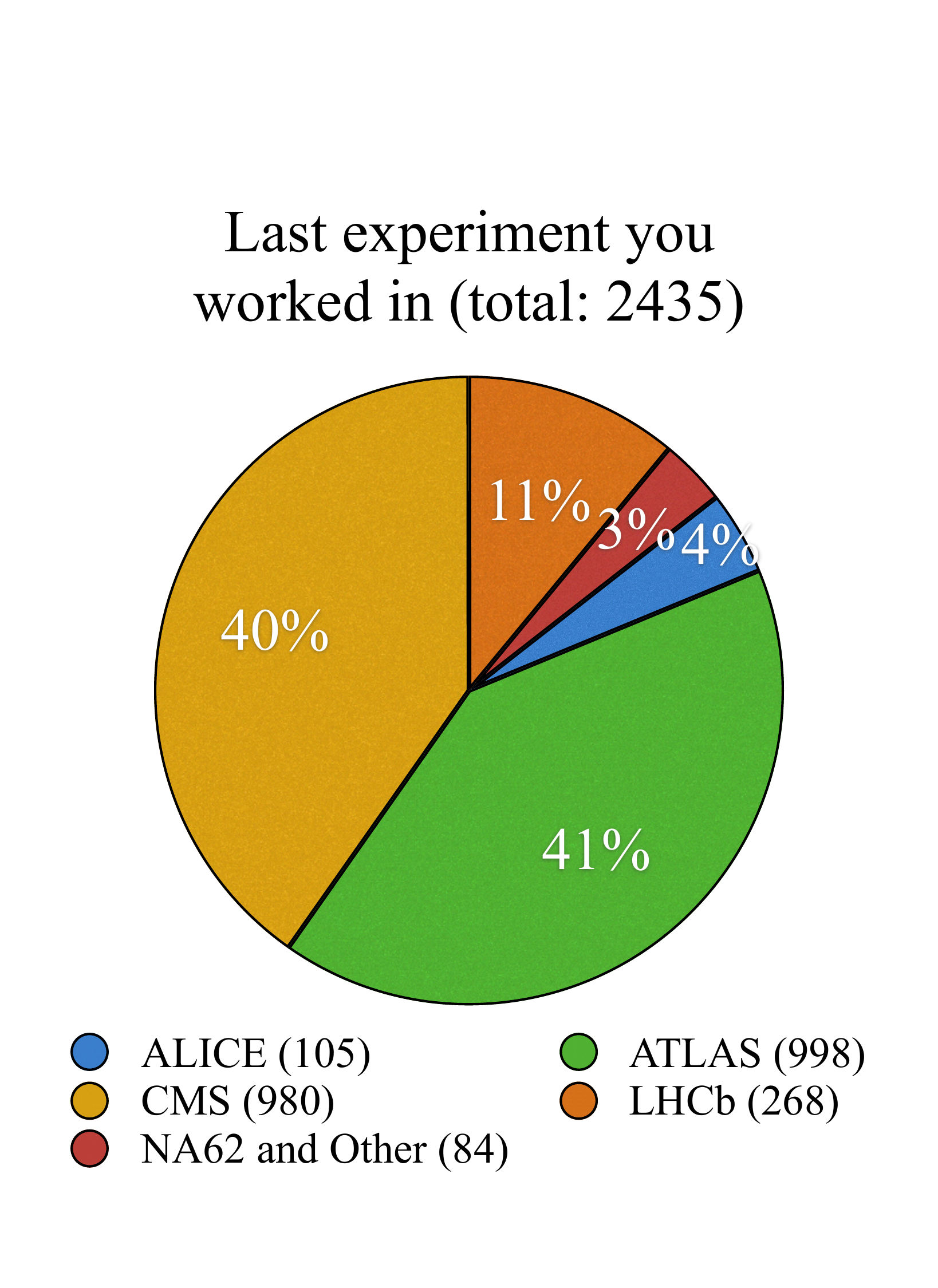}
\centering\includegraphics[trim={20cm 0 20cm 0},clip,page=59,width=.45\linewidth]{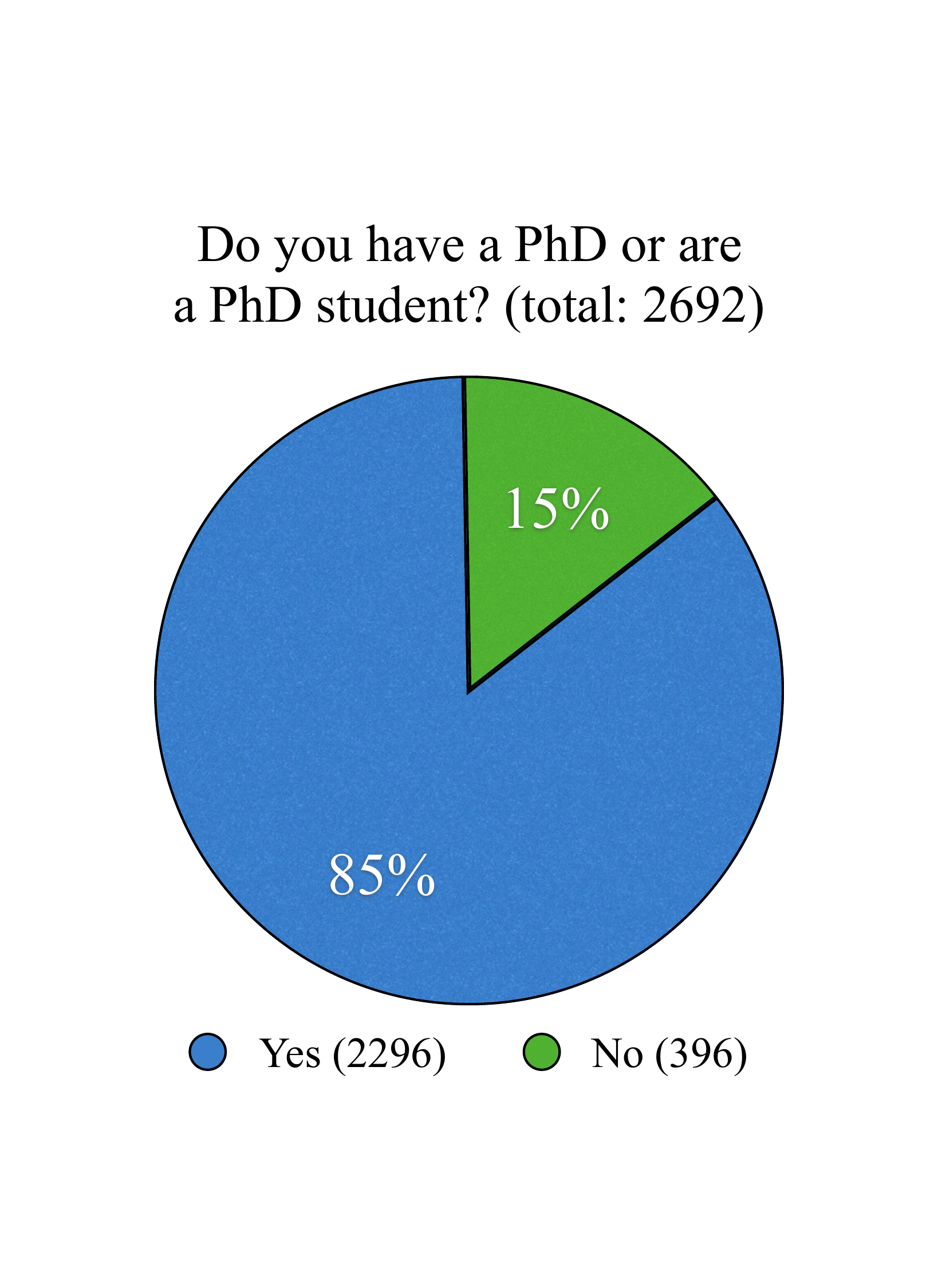}
\caption{Left: last experiment in which a respondent worked. Some respondents had not worked in any experiment (mostly administrative and technical personnel). Right: number of respondents who had obtained a PhD. }
\label{fig:experiment_phd}
\end{figure}




\newpage

\begin{table}[h]
\caption{Answers from people who left HEP, subdivided by country of citizenship and residence}
\centering
\begin{tabular}{llllll}
\hline \hline
Country & PhD & Residence & Nationality & Residence $-$ PhD & (Residence $-$ PhD) / PhD \\
\hline
Argentina & \phantom{$10$}$1$ & \phantom{$10$}$1$ & \phantom{$10$}$1$ & \phantom{$-1$}$0$ & \phantom{$-$}$0.0$ \\
Armenia & \phantom{$10$}$2$ & \phantom{$10$}$0$ & \phantom{$10$}$4$ & \phantom{$1$}$-2$ & $-1$ \\
Australia& \phantom{$10$}$4$ & \phantom{$10$}$5$ & \phantom{$10$}$3$ & \phantom{$-1$}$1$   & \phantom{$-$}$0.25$ \\
Austria  & \phantom{$10$}$9$ & \phantom{$1$}$12$ & \phantom{$10$}$8$ & \phantom{$-1$}$3$  & \phantom{$-$}$0.33$ \\
Belgium  & \phantom{$1$}$12$ & \phantom{$1$}$18$ & \phantom{$1$}$14$ & \phantom{$-1$}$6$ & \phantom{$-$}$0.5$ \\
Brazil   & \phantom{$10$}$8$ & \phantom{$1$}$20$ & \phantom{$1$}$21$ & \phantom{$-$}$12$ & \phantom{$-$}$1.5$  \\
Canada   & \phantom{$1$}$18$ & \phantom{$1$}$26$ & \phantom{$1$}$27$ & \phantom{$-1$}$8$ & \phantom{$-$}$0.44$ \\
China    & \phantom{$10$}$1$ & \phantom{$10$}$0$ & \phantom{$10$}$4$ & \phantom{$1$}$-1$ & $-1.0$ \\
Czech Republic & \phantom{$10$}$8$ & \phantom{$1$}$13$ & \phantom{$1$}$11$ & \phantom{$-1$}$5$ & \phantom{$-$}$0.625$ \\
Denmark  & \phantom{$10$}$1$ & \phantom{$10$}$5$ & \phantom{$10$}$2$ & \phantom{$-1$}$4$ & \phantom{$-$}$4.0$ \\
Finland  & \phantom{$10$}$1$ & \phantom{$10$}$3$ & \phantom{$10$}$1$ & \phantom{$-1$}$2$ & \phantom{$-$}$2.0$ \\
France   & \phantom{$1$}$51$ & \phantom{$1$}$57$ & \phantom{$1$}$58$ & \phantom{$-1$}$6$ & \phantom{$-$}$0.12$ \\
Germany  & $110$ & $110$ & $112$      & \phantom{$-1$}$0$ & \phantom{$-$}$0.0$ \\
Greece   & \phantom{$10$}$3$ & \phantom{$10$}$3$ & \phantom{$1$}$14$ & \phantom{$-1$}$0$ & \phantom{$-$}$0.0$ \\
Hungary  & \phantom{$10$}$1$ & \phantom{$10$}$6$ & \phantom{$10$}$5$ & \phantom{$-1$}$5$ & \phantom{$-$}$5.0$ \\
India    & \phantom{$10$}$2$ & \phantom{$10$}$5$ & \phantom{$1$}$16$ & \phantom{$-1$}$3$ & \phantom{$-$}$1.5$ \\
Ireland  & \phantom{$10$}$3$ & \phantom{$10$}$6$ & \phantom{$10$}$4$ & \phantom{$-1$}$3$ & \phantom{$-$}$1.0$ \\
Israel   & \phantom{$10$}$1$ & \phantom{$10$}$2$ & \phantom{$10$}$2$ & \phantom{$-1$}$1$ & \phantom{$-$}$1.0$ \\
Italy    & \phantom{$1$}$57$ & \phantom{$1$}$52$ & $105$ & \phantom{$1$}$-5$ & $-0.09$ \\
Japan    & \phantom{$10$}$3$ & \phantom{$10$}$7$ & \phantom{$10$}$4$ & \phantom{$-1$}$4$ & \phantom{$-$}$1.33$ \\ 
Luxembourg & \phantom{$10$}$1$ & \phantom{$10$}$2$ & \phantom{$10$}$1$ & \phantom{$-1$}$1$ & \phantom{$-$}$1.0$ \\
Netherlands & \phantom{$1$}$20$ & \phantom{$1$}$18$ & \phantom{$1$}$14$ & \phantom{$1$}$-2$ & $-0.1$ \\
Norway   & \phantom{$1$}$11$ & \phantom{$1$}$14$ & \phantom{$1$}$13$ & \phantom{$-1$}$3$ & \phantom{$-$}$0.27$ \\
Poland   & \phantom{$10$}$9$ & \phantom{$1$}$11$ & \phantom{$1$}$18$ & \phantom{$-1$}$2$ & \phantom{$-$}$0.22$ \\
Portugal & \phantom{$10$}$4$ & \phantom{$1$}$10$ & \phantom{$1$}$20$ & \phantom{$-1$}$6$ & \phantom{$-$}$1.5$ \\
Romania  & \phantom{$10$}$1$ & \phantom{$10$}$1$ & \phantom{$10$}$6$ & \phantom{$-1$}$0$ & \phantom{$-$}$0.0$ \\
Russia   & \phantom{$10$}$4$ & \phantom{$10$}$9$ & \phantom{$1$}$10$ & \phantom{$-1$}$5$ & \phantom{$-$}$1.25$ \\
Serbia   & \phantom{$10$}$1$ & \phantom{$10$}$1$ & \phantom{$10$}$1$ & \phantom{$-1$}$0$ & \phantom{$-$}$0.0$ \\
Slovakia & \phantom{$10$}$1$ & \phantom{$10$}$1$ & \phantom{$10$}$6$ & \phantom{$-1$}$0$ & \phantom{$-$}$0.0$ \\
South Korea & \phantom{$10$}$1$ & \phantom{$10$}$5$ & \phantom{$10$}$5$ & \phantom{$-1$}$4$ & \phantom{$-$}$4.0$ \\
Spain    & \phantom{$1$}$10$ & \phantom{$1$}$11$ & \phantom{$1$}$17$ & \phantom{$-1$}$1$ & \phantom{$-$}$0.1$ \\
Sweden   & \phantom{$10$}$8$ & \phantom{$10$}$7$ & \phantom{$1$}$10$ & \phantom{$1$}$-1$ & $-0.125$ \\
Switzerland & \phantom{$1$}$27$ & \phantom{$1$}$95$ & \phantom{$1$}$22$ & \phantom{$-$}$68$ & \phantom{$-$}$2.52$ \\
Turkey   & \phantom{$10$}$2$ & \phantom{$10$}$5$ & \phantom{$10$}$6$ & \phantom{$-1$}$3$ & \phantom{$-$}$1.5$ \\
United Kingdom & \phantom{$1$}$61$ & \phantom{$1$}$61$ & \phantom{$1$}$42$ & \phantom{$-1$}$0$ & \phantom{$-$}$0.0$ \\
United States  & $121$ & $131$ & \phantom{$1$}$99$ & \phantom{$-$}$10$ & \phantom{$-$}$0.08$ \\
\hline \hline
\end{tabular}
\end{table}
\begin{table}
\caption{Answers from people  still in HEP subdivided by country of citizenship and of residence.}
\centering \footnotesize
\begin{tabular}{llllll}
\hline \hline
Country & PhD & Residence & Nationality & Residence $-$ PhD & (Residence $-$ PhD) / PhD \\
\hline
Argentina& \phantom{$20$}$7$ & \phantom{$20$}$5$ & \phantom{$2$}$10$ & \phantom{$20$}$-2$ & $-0.29$ \\
Armenia & \phantom{$20$}$2$ & \phantom{$20$}$1$ & \phantom{$20$}$3$ & \phantom{$20$}$-1$ & $-0.5$\\
Australia & \phantom{$20$}$4$ & \phantom{$20$}$7$ &     \phantom{$20$}$8$ & \phantom{$-20$}$3$ & \phantom{$-$}$0.75$ \\
Austria & \phantom{$2$}$27$ & \phantom{$2$}$16$ & \phantom{$2$}$26$ & \phantom{$2$}$-11$ & $-0.41$\\
Belarus & \phantom{$20$}$3$ & \phantom{$20$}$3$ & \phantom{$20$}$5$ &   \phantom{$-20$}$0$ & \phantom{$-$}$0.0$ \\
Belgium & \phantom{$2$}$43$ & \phantom{$2$}$33$ & \phantom{$2$}$37$ & \phantom{$0$}$-10$ & $-0.23$\\
Brazil & \phantom{$2$}$30$ & \phantom{$2$}$34$ & \phantom{$2$}$43$ & \phantom{$-20$}$4$ & \phantom{$-$}$0.13$\\
Bulgaria & \phantom{$20$}$8$ & \phantom{$20$}$9$ & \phantom{$2$}$14$ & \phantom{$-20$}$1$ & \phantom{$-$}$0.125$\\
Canada & \phantom{$2$}$39$ & \phantom{$2$}$32$ & \phantom{$2$}$38$ & \phantom{$20$}$-7$ & $-0.18$\\
Chile  & \phantom{$20$}$4$ & \phantom{$20$}$3$ & \phantom{$20$}$8$ & \phantom{$20$}$-1$ & $-0.25$\\
China & \phantom{$2$}$17$ & \phantom{$2$}$16$ & \phantom{$2$}$39$ & \phantom{$20$}$-1$ & $-0.06$\\
Croatia  & \phantom{$20$}$4$ & \phantom{$20$}$7$ & \phantom{$2$}$11$ & \phantom{$-20$}$3$ & \phantom{$-$}$0.75$ \\
Cyprus & \phantom{$20$}$1$ & \phantom{$20$}$2$ &  \phantom{$20$}$5$ & \phantom{$-20$}$1$ & \phantom{$-$}$1.0$ \\
Czech Republic & \phantom{$2$}$15$ & \phantom{$2$}$15$ & \phantom{$2$}$17$ & \phantom{$-20$}$0$ & \phantom{$-$}$0.0$\\
Denmark & \phantom{$20$}$6$ & \phantom{$20$}$8$ & \phantom{$2$}$12$ & \phantom{$-20$}$2$ & \phantom{$-$}$0.33$ \\
Egypt & \phantom{$20$}$4$ & \phantom{$20$}$7$ & \phantom{$20$}$7$ & \phantom{$-20$}$3$ & \phantom{$-$}$0.75$\\
Estonia & \phantom{$20$}$1$ & \phantom{$20$}$2$ & \phantom{$20$}$3$ &   \phantom{$-20$}$1$ & \phantom{$-$}$1.0$ \\
Finland & \phantom{$2$}$12$ & \phantom{$20$}$8$ & \phantom{$2$}$14$ & \phantom{$20$}$-4$ & $-0.33$ \\
France  & \phantom{$2$}$15$ & $169$ & $130$ & \phantom{$-2$}$15$ &  \phantom{$-$}$0.097$ \\
Georgia & \phantom{$20$}$3$    & \phantom{$20$}$3$ & \phantom{$2$}$11$ & \phantom{$-20$}$0$ & \phantom{$-$}$0.0$ \\
Germany & $206$ & $203$ & $204$ & \phantom{$20$}$-3$ & $-0.015$ \\
Greece  & \phantom{$2$}$25$ & \phantom{$2$}$14$ & \phantom{$2$}$43$ & \phantom{$0$}$-11$ & $-0.44$ \\
Hungary & \phantom{$20$}$8$     & \phantom{$2$}$13$ & \phantom{$2$}$14$ & \phantom{$-20$}$5$ & \phantom{$-$}$0.625$ \\
India   & \phantom{$2$}$55$ & \phantom{$2$}$54$ & \phantom{$2$}$83$ & \phantom{$20$}$-1$ & $-0.02$ \\
Iran    & \phantom{$20$}$3$ & \phantom{$20$}$3$ & \phantom{$20$}$9$ & \phantom{$-20$}$0$ & \phantom{$-$}$0.0$ \\
Ireland & \phantom{$20$}$3$ & \phantom{$20$}$2$ & \phantom{$20$}$3$ & \phantom{$20$}$-1$ & $-0.33$ \\
Israel  & \phantom{$20$}$5$ & \phantom{$2$}$13$ & \phantom{$2$}$12$ & \phantom{$-20$}$8$ & \phantom{$-$}$1.6$ \\
Italy   & $240$ & $210$ & $373$ & \phantom{$0$}$-30$ &       $-0.125$ \\
Japan   & \phantom{$2$}$13$ & \phantom{$20$}$6$ & \phantom{$2$}$13$ & \phantom{$20$}$-7$ & $-0.54$ \\
Lebanon & \phantom{$20$}$1$ & \phantom{$20$}$1$ & \phantom{$20$}$2$ & \phantom{$-20$}$0$ &         \phantom{$-$}$0.0$ \\ 
Malaysia        & \phantom{$20$}$1$ & \phantom{$20$}$1$  & \phantom{$20$}$3$ & \phantom{$-20$}$0$ & \phantom{$-$}$0.0$ \\
Mexico  & \phantom{$20$}$2$ & \phantom{$20$}$8$ & \phantom{$20$}$7$ & \phantom{$-20$}$6$ & \phantom{$-$}$3.0$ \\
Morocco & \phantom{$20$}$2$ & \phantom{$20$}$2$ & \phantom{$20$}$2$ & \phantom{$-20$}$0$ & \phantom{$-$}$0.0$ \\
Netherlands     & \phantom{$2$}$41$ & \phantom{$2$}$33$ & \phantom{$2$}$47$ & \phantom{$20$}$-8$ &     $-0.2$ \\
Norway  & \phantom{$2$}$10$ & \phantom{$20$}$7$ &      \phantom{$20$}$8$ &     \phantom{$20$}$-3$ &    $-0.3$ \\
Poland  & \phantom{$2$}$33$     & \phantom{$2$}$29$ &  \phantom{$2$}$38$ &    \phantom{$20$}$-4$ &     $-0.12$ \\
Portugal        & \phantom{$2$}$17$ & \phantom{$2$}$23$ &      \phantom{$2$}$28$ &     \phantom{$-20$}$6$ &      \phantom{$-$}$0.35$ \\
Romania & \phantom{$20$}$4$     & \phantom{$20$}$5$ &   \phantom{$20$}$7$ &     \phantom{$-20$}$1$ &      \phantom{$-$}$0.25$ \\
Russia  & \phantom{$2$}$56$ & \phantom{$2$}$52$ &     \phantom{$2$}$69$ &   \phantom{$20$}$-4$ &    $-0.07$ \\
Serbia  & \phantom{$20$}$1$ & \phantom{$20$}$1$ & \phantom{$20$}$5$ & \phantom{$-20$}$0$ &       \phantom{$-$}$0.0$ \\
Slovakia &\phantom{$20$}$3$ & \phantom{$20$}$3$ &  \phantom{$20$}$8$ &  \phantom{$-20$}$0$ &     \phantom{$-$}$0.0$ \\
Slovenia & \phantom{$20$}$5$     & \phantom{$20$}$4$ & \phantom{$20$}$5$ &       \phantom{$20$}$-1$ &    $-0.2$ \\
South Africa &  \phantom{$20$}$2$ & \phantom{$20$}$4$ & \phantom{$20$}$5$ &     \phantom{$-20$}$2$ &      \phantom{$-$}$1.0$ \\
South Korea & \phantom{$20$}$7$ &\phantom{$20$}$7$ &  \phantom{$2$}$11$ & \phantom{$-20$}$0$ & \phantom{$-$}$0.0$ \\
Spain   & \phantom{$2$}$81$ & \phantom{$2$}$61$ &  \phantom{$2$}$95$ &       \phantom{$0$}$-20$ &   $-0.25$ \\
Sweden  & \phantom{$2$}$31$ & \phantom{$2$}$23$ & \phantom{$2$}$20$ &        \phantom{$20$}$-8$ &    $-0.26$ \\
Switzerland     & \phantom{$2$}$73$ & $428$ &    \phantom{$2$}$37$ &    \phantom{$-$}$355$ &   \phantom{$-$}$4.86$ \\
Taiwan  & \phantom{$20$}$1$     & \phantom{$20$}$9$ & \phantom{$20$}$7$ & \phantom{$-20$}$8$ &   \phantom{$-$}$8.0$ \\
Turkey  & \phantom{$2$}$10$ & \phantom{$2$}$13$ &     \phantom{$2$}$25$ &    \phantom{$-20$}$3$ &     \phantom{$-$}$0.3$ \\
Ukraine & \phantom{$20$}$2$     & \phantom{$20$}$1$ &   \phantom{$2$}$19$ &    \phantom{$20$}$-1$ &    $-0.5$ \\
United Arab Emirates &  \phantom{$20$}$1$ & \phantom{$20$}$0$ & & \phantom{$20$}$-1$ &    $-1.0$  \\
United Kingdom  & $162$ & $130$ &   $120$ &   \phantom{$0$}$-32$ & $-0.2$ \\
United States   & $350$ & $296$ &   $245$ &   \phantom{$0$}$-54$ & $-0.15$ \\
\hline \hline
\end{tabular}
\end{table}

\newpage
\section{Appendix: Being a theorist at CERN}

\begin{figure}[h]
\centering{\includegraphics[page=31,width=.9\linewidth]{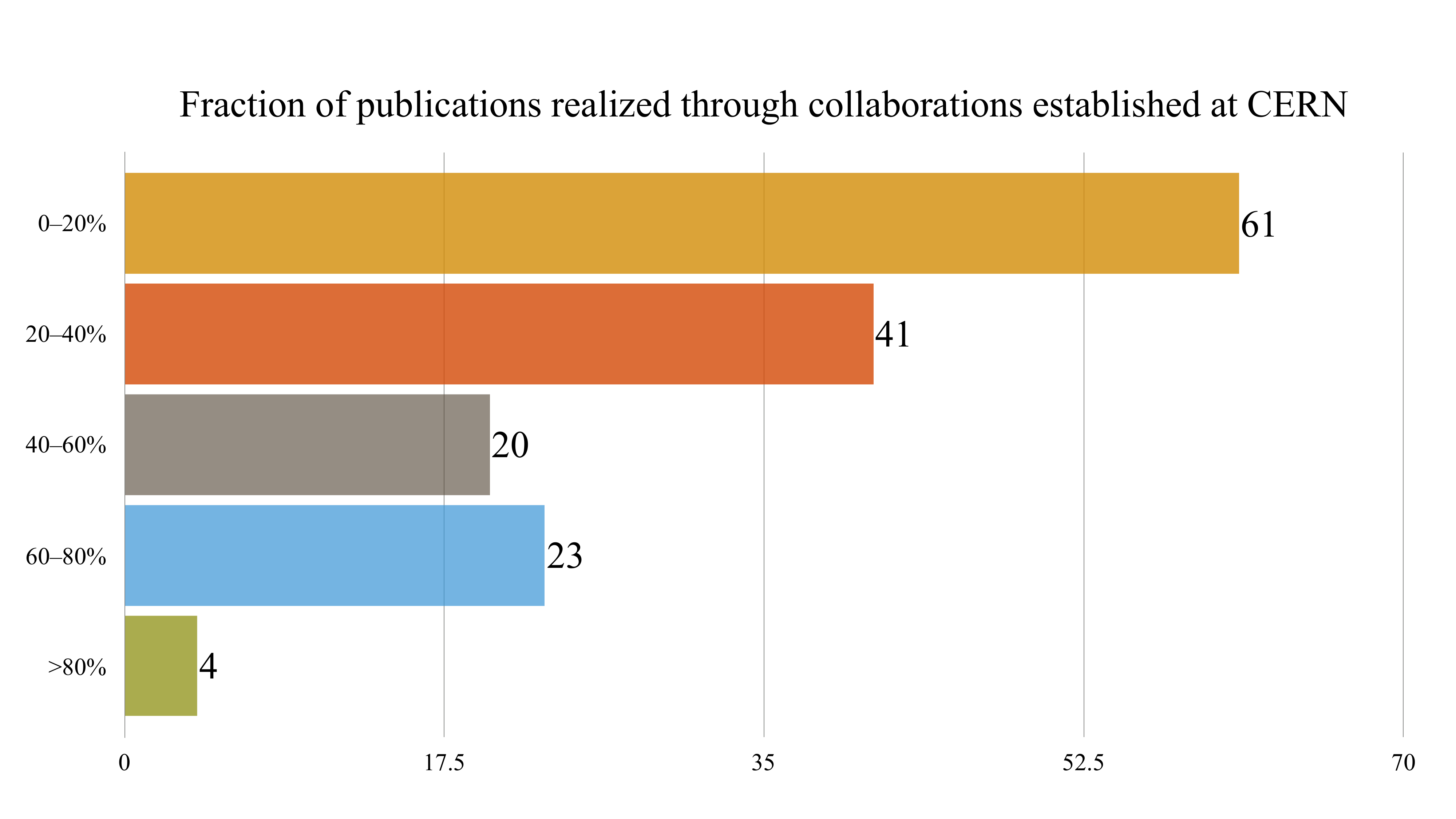}}
\caption{Fraction of publications developed through collaborations started at CERN.}
\label{fig:Theory_publications}
\end{figure}

Additional information from the questionnaire developed for Theorists is reported here. 
The impact on the scientific output was investigated by studying the fraction of publications achieved through collaborations started at CERN. Figure~\ref{fig:Theory_publications} shows that, on average, almost 35-40\% of the publications of the respondents originated from collaborations launched at CERN. 

\begin{figure}[h]
\centering\includegraphics[page=27,width=.9\linewidth]{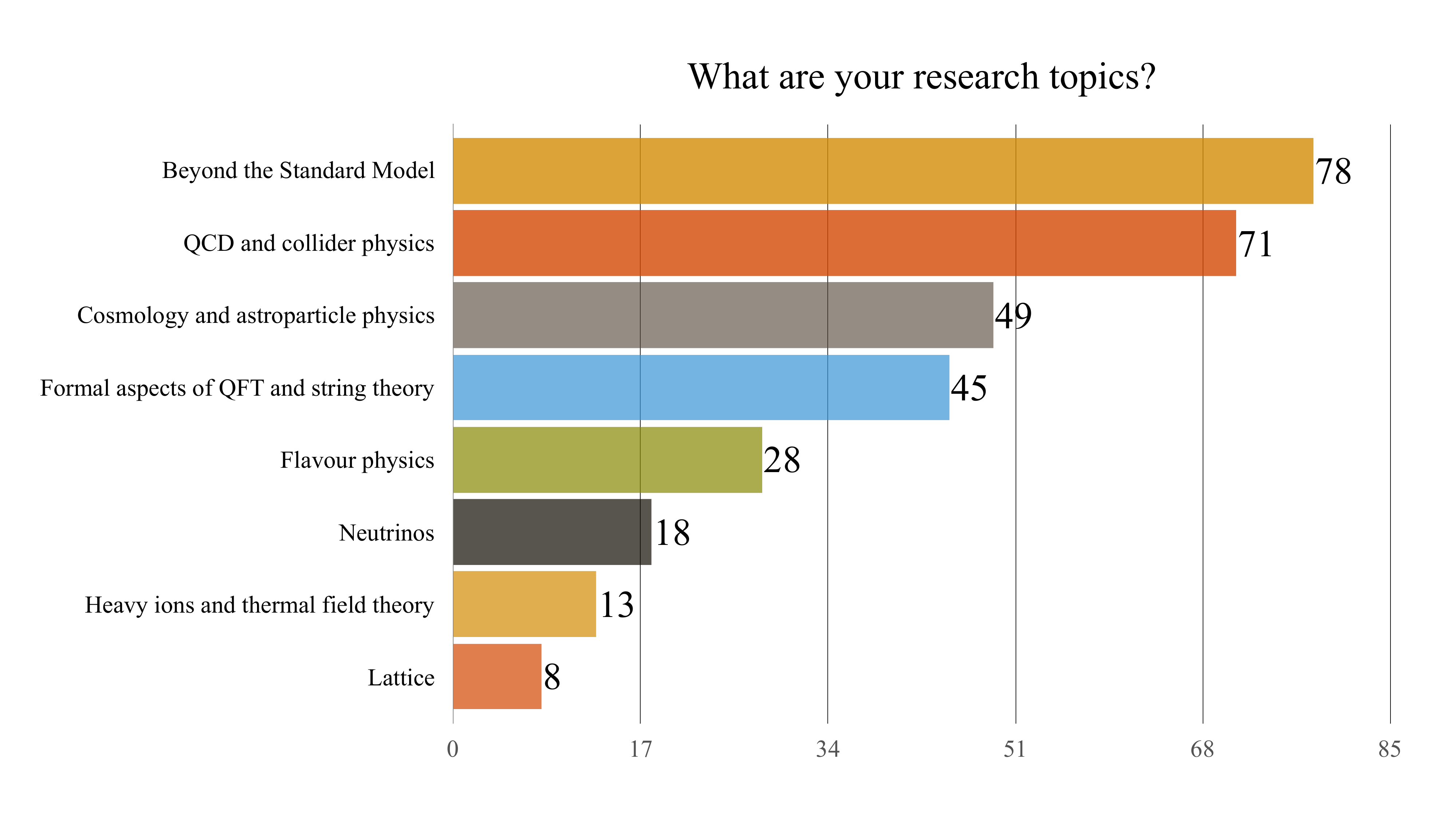}
\caption{Topics of interest of the respondents: several answers were allowed for this question}
\label{fig:Theory_Topics}
\end{figure}

Theory Department users work on various aspects of theoretical high-energy physics, such as phenomenology, physics beyond the Standard Model, cosmology, astroparticle physics, and string and field theories, as shown in Figure~\ref{fig:Theory_Topics}. 

Since theoretical physicists are usually experts in more than one topic, several answers were allowed to the question about their preferred fields of interest. Topics that lie a little outside the experimental environment of CERN, such as neutrinos, astroparticle physics, and cosmology, were covered by a fifth of the answers. Searches for new physics phenomena, developments of heavy flavour and heavy ion physics, closer to the typical studies of the CERN-based experiments, made up about 60\% of the answers. Finally, more formal topics, such as quantum field and string theories, accounted for the last fifth of the answers.

\begin{figure}[h]
\centering\includegraphics[page=32,width=.9\linewidth]{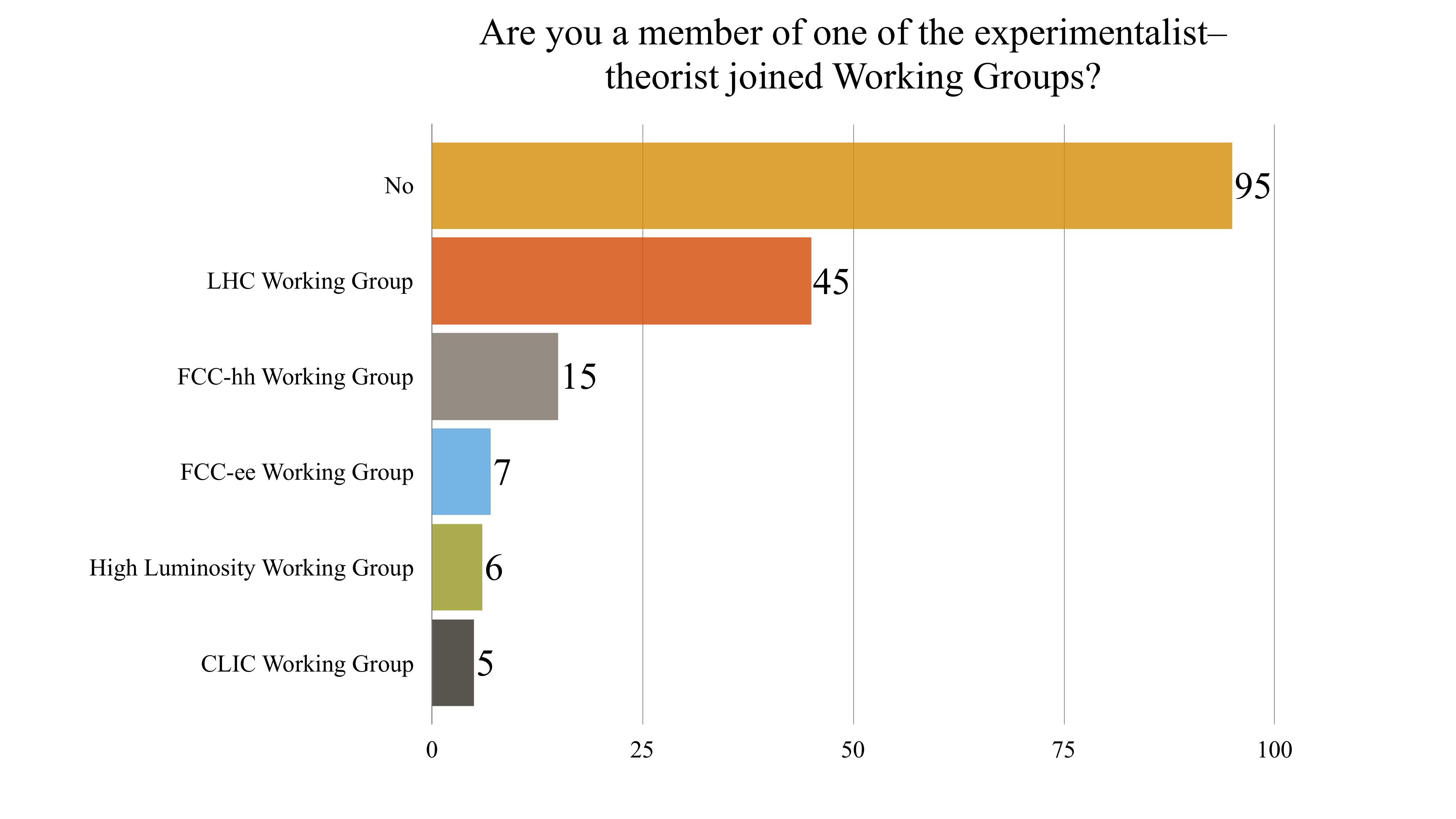}
\caption{Working groups: several answers were allowed.}
\label{fig:Theory_WG}
\end{figure}

\begin{figure}
\centering\includegraphics[page=68,width=.9\linewidth]{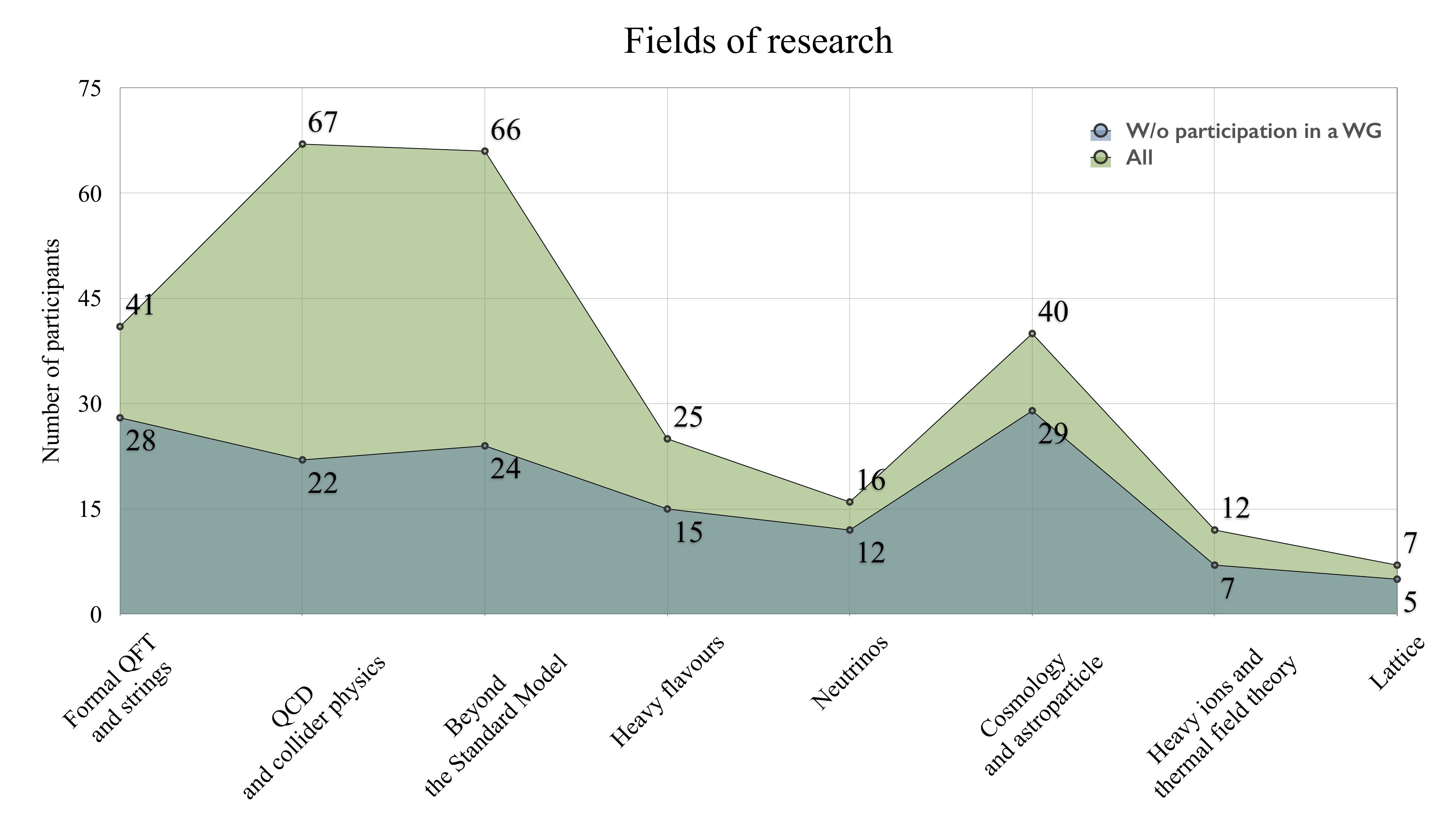}
\caption{Fields of interest for all the theorists compared with the subsample of theorists who do not participate in CERN combined working groups.}
\label{fig:AllvsNoWG}
\end{figure}

Since the start of the LHC, close, systematic and efficient co-operation between experimentalists and theorists had become necessary, in order to identify the most adequate tools to be used by the experiments and also to make faster progress by focusing theoretical efforts on well-defined topics. This experience turned out to be very successful and many such groups have been created through the years in various areas. Among them, several groups work on questions directly related to the LHC experiments (Higgs physics, dark matter, etc.) while others are investigating the future challenges at CERN, related to the \href{http://hilumilhc.web.cern.ch}{HL-LHC}~\cite{hl-lhc} and possible future accelerators such as the \href{http://tlep.web.cern.ch}{FCC-ee}~\cite{fcc,fcc-ee}, \href{https://fcc.web.cern.ch/Pages/fcc-hh.aspx}{FCC-hh}~\cite{fcc,fcc-hh} and \href{http://clic-study.web.cern.ch}{CLIC}~\cite{clic} facilities. 
Figure~\ref{fig:Theory_WG} shows the theorists' participation in these groups. A large fraction of theorists did not contribute to any group. The second, larger community are members of one of the \href{https://lpcc.web.cern.ch/lhc-working-groups}{LHC Working Groups}~\cite{lhc-wg}.
Figure~\ref{fig:AllvsNoWG} shows the fields of interest for all theorists and for those who are not members of any working group. This category contains people mainly involved in research on formal aspects (QFT, strings and lattice), but also in heavy flavour, neutrinos, cosmology and physics beyond the Standard Model .

\end{document}